\documentclass{elsart}
\usepackage{epsfig}
 
\newcommand{\req}[1]{(\ref{#1})}              % numbering equations
\newcommand{\be}{\begin{equation}}
\newcommand{\ee}{\end{equation}}
\newcommand{\Ka}{K$\alpha$ }
\newcommand{\Kb}{K$\beta$ }
\newcommand{\La}{L$\alpha$ }

\begin{document}

%\runauthor{Townsley {\em et al.}}
\begin{frontmatter}
\title{Modeling Charge Transfer Inefficiency in the {\em Chandra} Advanced CCD Imaging Spectrometer}

\author{L.K. Townsley},
\ead{townsley@astro.psu.edu}
\author{P.S. Broos},
\author{J.A. Nousek}, \and
\author{G.P. Garmire}

\address{Department of Astronomy \& Astrophysics, The Pennsylvania State 
University, 525 Davey Lab, University Park, PA 16802\\
http://www.astro.psu.edu}

%==========================================================================

\begin{abstract}
The front-illuminated (FI) CCDs in the Advanced CCD Imaging
Spectrometer (ACIS) on the Chandra X-ray Observatory ({\em Chandra})
were damaged in the extreme environment of the Earth's radiation belts,
causing charge traps that result in enhanced charge transfer
inefficiency (CTI) during parallel readout.  This causes row-dependent
gain, event grade `morphing' (spatial redistribution of charge) and energy resolution degradation.

The ACIS back-illuminated (BI) CCDs also exhibit pronounced CTI due to
their manufacturing.  It is mild enough that position-dependent energy
resolution is not seen, but it is present in both parallel and serial
registers.  This CTI also changes the gain and event grades, in a
spatially complicated way as parallel and serial CTI interact.

Given these realities, we have developed and tuned a phenomenological
model of CTI for both FI and BI CCDs and incorporated it into our Monte
Carlo simulations of the ACIS CCDs.  It models charge loss and the
spatial redistribution of charge (trailing), thus reproducing the
spatially-dependent gain and grade distribution seen in all ACIS CCDs
and the row-dependent energy resolution seen in the FI devices.  Here we
explore the evidence for CTI, compare our simulations to data, and
present a technique for CTI correction based on forward modeling.

\end{abstract}

%==========================================================================

\end{frontmatter}

%==========================================================================
\section{Introduction} \label{sec:intro}
%==========================================================================

This paper is the third in a suite of articles outlining recent efforts
to improve the modeling of CCDs used in the Chandra X-ray Observatory
(hereafter {\em Chandra}).  The Advanced CCD Imaging Spectrometer
(ACIS) instrument employs bulk front-illuminated (FI) CCDs and
back-illuminated (BI) CCDs to give good spectral resolution and good
0.2--10~keV quantum efficiency \cite{bautz98}.  These devices, designed
and manufactured at MIT's Lincoln Laboratories \cite{burke97}, couple
with the {\em Chandra} mirrors to give users the ability to do
spatially-resolved X-ray spectroscopy on spatial scales comparable to
ground-based visual astronomy.

In the first paper in this series \cite{PN99},
Pavlov and Nousek developed a solution to the diffusion equation,
including recombination, to describe the process of charge
diffusion in field-free regions of CCDs.  In the second paper
(\cite{townsley01a}, hereafter T01), Townsley {\em et al.} describe
the latest Pennsylvania State University (PSU) CCD simulator, which
includes the results from Pavlov and Nousek and many recent innovations in CCD
modeling made by our ACIS team colleagues at the Massachusetts
Institute of Technology (MIT) Center for Space Research (CSR)
\cite{prigozhin98a}, \cite{prigozhin98b}, \cite{prigozhin00}.  T01
also gives introductory material such as the geometry of ACIS devices
and the event grading scheme used; readers unfamiliar with the
operation of X-ray CCD cameras should see that paper and references
therein for details.  This paper narrows the focus of T01 to a
calibration complication with the ACIS CCDs -- charge transfer
inefficiency (CTI).

Below we discuss the pronounced CTI effects observed in ACIS flight
data.  CTI is a phenomenon in which some of the charge resulting from
an X-ray photon's interaction with the CCD is lost as that charge is
transferred across the device to the readout nodes.  Thus CTI changes
the gain across the device.  It also causes spatial redistribution of
the charge in the pixel neighborhood of an event, as charge traps
capture electrons and release them on short timescales during the
readout process.  This results in charge being redistributed in
surrounding pixels within the region recognized as a single event.
Charge loss can be substantial (equivalent to the reduction of apparent
X-ray energy of over 100 eV), depending on photon energy and the
event's position with respect to the CCD readout nodes.

Due to the unanticipated forward scattering of charged particles
(probably mainly $\sim$100 keV protons \cite{prigozhin00a},
\cite{prigozhin00b}) by the {\em Chandra} mirrors onto the ACIS CCDs,
the FI devices suffered degraded performance on-orbit, most pronounced
at the top of the devices, near the aimpoint of the imaging array.
This radiation damage created charge traps with a variety of time
constants that increased the parallel CTI of these devices markedly,
resulting in the equivalent of several years of damage in only a few
weeks of operation.

As a result of their manufacturing process, the ACIS BI devices have
always shown modest serial and parallel CTI effects, but their geometry
has protected them from severe radiation damage on-orbit so they show
no additional CTI.  Although operational changes have halted the rapid
degradation of the FI devices, the space environment is likely to
increase CTI effects slowly on all chips throughout the life of the
mission.  Prior to launch we began development of a CTI model for BI
devices and a data correction process to remove CTI effects.  We have
extended this model and corrector to address both BI and FI CTI effects
at two operating temperatures.

To ensure that no water ice could collect on {\em Chandra}'s
instruments, the focal plane temperature of the Observatory was kept at
the relatively warm setting of $-$110C from mid-September 1999 to the end
of January 2000.  Thereafter, it was determined that the Observatory
structures had outgassed enough that it was safe to lower the focal
plane temperature to $-$120C.  At this new colder temperature, the trap
population that causes the row-dependent energy resolution in FI
devices is partly suppressed \cite{prigozhin00b}, \cite{gallagher98}.
This step greatly improved the performance of the FI devices by
ameliorating the effects of CTI.  The BI CTI is very slightly worse at
$-$120C than at $-$110C but the improvements in the FI performance
completely justify operating at $-$120C.

Given the existing and potential complication CTI causes for ACIS
calibration, we have developed a model for CTI and incorporated it into
our Monte Carlo CCD simulator.  CTI is parameterized as a function of
the charge in each pixel, including the effects of de-trapping (charge
trailing), shielding within an event (charge in the leading pixels of
the 3$\times$3 pixel event island protect the rest of the island by
filling traps), and non-uniform spatial distribution of traps.  This
technique partially recovers the degraded energy resolution near the
top of the FI chips and it reduces the position dependence of gain and
grade distributions.  By correcting the grade distributions as well as
the event amplitudes, we can improve the response of the FI chips, both
in quantum efficiency at high energies and in background rejection.

First we will illustrate the effects of CTI on ACIS performance.  Our
CTI model is described in Section~\ref{sec:model}; simulations showing
its ability to reproduce CTI are then presented.  Next we describe the
program that we have developed to partially remove CTI effects from
ACIS data and examine the performance of this ``CTI corrector.'' A
preliminary version of the CTI model and corrector was described by
Townsley {\em et al.} \cite{townsley00}.

%==========================================================================
\section{Illustrating CTI with ACIS Data} \label{sec:data}
%==========================================================================

Due to crystal defects caused in the manufacturing process or by
radiation damage, silicon-based detectors such as CCDs can have charge
traps, where charge is inefficiently transferred.  There is a large
body of work on the reasons for and results of charge traps in
silicon:  see Janesick {\em et al.} \cite{janesick91} for more details
on bulk and radiation-induced traps in X-ray astronomy CCDs; see
Stetson \cite{stetson98} for an illustration of CTI effects on CCDs
used for visual astronomy.  For our purposes, we can think of charge
traps simply as potential wells that trap charge on a timescale short
compared to the timescale for charge transfer from one pixel to the
next.  The traps then release charge on some exponential timescale that
may be comparable to the charge transfer time (see Gendreau {\em et
al.} \cite{gendreau95} for a mathematical development of these
concepts).  Prigozhin {\em et al.} have shown that the charge traps in
ACIS FI devices consist of multiple trap species characterized by
different decay time constants \cite{prigozhin00b}.

Charge release from traps with short time constants results in a
phenomenon that we call ``charge trailing'' -- charge in the final
detected event appears shifted from its native pixel into the adjacent
pixel farther from the readout node.  Since events are graded based on
the distribution of charge in a $3 \times 3$ pixel neighborhood around the
brightest pixel (see Figure 2 in T01), charge trailing causes events to
be recognized with different grades than their original charge
distribution would have generated -- we call this phenomenon ``grade
morphing.''  Only part of the trapped charge appears trailed into the
adjacent pixel(s); the rest of it is lost to the event because the
charge in that event was transferred away from the interaction site
before the other traps released all of their captured charge.

Our data show that the charge loss and trailing vary with photon energy
in a complicated way.  This is due to the nature of the traps convolved
with the size of the charge clouds.  In the BI device, it is also the
result of the device geometry.  For a BI device, low-energy photons can
create large-diameter charge clouds, since many of those photons
interact near the surface of the device and suffer substantial charge
diffusion as the charge cloud transits the depletion region.
High-energy photons create large-diameter initial charge clouds, which
also can suffer substantial diffusion, depending on the interaction
depth.  As charge clouds get bigger, they spread over more pixels, so
they can encounter more charge traps as they are transferred across the
device.

%====================================
\subsection{On-orbit Calibration Data}

During radiation belt passage, the {\em Chandra} Science Instrument
Module is configured so that ACIS is out of the telescope's field of
view.  This protects ACIS from further radiation damage and orients the
instrument under its calibration source \cite{pog} (called the
``External Calibration Source'' or ECS here).  As a way to monitor the
CTI, ECS data are collected and telemetered every orbit.  We have
combined these ECS data to calibrate our model for CTI and to check the
fidelity of our CTI corrector that is based on that model.

The ECS spectrum is well-understood and was described in detail by the
MIT/ACIS team in the ACIS Calibration Report \cite{calreport}.  Using
their line identifications, Table~\ref{table:lineids} shows the main
spectral features apparent in our composite ECS dataset and notes the ones
used in this analysis.

\protect \footnotesize
%-------------------------------------------------------------------------
\begin{table}[htb] \centering
\begin{tabular}{||l|r|c|l||} \hline
 
Spectral Line  & Energy (eV) & Used here? & Comments \\ \hline \hline
 
Mn, Fe L complex & $\sim$680 	& FI, BI &	treated as single line	\\
Al \Ka		&	1486	& FI, BI &	major calibration line	\\
Si \Ka		&	1740	& BI &	blurred with Al \Ka in FI devices\\
Au M complex	& $\sim$2112	& FI &		treated as single line \\
Ti \Ka escape	&	2771    & FI, BI &				\\
Ti \Kb escape	&	3192    &   &		faint			\\
Ca \Ka ?	&	3690	&   &   	faint, line ID uncertain\\
Mn \Ka escape	&	4155	& FI ($-$120C), BI &	blurred with Ti \Ka in FI devices at $-$110C\\
Ti \Ka		&	4511	& FI, BI &	major calibration line	\\
Ti \Kb		&	4932	& FI, BI &				\\
Mn \Ka		&	5895	& FI, BI &	major calibration line	\\
Mn \Kb		&	6490	& FI, BI &				\\
Mn \Ka + Al \Ka pile-up &  7381	&   &   	faint, blurred with Mn \Kb \\
Ni \Ka		&	7470	&   &		faint			\\
Au \La		&	9711	& FI, BI &   	not used in trailing model\\
Mn \Ka + Ti \Ka pile-up & 10406	&   &		faint		\\ \hline
 
\end{tabular}
\caption{\protect \footnotesize Spectral features discernable in our ACIS
External Calibration Source.  Energies are from the ACIS Calibration
Report \cite{calreport} and Bearden \cite{bearden67}.  The third column notes whether this
feature was employed in the charge loss and trailing calculations that
form the fundamental CTI model described below.}

\label{table:lineids}
\end{table}
%-------------------------------------------------------------------------
\normalsize

In order to achieve the highest fidelity in our model calibration, we
have combined all available data (as of early 2001) from the ACIS ECS
at a given focal plane temperature to generate a corrector appropriate
for any dataset taken at that temperature.  We have checked that the
CTI is stable enough over the time periods in question that this
approach is sound and that the resulting corrector performs acceptably
on all ECS datasets in the sample.

\begin{figure}[htb]
\centerline{\mbox{
         \epsfig{file=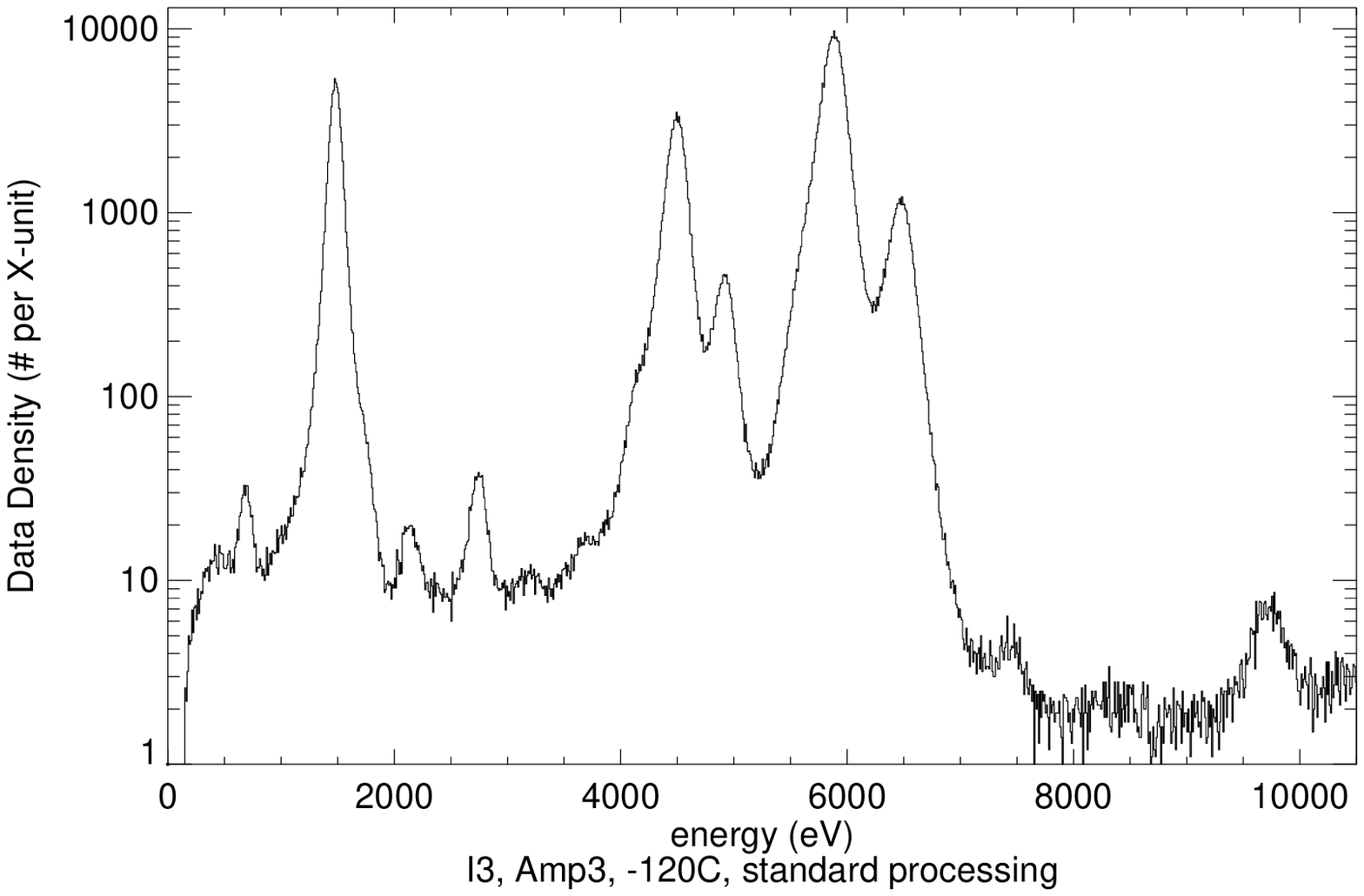,width=3.0in }
         \hspace{0.25in}
         \epsfig{file=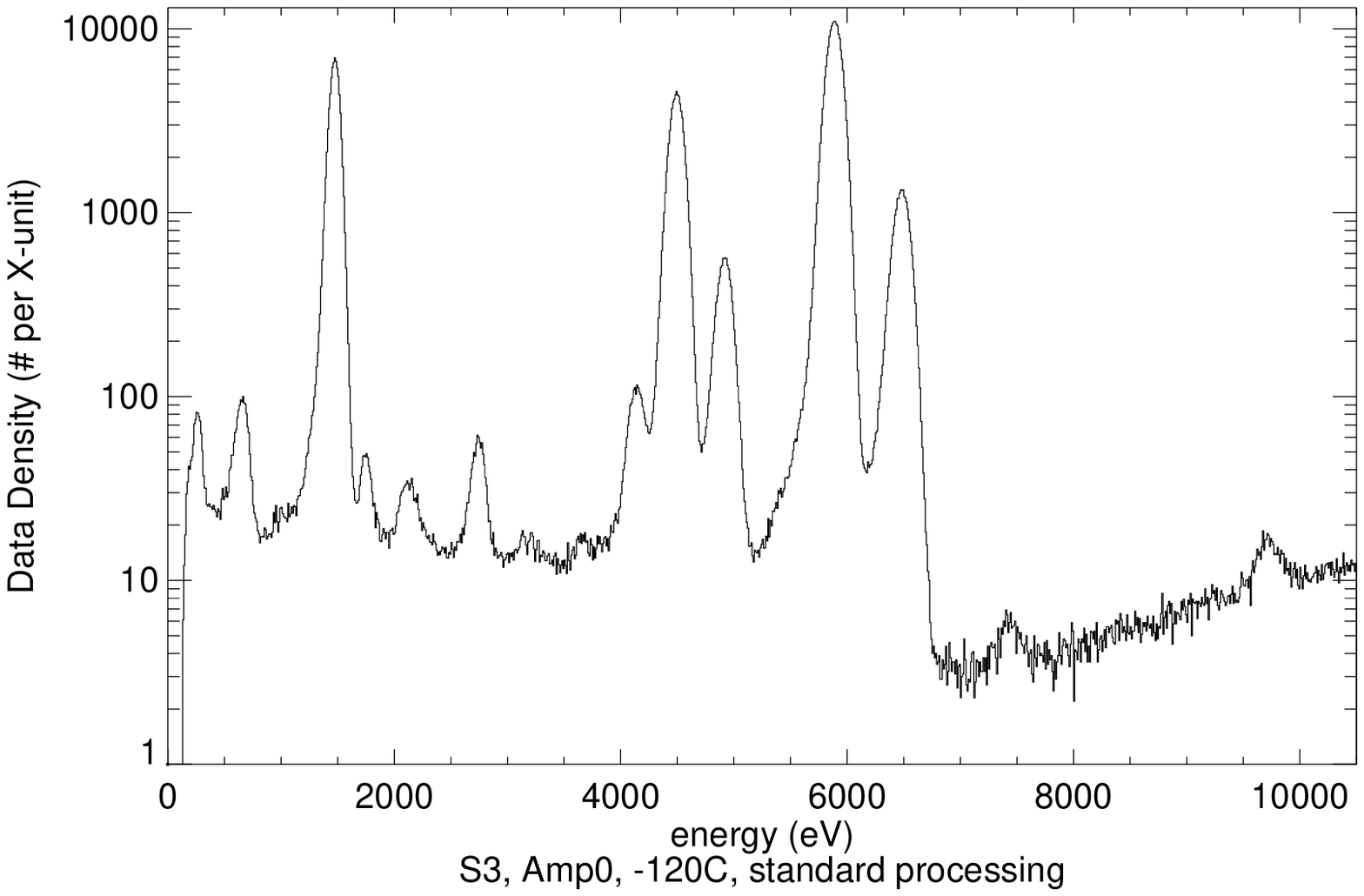,width=3.0in }}}
\caption{\protect \small ECS
spectra, made by combining over 100 observations to give $\sim 4 \times 10^6$ events.  These spectra use standard ASCA g02346 grade selection
and standard data processing.  Left:  all rows of I3, Amplifier 3, at $-$120C.  
Right: all rows of S3, Amplifier 0, at $-$120C.} 

%yrange (log) 1:13000, xrange 0:10500.  Aspect ratio 0.7.  Left margin 9.
 
\normalsize
\label{fig:stanspec}
\end{figure}

Example spectra for Amplifier 3 (Node D) on the FI device I3 and 
Amplifier 0 (Node A) on the BI
device S3, using a large sample of the appropriate ECS data taken at
$-$120C and all rows of the devices, are shown in
Figure~\ref{fig:stanspec}.  These spectra reflect the work of the
{\em Chandra} X-ray Center (CXC) to remove the largest effect of CTI
(illustrated in Figures~\ref{fig:allenergy-cti-fi} and
\ref{fig:allenergy-cti-bi} below), the row-dependent gain.  They
represent the standard processing applied to all ACIS data and form the
fundamental comparitors for our CTI corrector's performance.

%-------------------------------------------------------------------------  
%-------------------------------------------------------------------------

The data were filtered to keep only events with certain ``standard''
grades.  The ACIS event grading scheme can be related to that of the
Japanese X-ray satellite ASCA \cite{calreport}, which is familiar to
X-ray astronomers.  The standard ACIS grade filtering scheme includes
ASCA-like Grades 0, 2, 3, 4, and 6 (``g02346'', \cite{yamashita}) and
was developed (in the absence of CTI) to maximize both sensitivity and
spectral resolution (see the ``{\em Chandra} Proposers' Observatory Guide,''
\cite{pog}).  Virtually all standard data products employ this grade filter.

%====================================
\subsection{FI CTI}

\begin{figure}[htb]
\centerline{\mbox{
         \epsfig{file=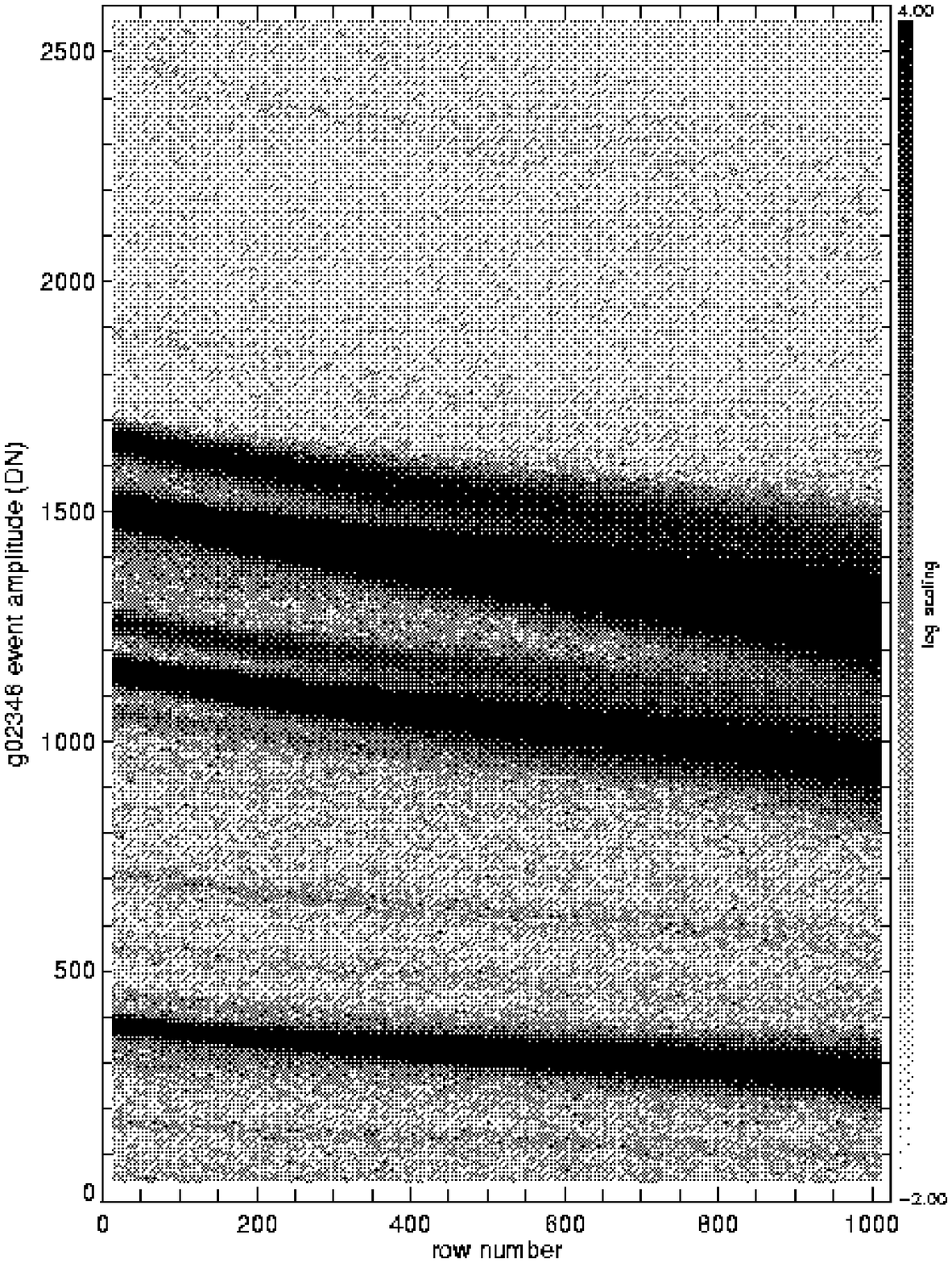,height=4.0in }
         \hspace{0.5in}
         \epsfig{file=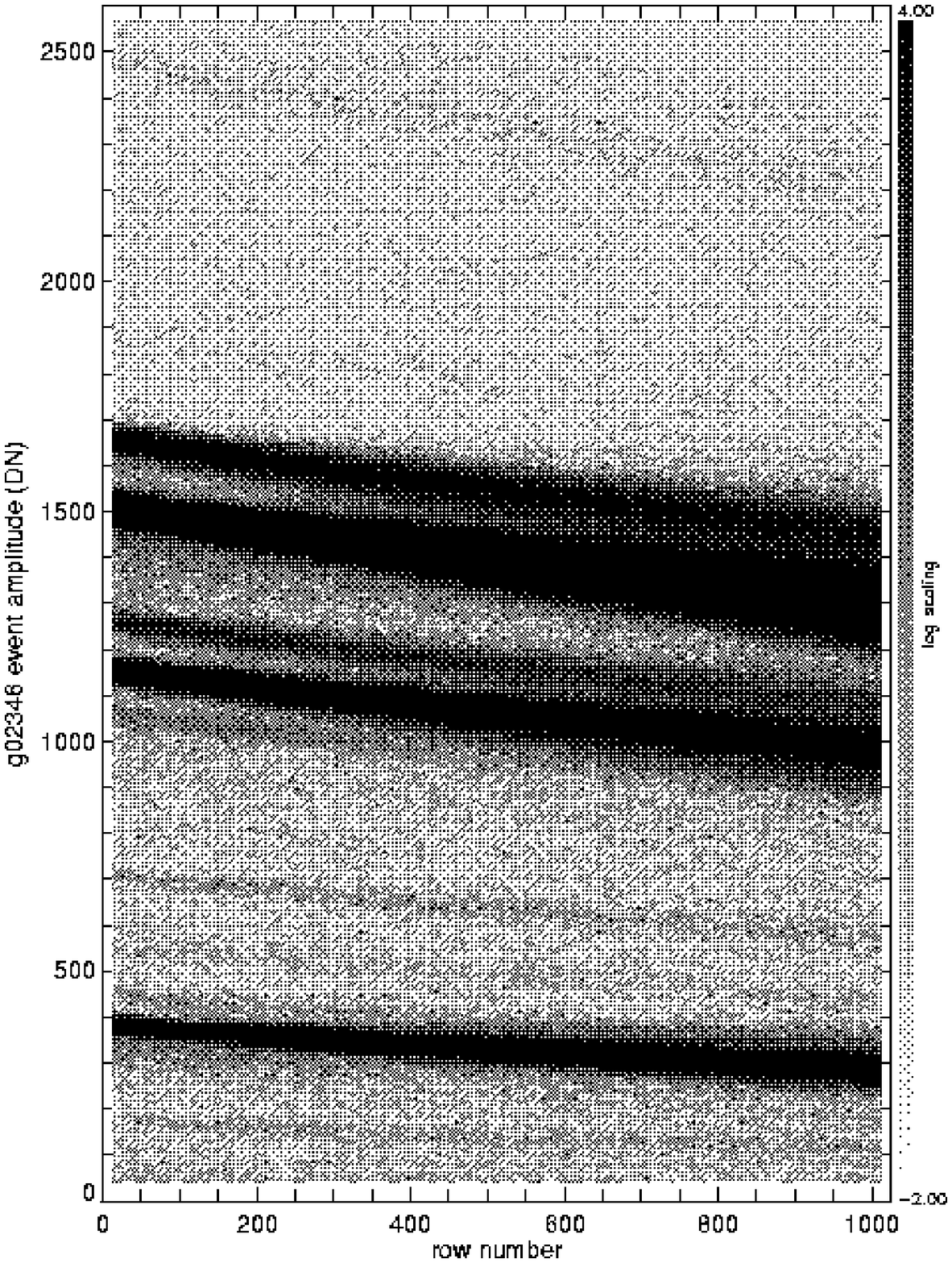,height=4.0in }}}
\centerline{\mbox{
         \epsfig{file=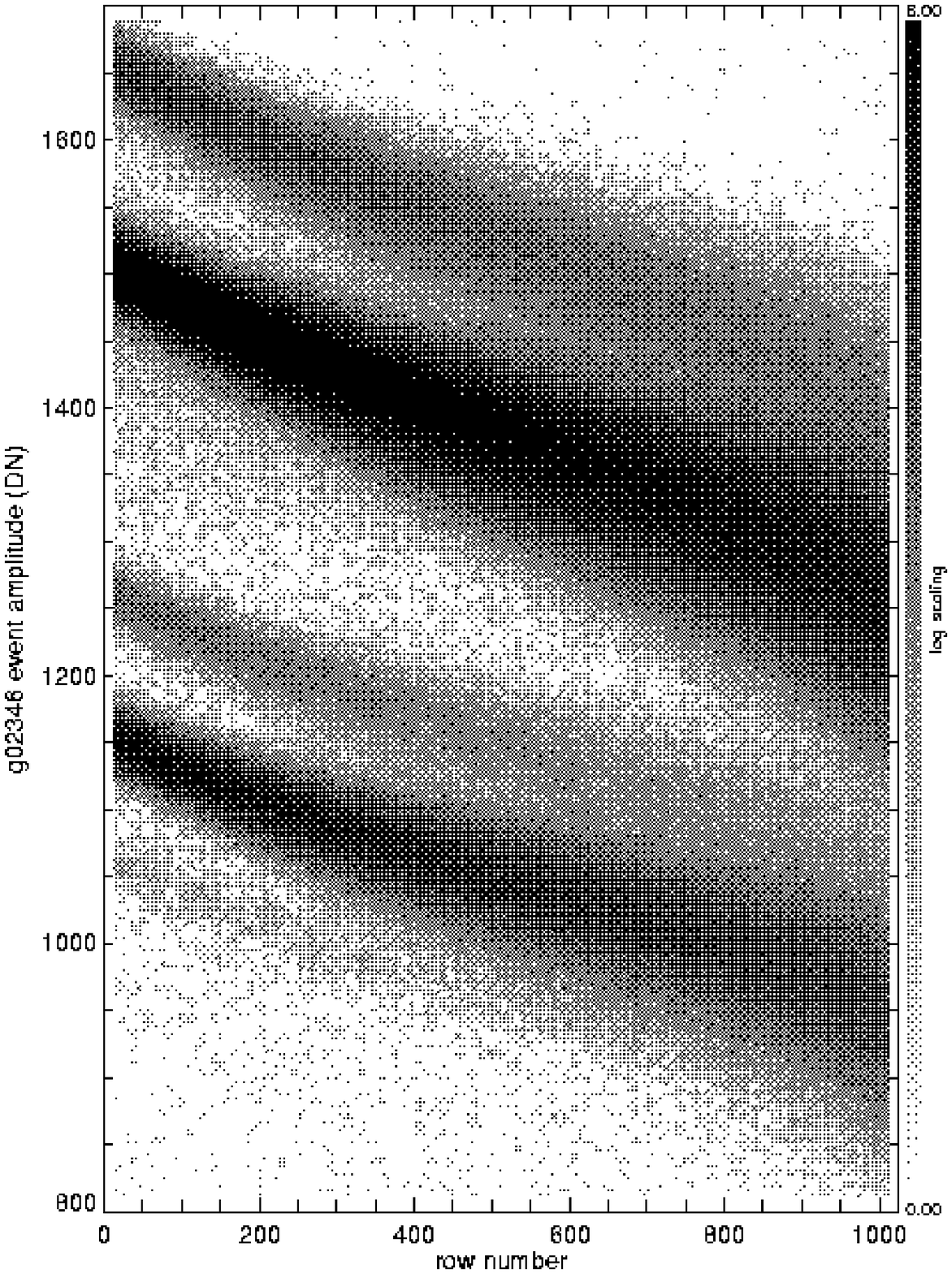,height=4.0in }
         \hspace{0.5in}
         \epsfig{file=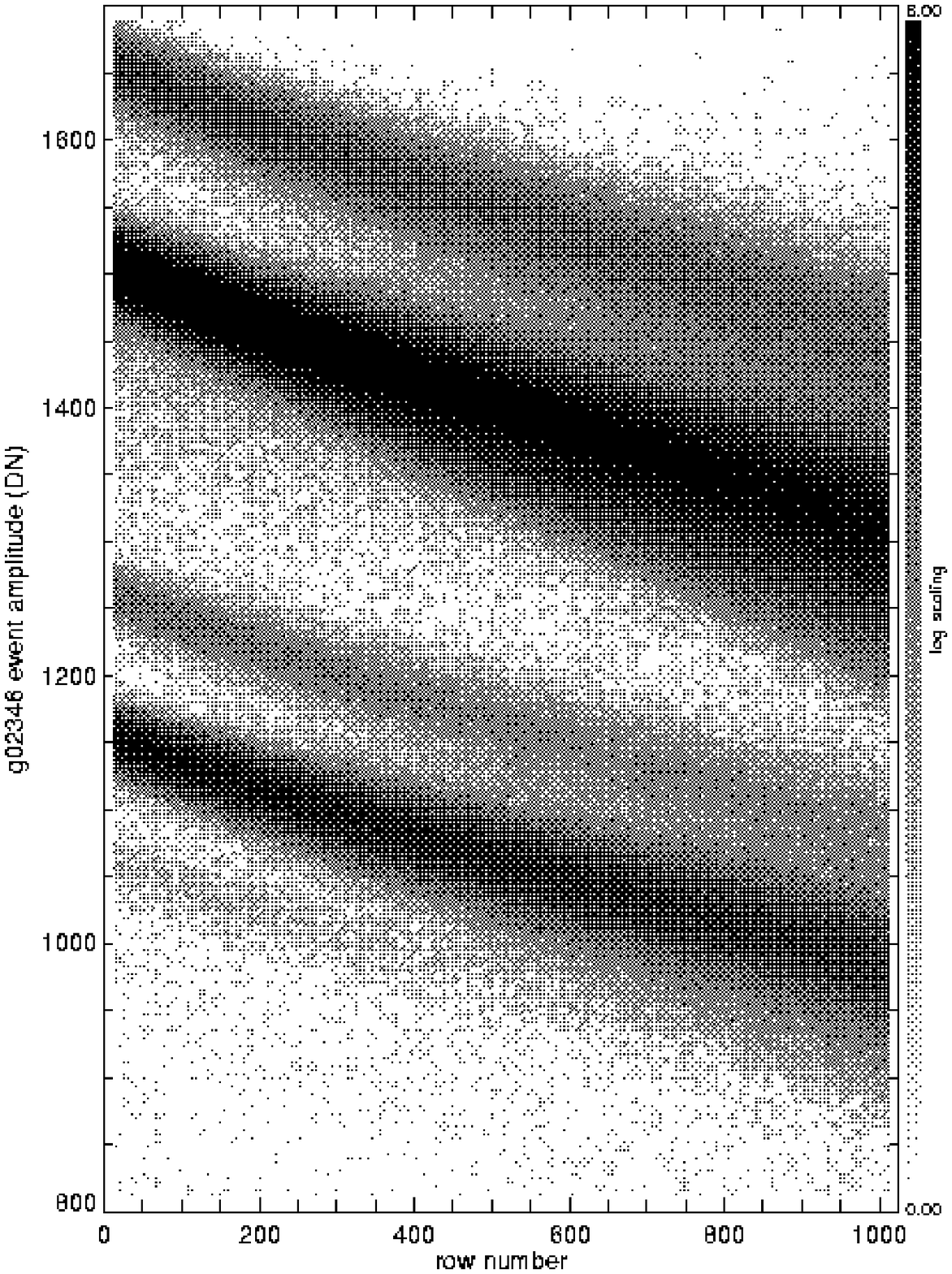,height=4.0in }}}
\caption{\protect \small Parallel CTI in the FI chip I3, Amplifier 3,
illustrated with data from the ECS.  Standard
grades (``g02346'') are included.  Note the row-dependent gain and
energy resolution and the energy-dependent charge loss.  Left panels:
FI CTI at a focal plane temperature of $-$110C.  The upper panel shows
the full energy range; note the faint but useful Au L$\alpha$ line at
$\sim$2500 DN.  The lower panel is an expanded view of the Ti and Mn
calibration lines.  Right panels:  the same plots for a focal plane
temperature of $-$120C.  Both $-$110C and $-$120C datasets contain about
$1.8 \times 10^6$ events to facilitate comparison. }

% Aspect ratio for these figures is 1.4.
% Binsize used for the image was 8x4.  Y-range 0-2600 for all, 800-1700 
% for zoom.

\normalsize
\label{fig:allenergy-cti-fi}
\end{figure}

Figure~\ref{fig:allenergy-cti-fi} illustrates the effects of CTI on the
ACIS I3 chip at focal plane
temperatures of $-$110C (left) and $-$120C (right) by showing the
row-dependent charge loss and energy resolution for the spectral lines
in the ECS.  The events are binned by position and energy using the
Event Browser software \cite{tara00} into an ``image'' to make the
features easier to see.

%-------------------------------------------------------------------------
%-------------------------------------------------------------------------

Note that the charge loss gets more severe at higher event energies
(the loci in Figure~\ref{fig:allenergy-cti-fi} steepen with line
energy, or larger DN values).  The data are displayed using the digital
value (``Data Number,'' DN) of the event amplitude because the standard
data processing pipeline applies a correction factor to the event
energy in eV (a quantity derived from the event amplitude) as a
first-order correction for CTI.  Since we wish to illustrate this
charge loss effect here, we display the data in their native units.
For a rough approximation of a line's event amplitude in DN, divide its
energy in eV by 4.  We use I3 as the example FI CCD because it contains
the aimpoint for the ACIS imaging array.  The other I3 amplifiers and
other FI devices show similar charge loss effects.

The left-hand panels of Figure~\ref{fig:allenergy-cti-fi} are
representative of the state of the CTI at $-$110C and apply to data
gathered in the first five months of the mission.  The improvement
gained by operating at $-$120C is apparent in the right-hand panels of
Figure~\ref{fig:allenergy-cti-fi} which show less line blurring at
large row numbers.

The lower panels of Figure~\ref{fig:allenergy-cti-fi} show expanded
views of the Mn and Ti lines.  Note the non-linearity of the charge
loss with row number that lends a slight S-shape to the curve:  the
slope is flatter in the middle third of the CCD than at low and high
row numbers.  Here the increase in linewidth toward the top of the chip
(high row number) is easily seen.

\clearpage

\begin{figure}[htb]
\centerline{\epsfig{file=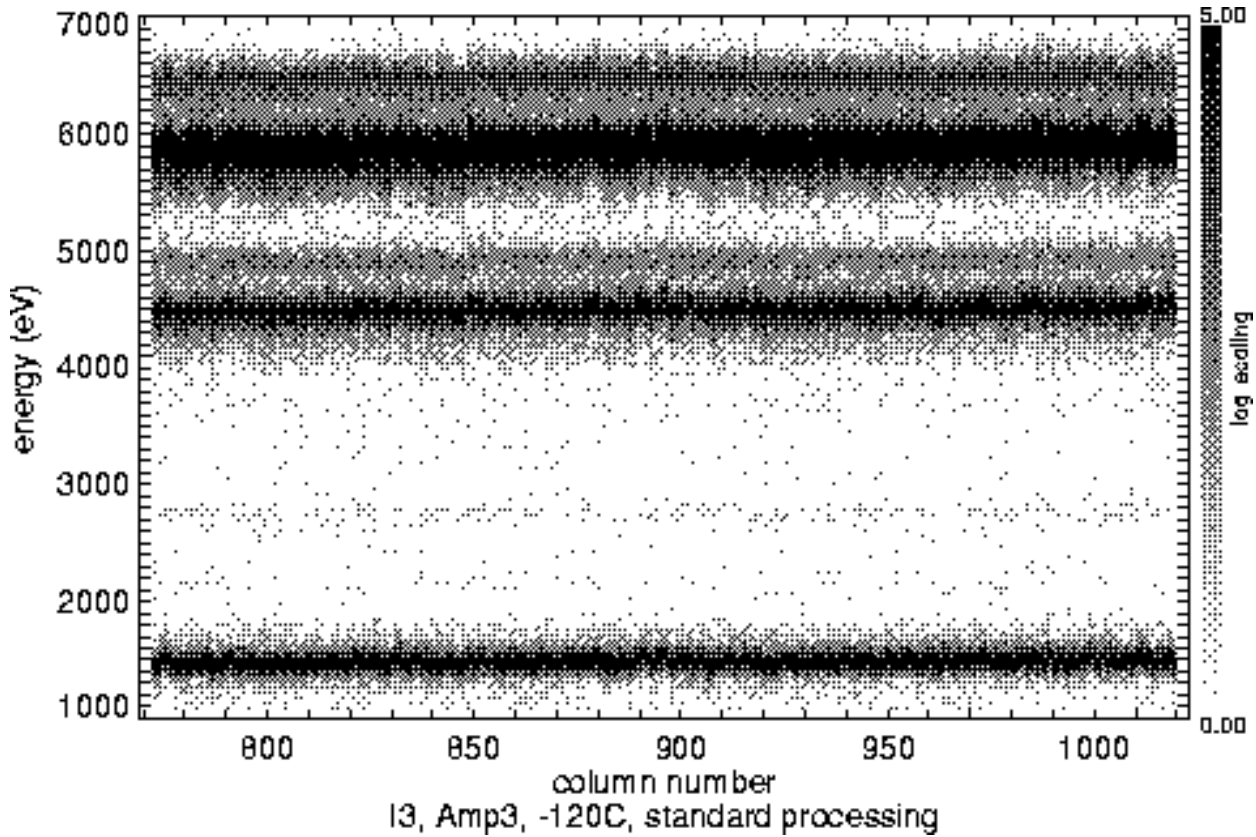, width=6.5in }}
\caption{\protect \small Image showing energy vs.\ column number at the
main calibration energies in the ECS, averaged over all rows on
I3, Amplifier 3 at $-$120C.  This illustrates the remaining
non-uniformities after standard data processing to remove the
row-dependent gain variations caused by CTI.}

% Aspect ratio for this figure is 0.7.  Binsize 1x10.  Used all g02346 events,
% about 4e6.
 
\normalsize
\label{fig:xgainvars}
\end{figure}

Due to the non-uniform spatial distribution of traps, ACIS CCDs also
exhibit column-to-column gain variations that increase in amplitude
with energy and row number.  These variations contribute substantially
to the spectral resolution degradation.  Figure~\ref{fig:xgainvars}
illustrates these variations using ECS data from Amplifier 3 of the I3
chip.  The data are displayed in energy space now, using the standard
processing techniques to remove the gross row-dependent gain variations
so that the column-wise variations are more apparent. 
	
%-------------------------------------------------------------------------  
%-------------------------------------------------------------------------

%====================================	
\subsection{BI CTI}

Although CTI in ACIS BI devices is caused by a different mechanism than in
the FI devices, it causes similar non-uniformities in the data.  The two
major differences are that BI devices show CTI in the serial as well as
the parallel registers and the framestore is also affected.  This means
that parallel CTI acts over 2048 parallel transfers in BI devices rather
than just 1024 parallel transfers as in the FI devices.  The amplitude
of serial CTI is much larger than parallel CTI in the BI devices, but
there are at most 256 serial transfers.  Thus both parallel and serial
CTI must be modeled in BI devices and the interaction of these two effects
results in spatially-complex charge loss and grade morphing behavior.

\begin{figure}[htb]
\centerline{\mbox{
         \epsfig{file=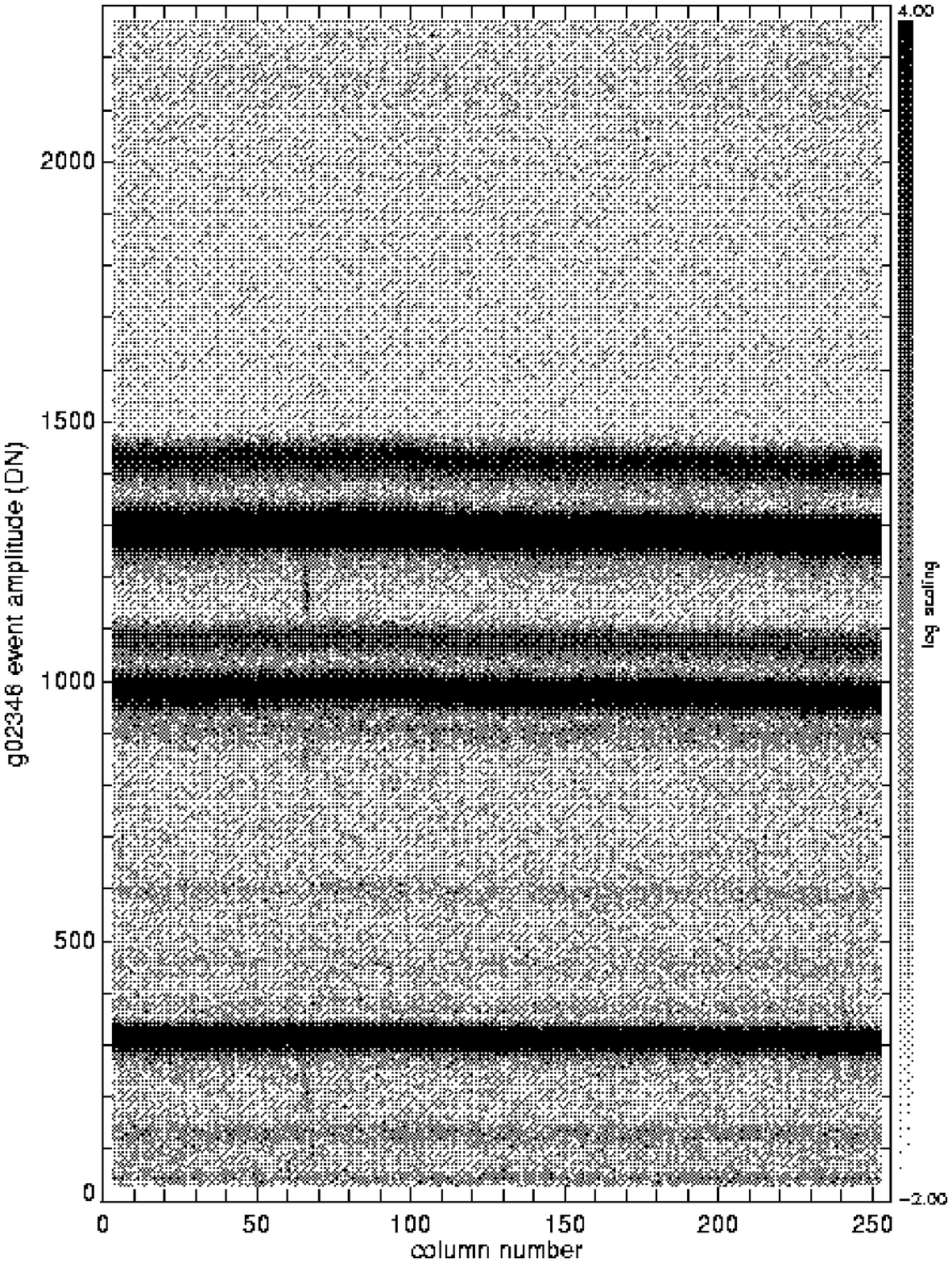,height=4.0in }
         \hspace{0.5in}
         \epsfig{file=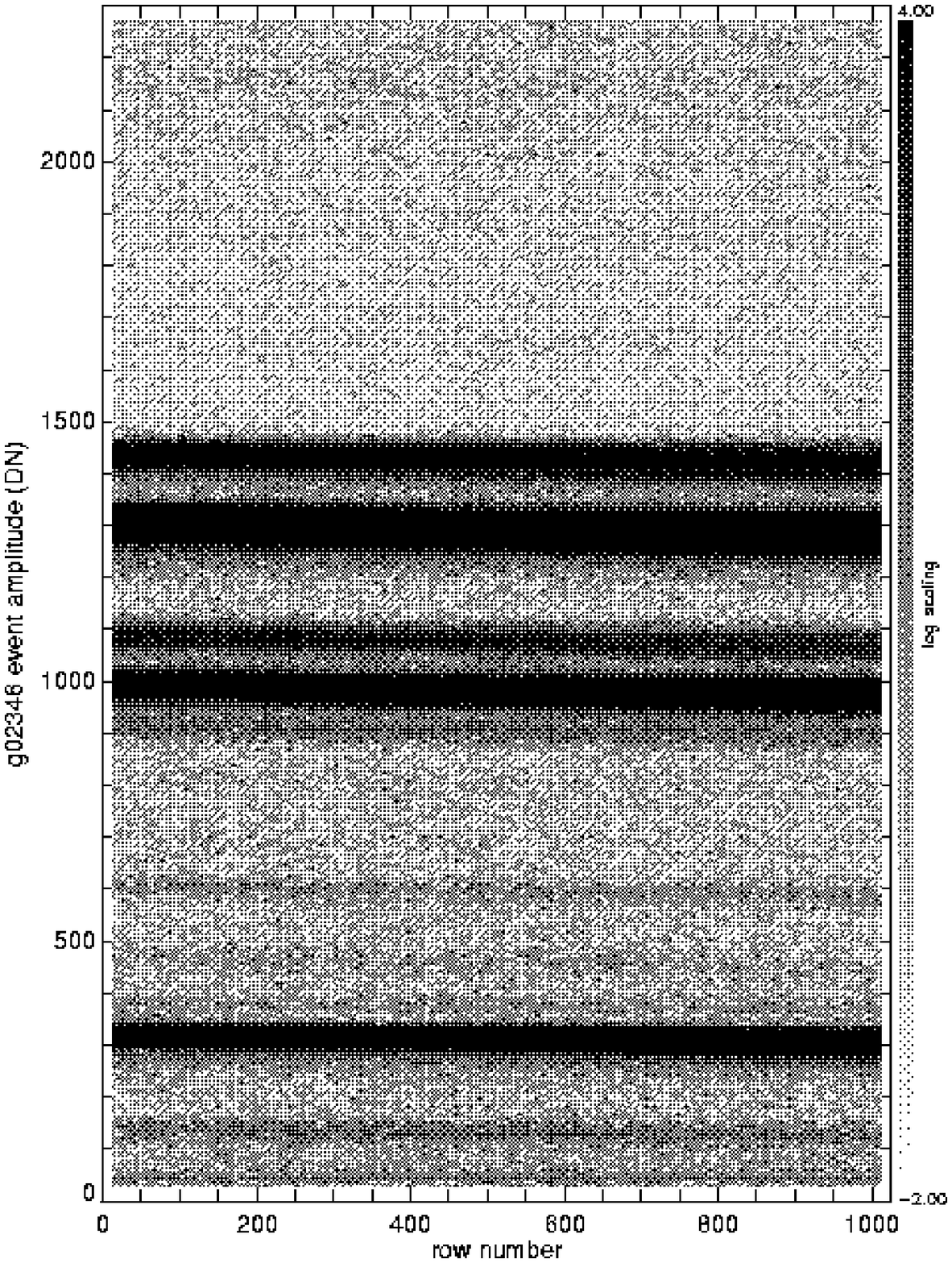,height=4.0in }}}
\centerline{\mbox{
         \epsfig{file=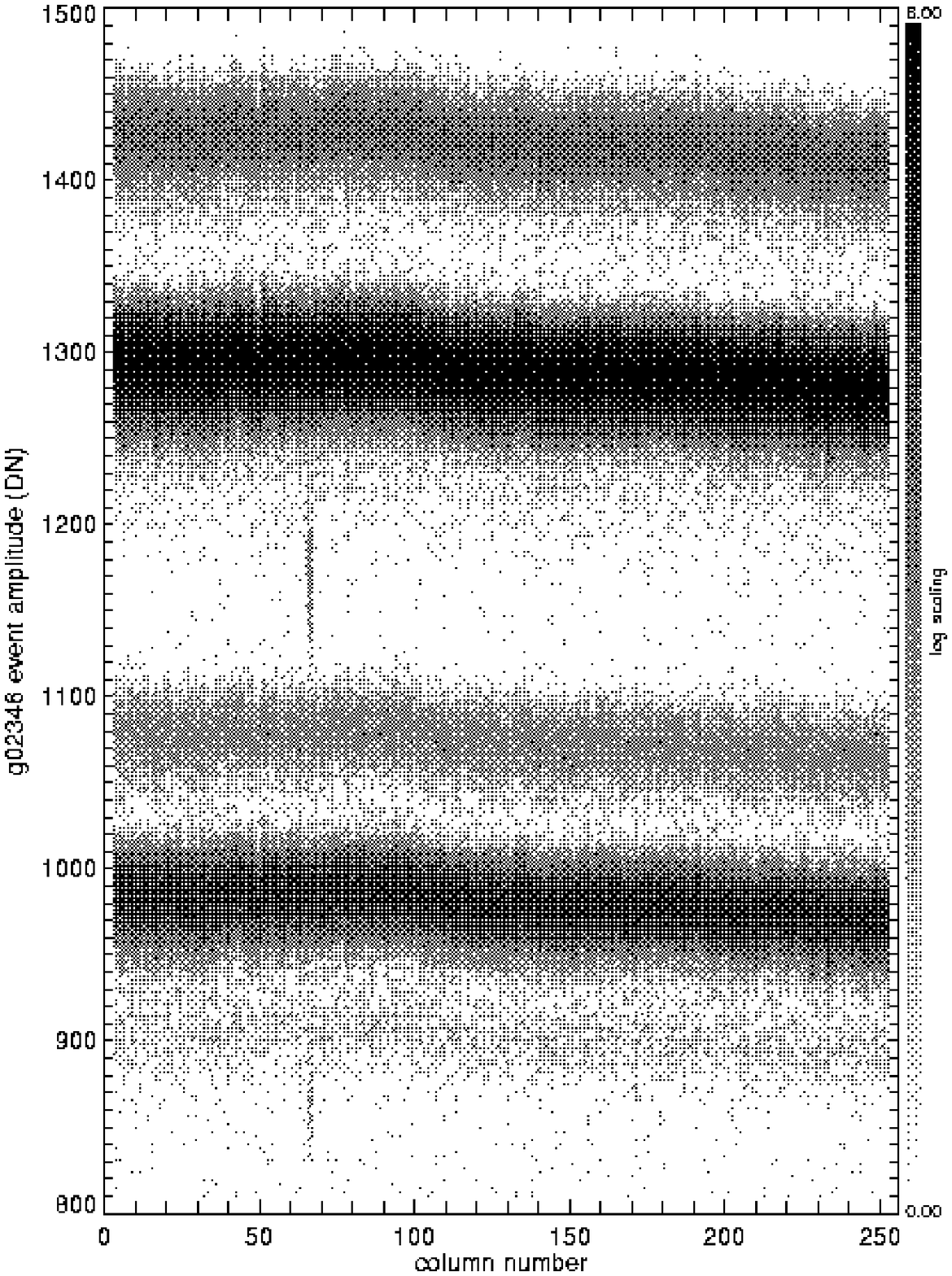,height=4.0in }
         \hspace{0.5in}
         \epsfig{file=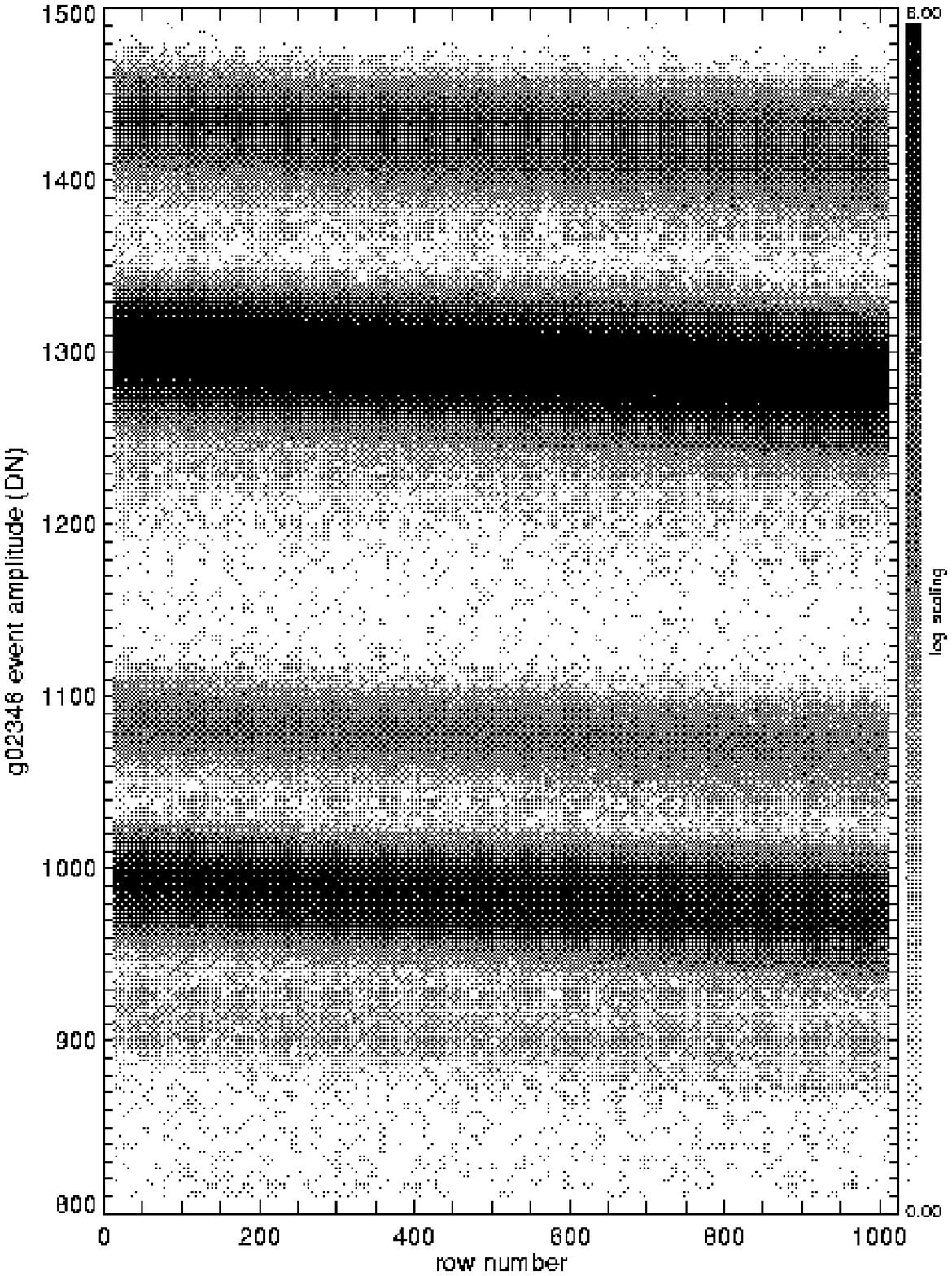,height=4.0in }}}
\caption{\protect \small Serial and parallel CTI at a focal plane
temperature of $-$120C in the BI chip S3, Amplifier 0, illustrated with
data from the ECS.  Standard grades (``g02346'') are included.  Note
the row-dependent gain and energy resolution and the energy-dependent
charge loss.  Left panels:  serial CTI.  The upper panel shows the full
energy range; note the faint but useful Au L$\alpha$ line at $\sim$2150
DN.  The lower panel is an expanded view of the Ti and Mn calibration
lines.  Right panels:  similar plots for parallel CTI.  Here the serial
CTI was suppressed by filtering spatially to keep only events in the 64
columns closest to the readout node.  Both datasets contain about $1.8
\times 10^6$ events to facilitate comparison with each other and with
the FI data in Figure~\ref{fig:allenergy-cti-fi}. }

% Aspect ratio for these figures is 1.4.
% Binsize used was 8x4 for parallel, 1x4 for serial.  Y-range 0-2300 for all,
% 800-1500 for zoom.

\normalsize
\label{fig:allenergy-cti-bi}
\end{figure}

Figure~\ref{fig:allenergy-cti-bi} illustrates CTI in the BI device S3,
using only data from Amplifier 0 at $-$120C.  BI CTI is only slightly
worse at $-$120C than at $-$110C, so only $-$120C data are presented here.
The left-hand columns illustrate serial CTI over the full range of the
ECS data (top) and for the major spectral lines (bottom).  The
amplifier readout is at Column 0.  The right-hand columns show parallel
CTI for the same spectral ranges.  Parallel CTI is illustrated using
only data from the 64 columns closest to the readout node in order to
minimize the effects of serial CTI.  As in
Figure~\ref{fig:allenergy-cti-fi}, the data are displayed in their
native units (DN) and only standard grades are used. 

%-------------------------------------------------------------------------
%-------------------------------------------------------------------------

Again due to the non-uniform spatial distribution of charge traps, BI
CCDs show column-to-column gain variations that increase in amplitude
with energy and degrade the spectral resolution.  These variations are
visible in the left-hand panels of Figure~\ref{fig:allenergy-cti-bi}.
Also apparent there are charge loss variations of lower spatial
frequency.

\clearpage

%====================================
\subsection{Grade Morphing}

\begin{figure}[htb]
\centerline{\mbox{
         \epsfig{file=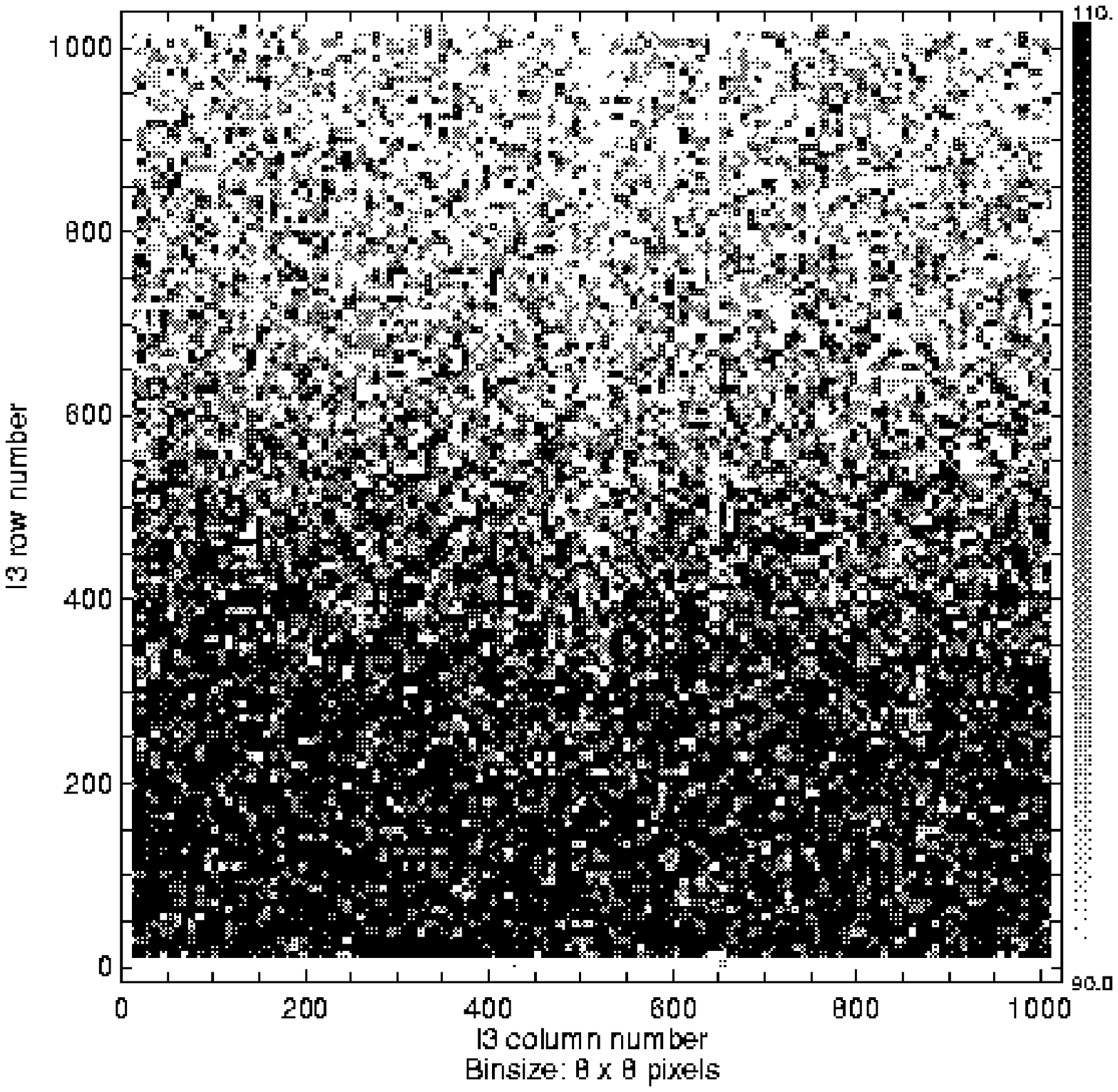,width=3.0in }
         \hspace{0.5in}
         \epsfig{file=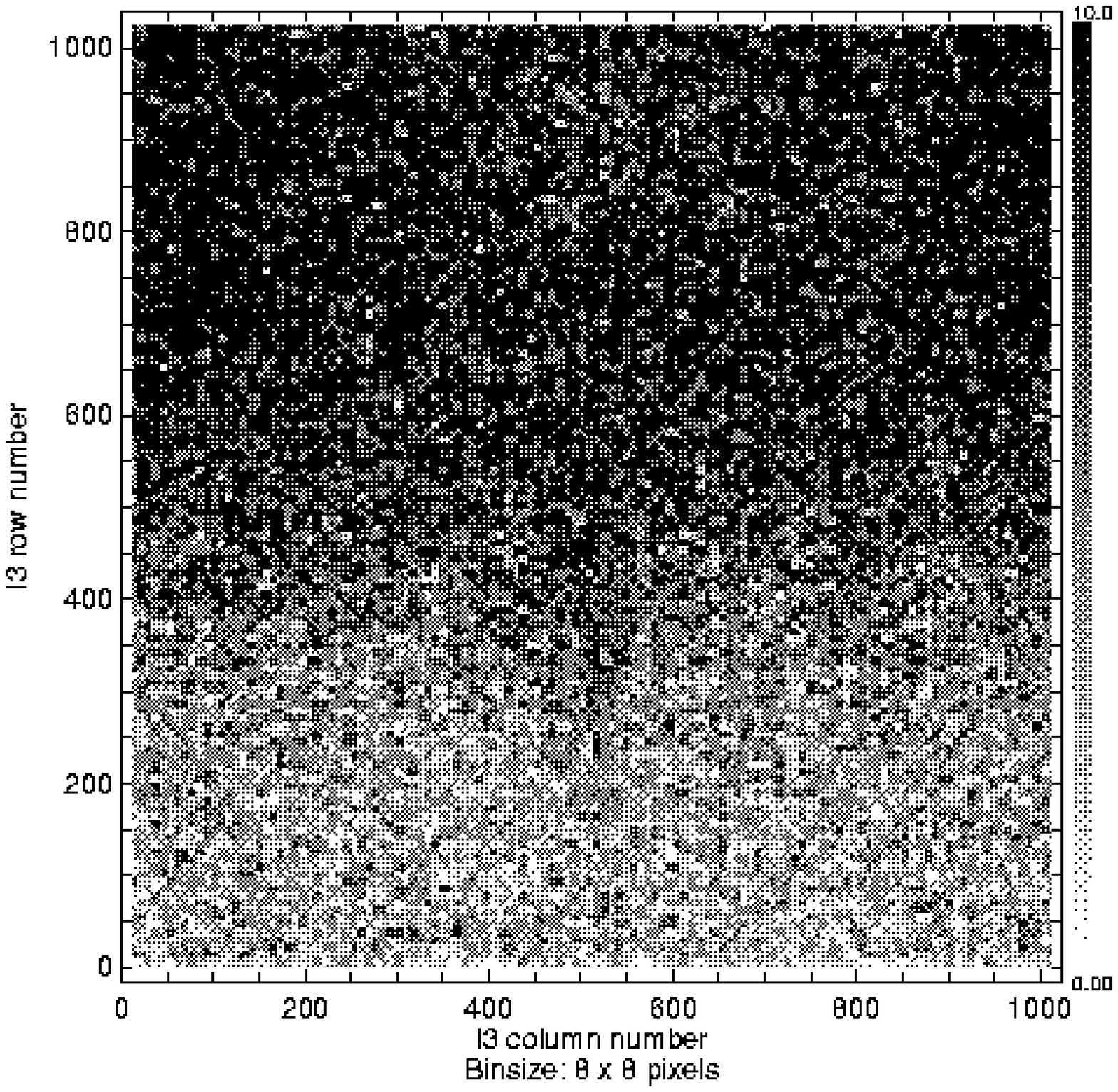,width=3.0in }}}
\centerline{\mbox{
         \epsfig{file=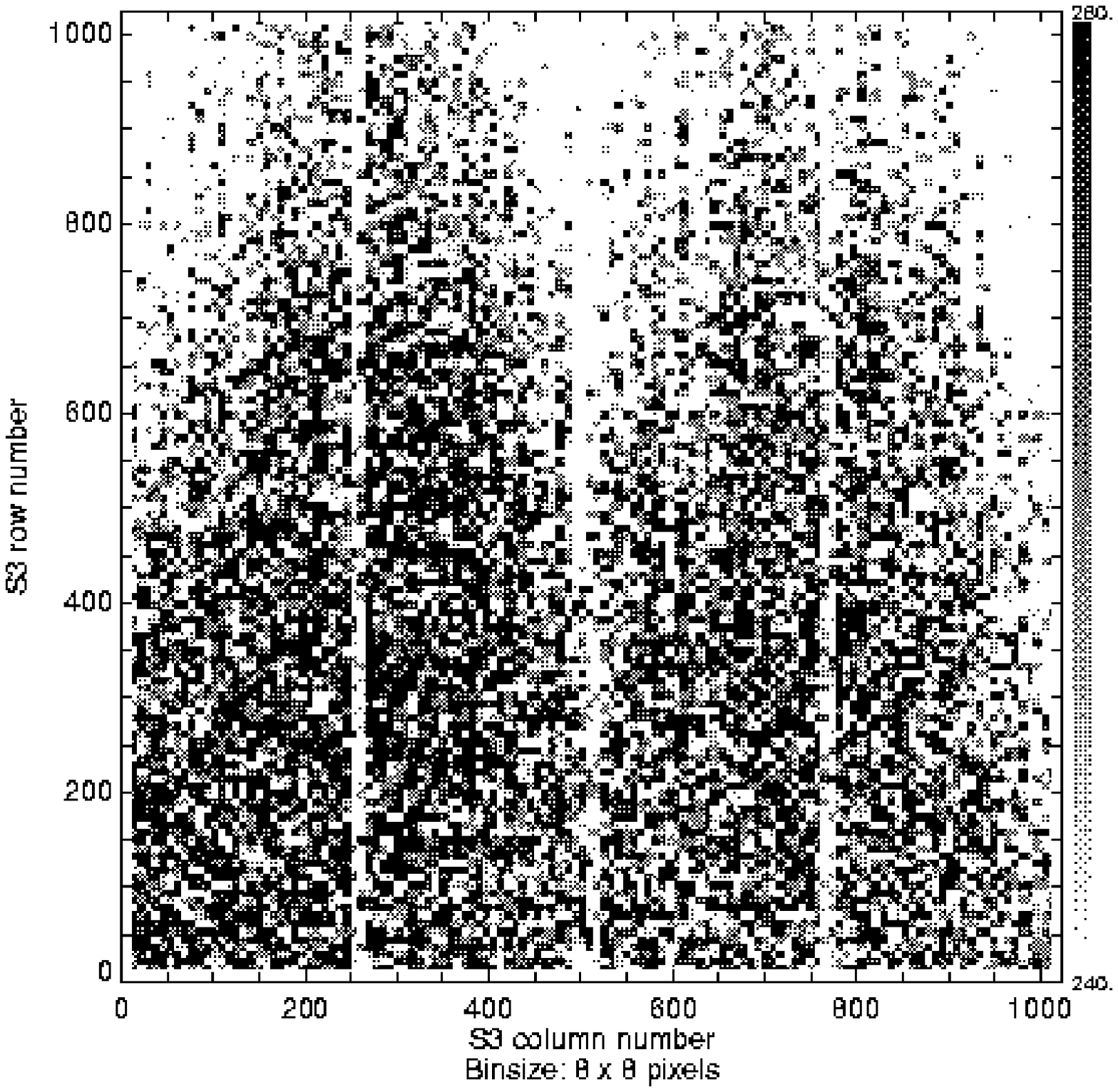,width=3.0in }
         \hspace{0.5in}
         \epsfig{file=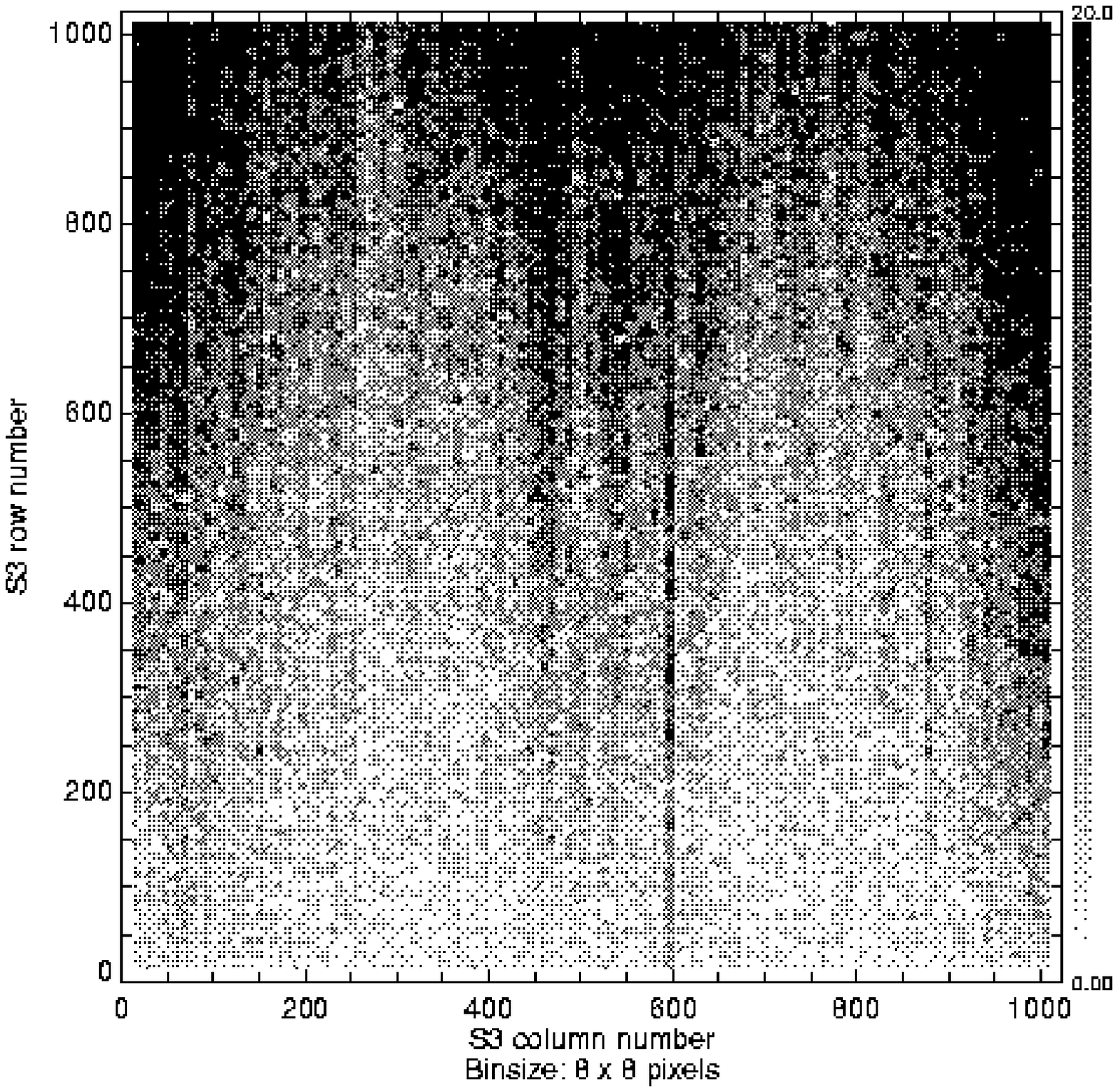,width=3.0in }}}

\caption{\protect \small Grade morphing due to CTI:  I3 at $-$110C (top);
S3 at $-$120C (bottom).  These images give the number of ASCA g02346
events (left) and the complementary Grades 1, 5, and 7 (right) as a
function of chip position for the Mn K lines ($\sim 6$keV) in ECS
data.  These effects become less pronounced in FI devices at lower
energies.  At $-$120C, CTI on FI devices is mitigated so that these
variations are not evident even at high energies. }

\normalsize
\label{fig:qeumap}
\end{figure}

Since CTI causes charge to be shifted into adjacent pixels, thus
causing grade morphing, stringent grade filtering results in
non-uniform quantum efficiency across the device.
Figure~\ref{fig:qeumap} shows the number of detected events in the
$\sim$6~keV Mn K lines in ECS data, using standard grades (g02346, left)
and the complementary Grades 1, 5, and 7 (right).  The top plots are
for the FI device I3 at $-$110C; at $-$120C the FI CTI is reduced and the
quantum efficiency is quite uniform across the device for the standard
grades.  The bottom plots are for the BI device S3 at $-$120C; the
behavior at $-$110C is similar.  Note the complex interaction between
parallel and serial CTI in the BI device:  events morph into Grade 7
due to parallel charge trailing ({\em e.g.} top left corner of image), but
they morph back out of Grade 7 due to serial charge loss ({\em e.g.} top of
image around column 300).  The degree of grade morphing is a
complicated function of event energy and the original event grade.  The
mirroring effect seen in the BI images is due to the alternating
direction of serial readout in the four amplifiers.

%-------------------------------------------------------------------------
%-------------------------------------------------------------------------

\clearpage

%==========================================================================
\section{The Model} \label{sec:model}
%==========================================================================

%====================================
\subsection{The Primary Model}

Our CTI model is phenomenological rather than physical, opting to
characterize the effects of CTI in the data rather than to model
directly the spatial distribution and time constants of the trap
population (see \cite{gallagher98}, \cite{bautz01}, and \cite{krause99}
for examples of such physical models).  It consists of a set of
amplifier-dependent piecewise-linear equations that relate the quantity
of charge lost from a pixel to the original quantity of charge in
that pixel.  In turn, the amount of charge trailed into the adjacent
pixel was determined empirically to be a piecewise-linear function of
the charge loss.  Trailing is not modeled as an amplifier-dependent
relation, rather data from all amplifiers can be combined to derive the fit
parameters. 

\begin{figure}[htb]
\centerline{\epsfig{file=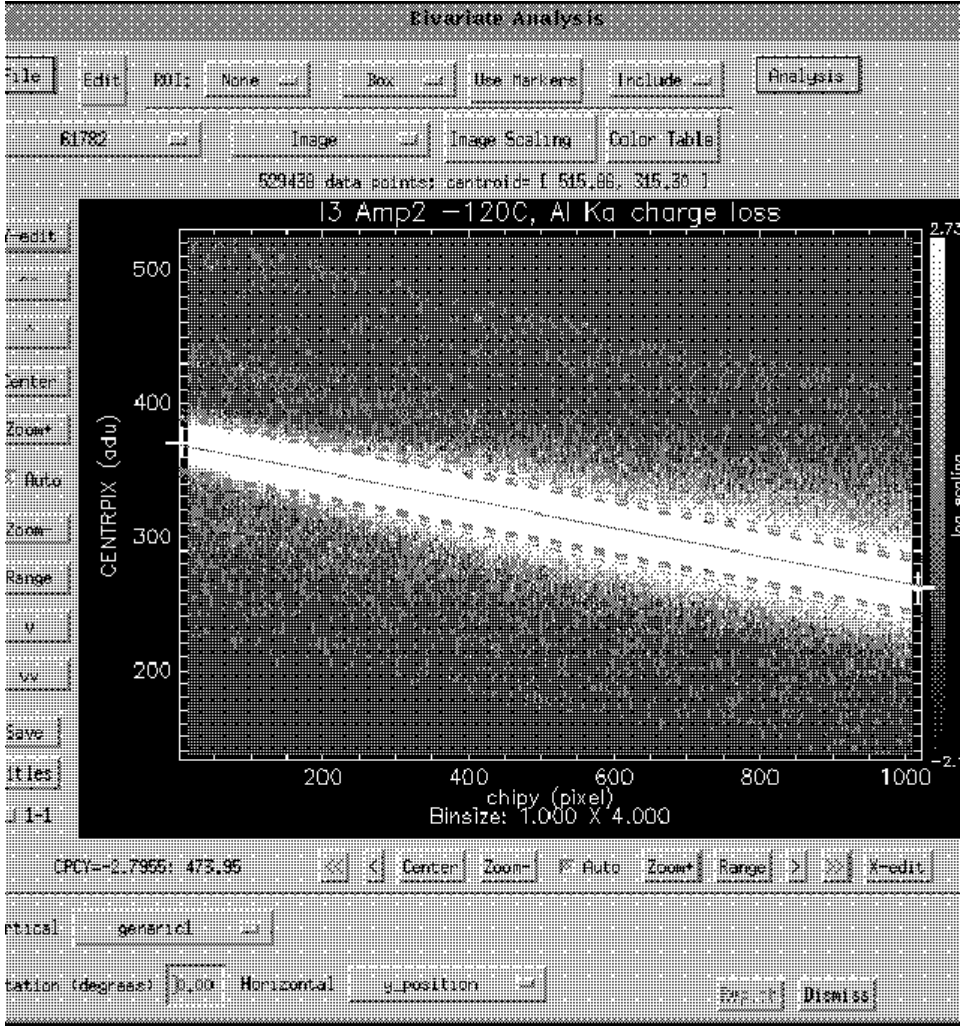, width=5.0in }}
\caption{\protect \small An Event Browser window demonstrating a fit
of Al \Ka events to measure parallel charge loss on I3, Amplifier 2, at $-$120C.
The final fit is shown as a solid line.}
 
\normalsize
\label{fig:ebloss}
\end{figure}

We measure charge loss at a given energy by fitting a line to a plot of
the central pixel value (in DN) for only Grade 0 (single-pixel) events
as a function of event position on the chip ({\em e.g.} row number,
called ``CHIPY,'' for FI devices).  All data are fit as an ensemble of
points, using sigma-clipping to remove outliers.  This is illustrated
in Figure~\ref{fig:ebloss}.  We use the Event Browser software
\cite{tara00} for display and fitting.  Here the event data are binned
into an image for display only; the actual fitting is done on the
events themselves, not on binned data.  

We show the Event Browser window here because it illustrates details of
the fitting process.  The $+$'s show the input to the fitting tool:
approximate endpoints of the line and a rough guess at the linewidth
(marked at CHIPY $\simeq$450).  The set of dashed lines at the edges of
the spectral line show iterations of the sigma-clipping.  The final fit
is shown as a solid line.

%-------------------------------------------------------------------------  
%-------------------------------------------------------------------------

\begin{figure}[htb]
\centerline{\epsfig{file=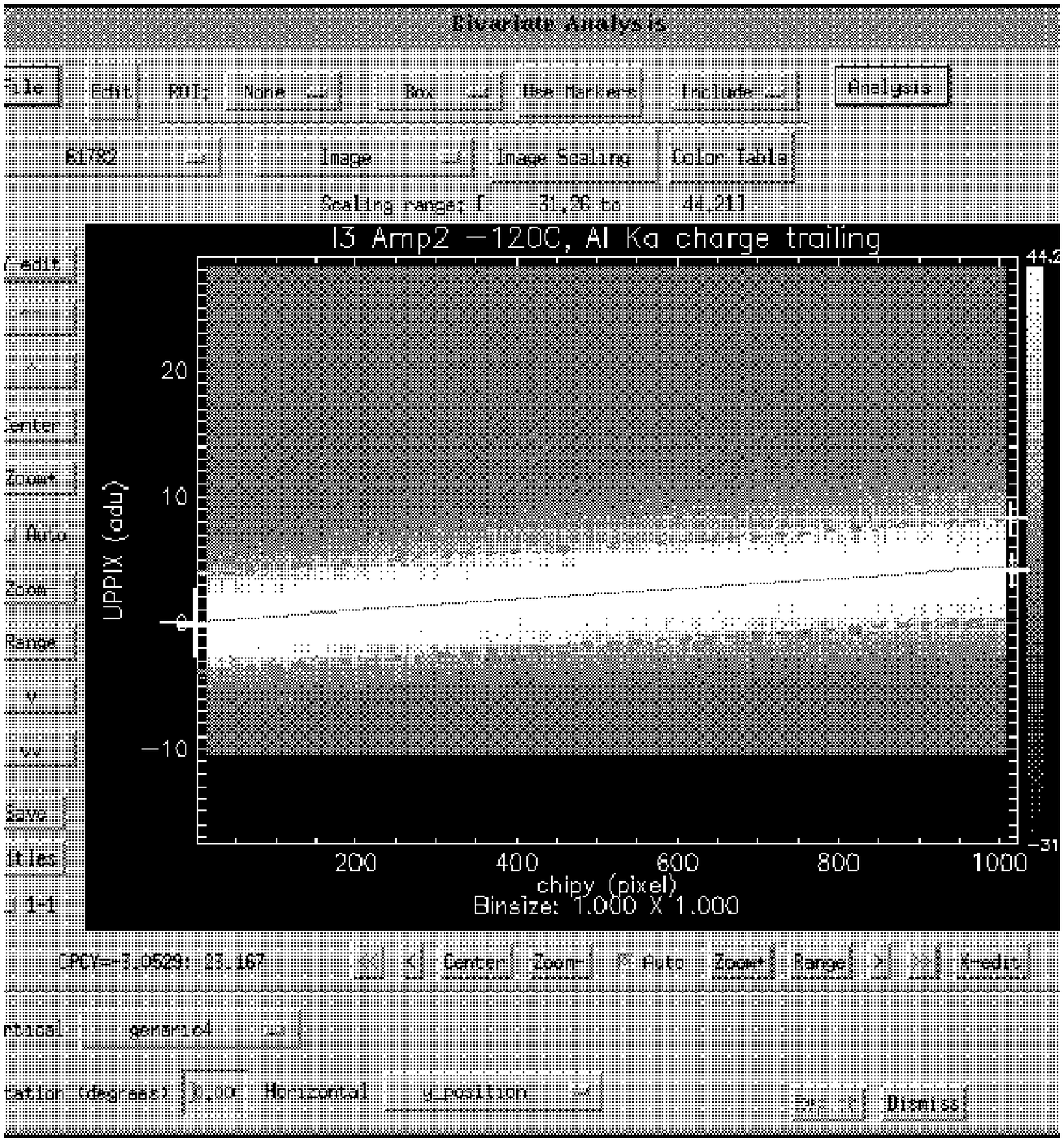, width=5.0in }}
\caption{\protect \small An Event Browser window demonstrating a fit
of Al \Ka events to measure parallel charge trailing on I3, Amplifier 2, at $-$120C.  The final fit is shown as a solid line.}
 
\normalsize
\label{fig:ebtrail}
\end{figure}

Charge trailing is measured similarly:  for FI devices, we plot the
``top center'' pixel of the $3 \times 3$ neighborhood for ACIS Grade 0
and Grade 64 (upward singly-split) events as a function of CHIPY, then
perform a sigma-clipped linear fit to the ensemble of events.  This is
illustrated in Figure~\ref{fig:ebtrail}, where again we show the Event
Browser window to illustrate the input to the fitting routine ($+$'s),
the sigma-clipping (dashed lines), and the final fit (solid line).

%-------------------------------------------------------------------------  
%-------------------------------------------------------------------------

Our linear characterization of the data leads to a linear model of the
charge lost (${\mathcal L}$) and trailed (${\mathcal T}$) for each
pixel, i.e. 
\be
{\mathcal L}  = N_t L(P)
\ee
\be
{\mathcal T} = N_t T(L)
\ee
where $N_t$ is the number of transfers that the charge packet must
undergo to reach the readout node, $P$ is the charge in a single pixel,
$L(P)$ is the charge lost per pixel transfer for an isolated charge
packet, and $T(L)$ is the charge trailed per pixel transfer for an
isolated charge packet.  Obviously the physical process of CTI occurs
one transfer at a time and the correct model would recalculate the
pixel's charge for each transfer separately ({\em e.g.} via a loop in the
code), but that is computationally expensive so we do not attempt such
a solution.

Accurate modeling of CTI requires accurate knowledge of the shape of
$L(P)$ and $T(L)$, especially for small charge packets (low energies) where
we have very little calibration information.  We have found that the
grades and energies of corrected events depend on the exact shape of
$L(P)$ and $T(L)$ at low energies.  For example, a single linear fit to
$L(P)$ yields a non-zero intercept, implying that there can be
substantial charge loss in pixels originally containing very little
charge -- this is not supported by the data.  This single-line model
causes the corrector to boost the charge in low-valued pixels, often
causing them to exceed the split threshold (13~DN for both BI and FI
devices).  This disrupts the grade distribution of a set of
monoenergetic events and causes the split events in that set to exhibit
higher energies than single-pixel events.  We have determined
empirically that $L(P)$ and $T(L)$ must pass through the origin to yield
realistic grades and energies for monochromatic input photons.

\begin{figure}[htb]
\centerline{\mbox{
         \epsfig{file=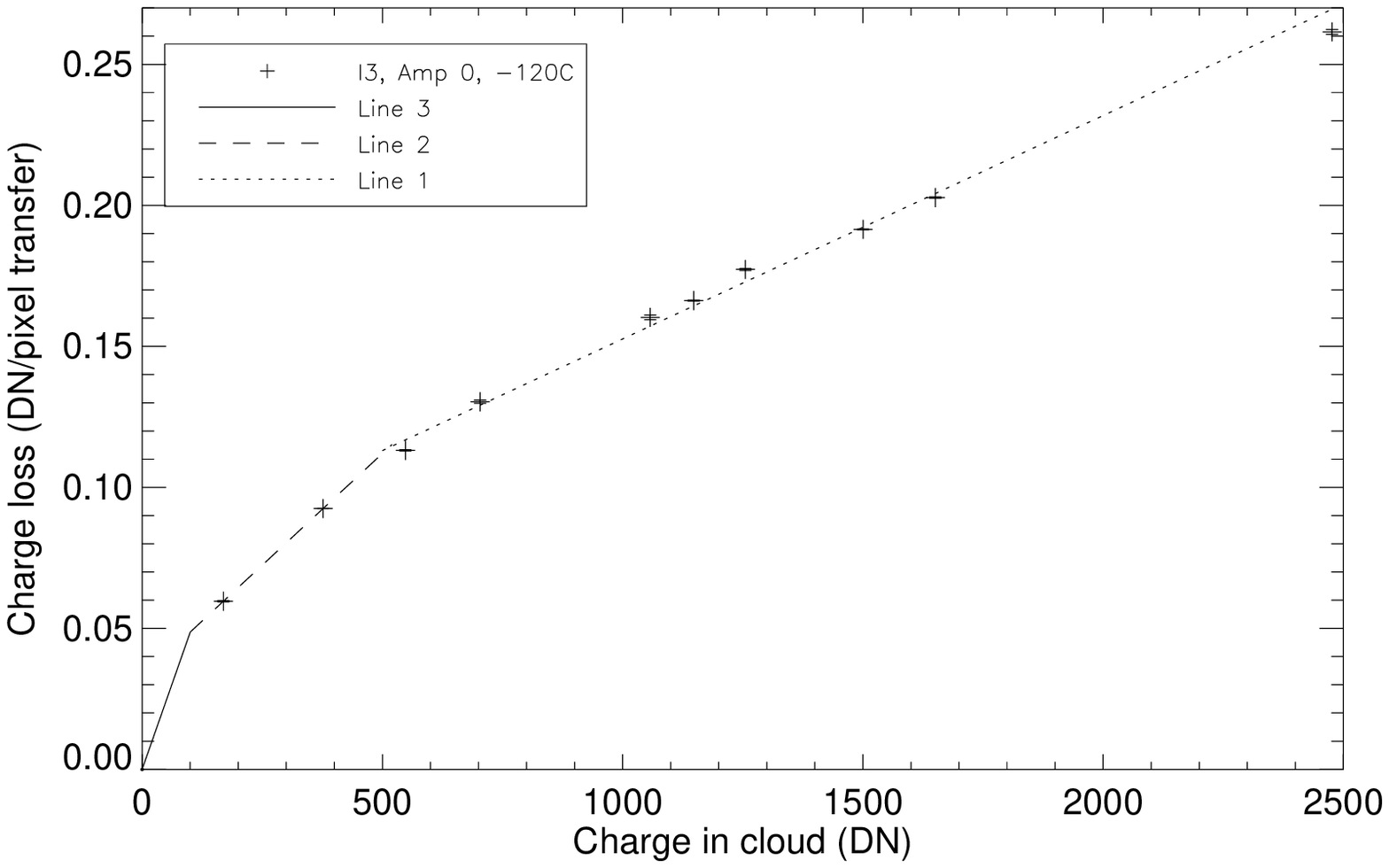,width=3.0in }
         \hspace{0.25in}
         \epsfig{file=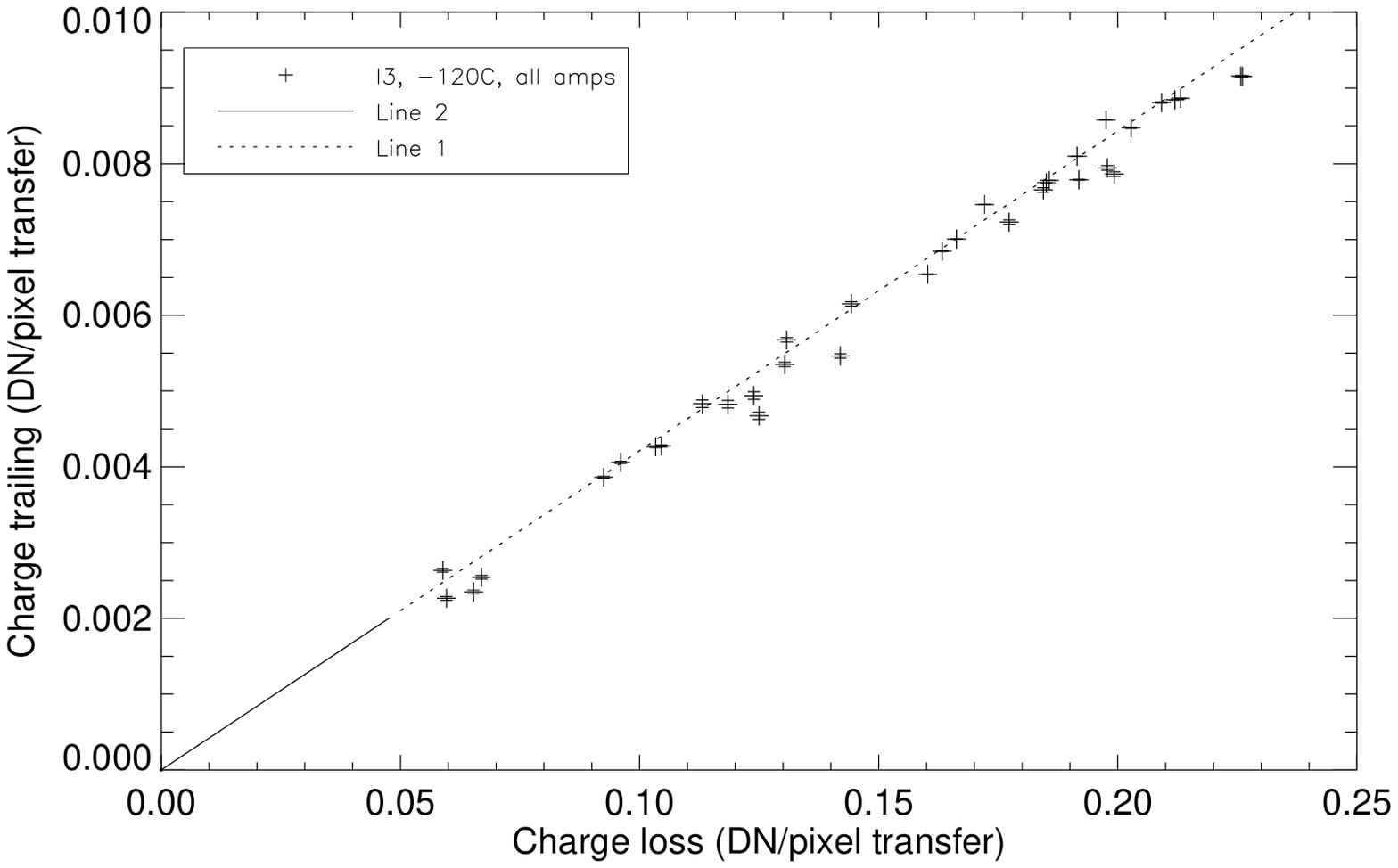,width=3.0in }}}
\caption{\protect \small Left panel:  The piecewise-linear model of
parallel charge loss per pixel transfer for I3, Amplifier 0, at
$-$120C.  Right panel:  The piecewise-linear model of parallel charge
trailing per pixel transfer for I3 at $-$120C (used for all
amplifiers).}
 
\normalsize
\label{fig:lossfits120}
\end{figure}

Thus we model $L(P)$ and $T(L)$ as piecewise linear, continuous curves
that are required to pass through the origin (see
Figure~\ref{fig:lossfits120}).  Three lines are needed for $L(P)$ and
two for $T(L)$.  The functions' values (for both parallel and serial
CTI) are stored as lookup tables that are accessed for each pixel of
each event.  We chose a set of lines as the functional form for these
relations because they adequately fit the data, they keep the model
simple, and because we can restrict the coefficients to ensure zero
charge loss or trailing when there is zero charge present.

%-------------------------------------------------------------------------  
%------------------------------------------------------------------------- 

Our original model for $-$110C FI data \cite{townsley00} employed a
power-law fit, which worked well for all measurable energies up to
$\sim$6~keV, but underfit the Au~\La line at 9.7~keV.  Application to
the first Galactic Center observation made by ACIS showed that this
underfitting might be affecting observations of astrophysical iron
lines (Y. Maeda, private communication), so the power-law model was
abandoned in favor of the piecewise linear model.  The new model
improves the $-$110C corrector performance at high energies without
compromising the low-energy accuracy.  For the more modest FI CTI
effects seen at $-$120C and for BI devices, the power law fit is
adequate \cite{grant01} but we choose to use the piecewise linear model
for all temperatures and devices for consistency.

For the BI devices, we must model parallel and serial charge loss and
trailing independently to match the data.  There are charge traps in
the imaging array, framestore array, and serial register.  Since the FI
devices were damaged by radiation that penetrates only a few microns
into the device \cite{prigozhin00}, their (shielded) framestore arrays
and serial registers are protected.  So, although the CTI is more
pronounced on FI devices, it is easier to model than in the BI
devices.

The primary FI CTI model consists of a piecewise linear fit to the
measurements of parallel charge loss per pixel transfer as a function
of line center (in DN units), measured using only the central pixel of
Grade 0 events and on a per-amplifier basis.  The components cover the
ranges 0--100~DN (0--0.4~keV), 100--550~DN (0.4--2.2~keV) and
550--2500~DN (2.2--10~keV) (see the left panel of
Figure~\ref{fig:lossfits120}).  The mid-energy line is necessary to
accomodate a known turnover around these energies (known from
calibrations with the soft supernova remnant E0102-72.3).  The break is
wherever the mid- and high-energy lines meet and is calculated in the
code.  The low-energy line takes over at 100~DN, a value determined
roughly by looking at the energies of corrected split events in the S3
(BI) device.  It forces the model through the origin.  As mentioned
above, this was necessary to avoid anomalous energies for low-energy
split events in S3 -- in effect, it forces the model to assume zero
charge loss for zero-energy events.

An example of the charge trailing model is shown in the right panel
of Figure~\ref{fig:lossfits120}.  For trailing, only two lines were
necessary to fit the data; the low-energy line is forced through
the origin as in the loss model.  The break between the lines is taken
as the loss value where the third line takes over in the loss model
(averaged over the four amplifiers).  For the example in Figure~\ref{fig:lossfits120}, the two lines in the fit have very similar
slopes.  Note that the Au \La points are omitted from the trailing model.
Due to the large charge cloud size and confusion with background events,
our method for measuring charge trailing proved unreliable for Au \La.
	
The (very small) error bars represent only measurement uncertainty
(estimated by fitting each line 3 times); we have not attempted to
quantify systematics.  The high-energy charge loss fit was made using
errors in both line centers and charge loss and used all points shown
on the plot above 500~DN.  The low-energy charge loss fit only includes
the Al \Ka and Mn/Fe L-complex points.

A similar approach is taken for the BI devices.  First the serial
charge loss and trailing model coefficients are determined for each
amplifier, also using a 3-component piecewise linear fit for loss and a
2-component piecewise linear fit for trailing.  Then the CTI corrector
(described below) is run on the data, with coefficients for parallel
CTI set to zero.  The resulting serial-corrected data are used to
measure the coefficients for parallel CTI.  Due to non-linearities in
the serial CTI, columns near the edges of each amplifier are excluded
when computing the linear parts of both the serial and parallel
models.

\begin{figure}[htb]
\centerline{\epsfig{file=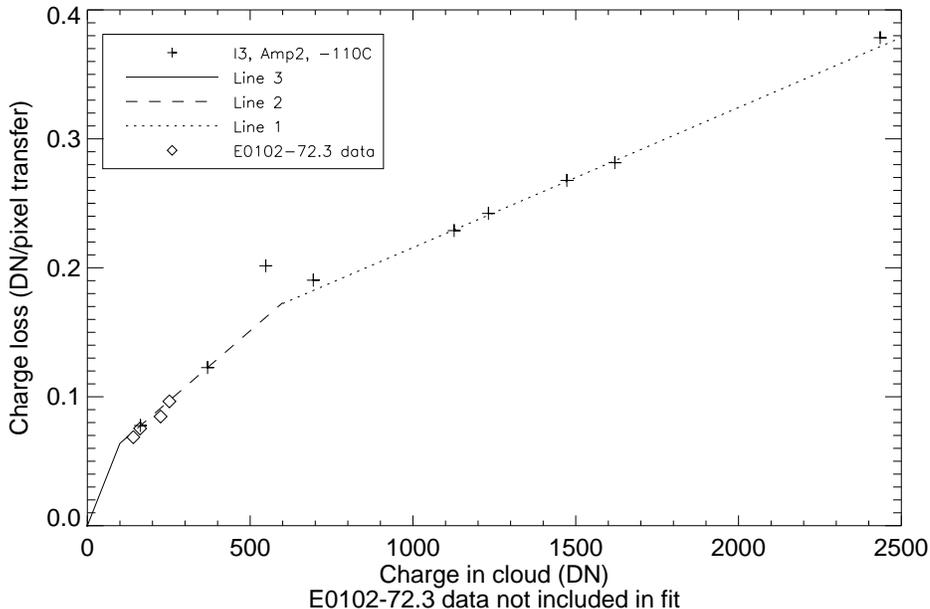, width=5.0in }}
\caption{\protect \small The model of parallel charge loss per pixel transfer for I3, Amplifier 2, at $-$110C.  The E0102 data are overplotted on the model just to confirm its validity and were not
used to determine the model parameters.}
 
\normalsize
\label{fig:lossfitse0102}
\end{figure}

As a check on the functional form of the model at low energies, we
measured the low-energy parallel charge loss using the soft supernova
remnant E0102-72.3 (``E0102''), a calibration target for {\em
Chandra}.  Figure~\ref{fig:lossfitse0102} shows the charge loss model
for I3, Amplifier 2, at a focal plane temperature of $-$110C, with the
E0102 charge loss values overplotted for the three soft lines
measurable with these data.

%-------------------------------------------------------------------------  
%-------------------------------------------------------------------------

The E0102 points were obtained from two datasets, Observation ID 49
(source placed at CCD row number $\sim 100$) and Observation ID 48
(source placed at CCD row number $\sim 500$).  The charge loss per
pixel transfer was calculated using only these data, then corrected for
the fact that the full-chip charge loss slope is more shallow than that
calculated when only using the bottom half of the chip by executing a
similar half-chip analysis on ECS data, then determining the correction
factor necessary to recover the full-chip slope. 

The E0102 data were not used to modify the model.  Their agreement with
the low-energy model components confirms that the 3-line fit chosen to
parameterize the charge loss is reasonable down to $\sim 0.5$~keV at
least.  This is adequate for characterizing FI devices but we have no
confirmation of our model in the 0.2--0.5~keV range accessible to BI
devices.

%=====================================
\subsection{Shielding Within an Event}

This CTI model is ``local'' -- prediction of the amount of charge
lost by a particular event does not consider the possibility that
another event falling in the same columns but closer to the amplifier
readout will
have filled some portion of the traps, reducing the charge loss from
its nominal value (see \cite{bautz01} and \cite{gendreau95} for
descriptions of such ``precursor'' models).  We do however consider
shielding effects that may occur within the nine pixels of a single
event.  Physical principles suggest that pixels which are not isolated
experience loss and trailing that are influenced by their neighbor
pixels -- what we call the ``self-shielding'' effect.  We hypothesize
that the loss mechanism essentially acts only on the portion of a
pixel's charge which {\em exceeds} the charge in the previous pixel;
similarly the trailing mechanism essentially acts only on the portion
of a pixel's charge which exceeds the charge in the {\em following}
pixel.  Thus a given pixel experiences {\em either} loss or trailing,
not both -- loss if its predecessor has less charge than the pixel in
question and trailing if its predecessor has more charge.

The physical process of self-shielding is of course governed by the
relationship between the physical volume of silicon occupied by one
pixel's charge packet compared with the physical volume of silicon
occupied by its predecessor or successor charge packet.  A physically-based
model of self-shielding would involve knowing the details of the
three-dimensional charge distribution \cite{bautz01}.  For our
phenomenological model, we instead take a very simple approach to
modeling this effect.

For a pixel with charge $P$ (``root'') whose predecessor (``lead'') has
a {\em smaller} charge $P_{\rm  lead}$, we compute:
\be
  {\mathcal L}_{\rm  root}  = N_t L(P)     < P
\ee
\be	
  {\mathcal L}_{\rm  lead}  = N_t L(P_{\rm  lead}) < P_{\rm  lead}
\ee
\be	
  {\mathcal L}_{\rm  eff} = ({\mathcal L}_{\rm  root} - {\mathcal L}_{\rm  lead}) > 0	
\ee
The effective charge loss ${\mathcal L}_{\rm  eff}$ is then {\em subtracted} from $P$.

For a pixel with charge $P$ (``root'') whose predecessor (``lead'') has
a {\em larger} charge $P_{\rm  lead}$, we compute:

\be  
  {\mathcal T}_{\rm  root}  = N_t T(P)     < P
\ee
\be	
  {\mathcal T}_{\rm  lead}  = N_t T(P_{\rm  lead}) < P_{\rm  lead}
\ee
\be	
  {\mathcal T}_{\rm  eff} = ({\mathcal T}_{\rm  lead} - {\mathcal T}_{\rm  root}) > 0	
\ee
The effective charge trailed ${\mathcal T}_{\rm  eff}$ is then {\em
added} to $P$.  In this model, events which are almost evenly split
between two pixels in the direction of transfer suffer CTI effects
almost identical to those experienced by a Grade~0 event with half the
energy of the split event.

%=====================================
\subsection{Deviations from Linearity}

Although ACIS FI devices do not exhibit serial CTI, they do show a
small charge loss effect which is proportional to the distance between
the interaction site and the left or right edge of the CCD (as opposed to the
distance to the amplifier readout).  This is believed to be caused by the
slight drop in clocking voltages necessarily suffered between the
middle of the CCD and the supply leads at the chip edges (M. Bautz,
private communication).  CTI scales the effect so that these gain
variations are row- and energy-dependent, with the bottom rows of the
chip showing virtually no variation and the top rows varying by up to
$\sim$100 DN at high energies ($\sim$6 keV).

ACIS BI devices also show low-spatial-frequency deviations from
linearity in their charge loss.  As shown above, by combining large
numbers of ECS observations, column-to-column gain variations can be
seen.  (Row-to-row variations are not seen in either FI or BI
devices.)  These effects are also energy-dependent, with high-energy
calibration lines in the ECS exhibiting larger-amplitude variations
than low-energy lines.

These nonlinearities result in broadened spectral lines for both FI
and BI devices even after correction for linear charge loss and
trailing.  Thus we have developed the concept of a ``deviation map'' to
make moderate gain adjustments to each event that help to regularize
the CCD response across the device and narrow the linewidths.  Our
techniques are reminiscent of flat-fielding techniques used to remove
instrumental non-uniformities in visual CCD data.

\begin{figure}[htb]
\centerline{\mbox{
         \epsfig{file=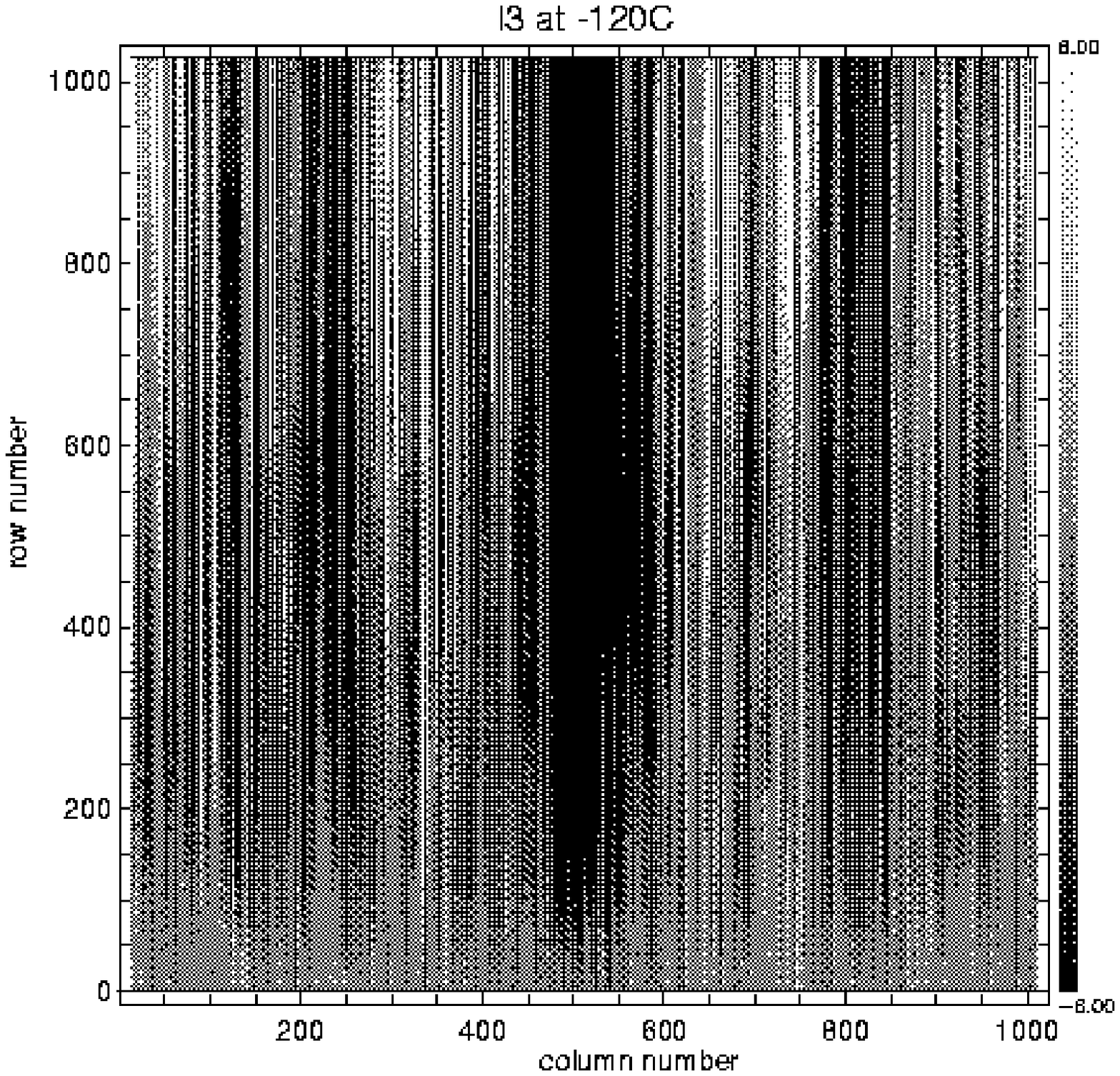,height=3.0in }
         \hspace{0.25in}
         \epsfig{file=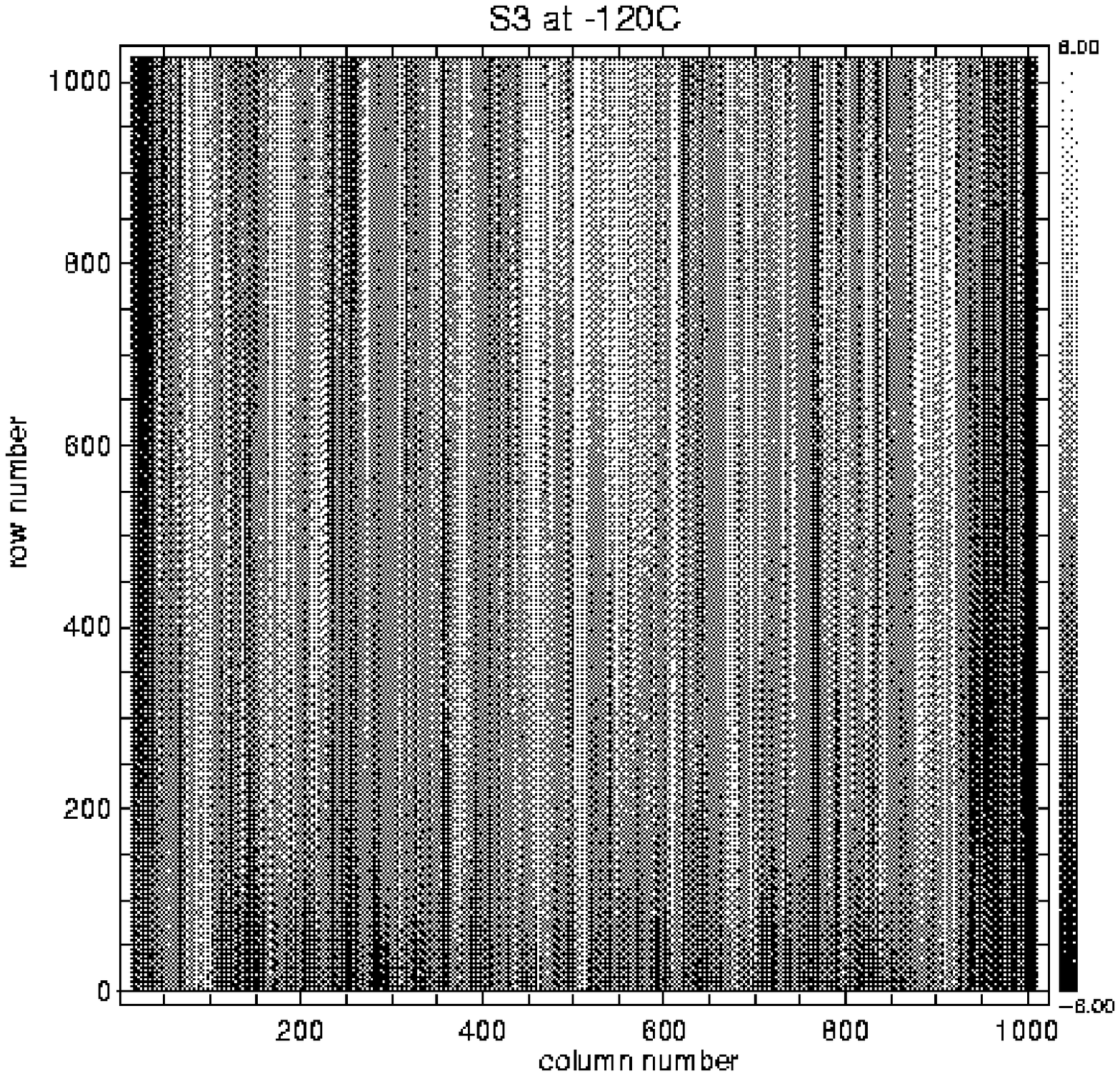,height=3.0in }}}
\caption{\protect \small $D_0$, the deviation maps for I3 at $-$120C (left)
and S3 at $-$120C (right), at 0~DN.}
 
\normalsize
\label{fig:devmap}
\end{figure}

\begin{figure}[htb]
\centerline{\epsfig{file=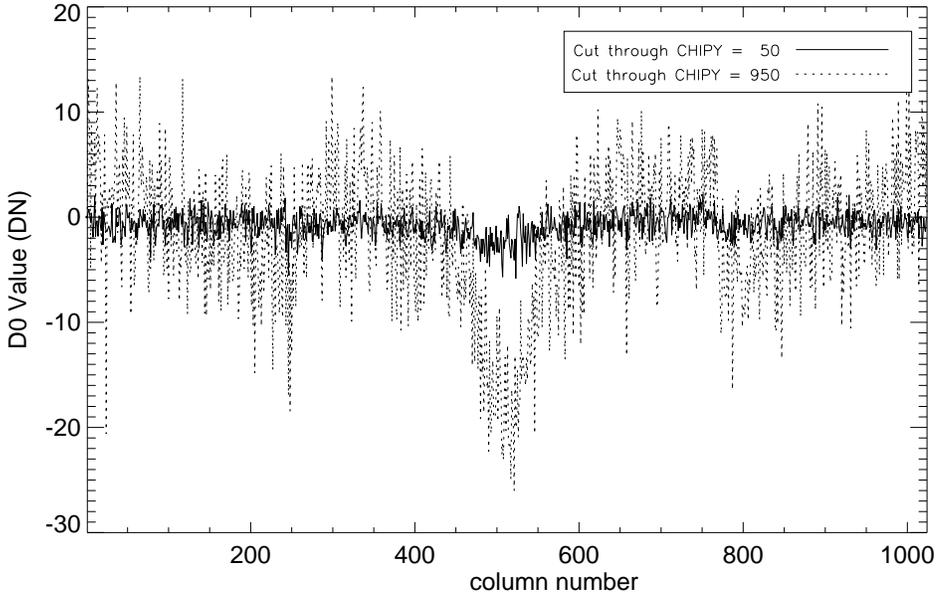, width=5.0in }}
\caption{\protect \small Cuts through D0 at low and high row numbers
(``CHIPY'') for I3 at $-$120C, illustrating that the amplitude of the
deviation applied to each event in an FI device is highly dependent on
that event's $y_c$ location.}
 
\normalsize
\label{fig:devmapcuts}
\end{figure}

We model the deviations via a 2-dimensional map ($D(x_c,y_c,E)$) with an
energy-dependent amplitude:
\be
		D(x_c,y_c,E) = D_0(x_c,y_c)(1 + gE)
\label{deviation}
\ee
where $x_c$ and $y_c$ are the position coordinates of the event in
units of CCD pixels, $E$ = event amplitude in DN, $g$ = deviation
amplitude, and $D_0$ is the two-dimensional map of deviations
extrapolated to 0~DN.  The value of this deviation is calculated for
each event based on its $(x_c,y_c)$ location and applied as an additive
adjustment to the central pixel value.  Figure~\ref{fig:devmap} shows
images of $D_0$ for I3 and S3 at $-$120C.  Cuts through the I3 map, given
in Figure~\ref{fig:devmapcuts}, show that the amplitude of the
deviations increases substantially with $y_c$ on FI devices, as noted
above.
	
%-------------------------------------------------------------------------  
%-------------------------------------------------------------------------

%-------------------------------------------------------------------------  
%-------------------------------------------------------------------------

The zero-energy deviation map $D_0$ is made for each amplifier in a
series of smoothing and averaging steps using constituent maps made in
the three strongest lines in the ECS: Al \Ka, Ti \Ka, and Mn \Ka.
First the ECS data are CTI-corrected using the linear model described
above.  The resulting event list is filtered spectrally to keep only
events in one of these strong lines.  We then make an image of these
event amplitudes in DN space, binned by 1 pixel in $x_c$ and by 16
pixels in $y_c$, with intensity computed as a sigma-clipped mean.  This
process is repeated for the other two lines, resulting in three images
of the deviations covering a substantial range in energy. 

For each energy, we then perform a trending analysis of gaussian fits
to the spectral line, noting the mean and standard deviation (width) of the
spectral line as a function of row number.  We perform a linear fit to
these trended properties to determine the line energy and width at row
zero, where the values are unaffected by CTI.  Then we plot the
row-zero standard deviations as a function of line energy (in DN) and fit a
line, noting its slope (S) and intercept ($\sigma_0$).  This gives the equation 
\be
\sigma(E) = \sigma_0 + SE = \sigma_0(1 + \frac{S}{\sigma_0}E)
\ee
which, by analogy to the definition of the deviation map above, yields
\be
g = \frac{S}{\sigma_0} .
\ee
We obtain an estimate of $S$ and $\sigma_0$ for each amplifier; these are 
averaged then divided to obtain $g$.

The maps at Al~\Ka, Ti~\Ka, and Mn~\Ka are then smoothed column-wise to
reduce the effects of shot noise.  Next we generate three estimates of
$D_0$ using the smoothed maps and average those three estimates to
obtain our final $D_0$ for that amp.  Lastly, the maps for each amplifier are
incorporated into an array to make one deviation map for the entire CCD. 

This deviation map is used to adjust the amplitude of an event.  We
compute $E$ in Equation~\req{deviation} above by summing the values
of pixels in the $3 \times 3$ event neighborhood that are above the
split threshold (13~DN).  The adjustment is then computed from
Equation~\req{deviation}, using the event's $(x_c,y_c)$ location to
access the correct element of $D_0$.  The adjustment is made by adding
the result to the central pixel of the event.  This preserves the event
grade; we determined by experimentation that this technique produces
the best reconstructed event.

\clearpage

%==========================================================================
\section{Simulator Results}  \label{sec:simulations}
%==========================================================================

The Monte Carlo simulator's ability to reproduce the effects of CTI is
illustrated in the following figures.  All data and simulations are for
Al \Ka (1.486~keV) with ASCA g02346 grade filtering.  In both figures,
the left panels show the data and the right give the matching
simulations.

\begin{figure}[htb]
\centerline{\mbox{
         \epsfig{file=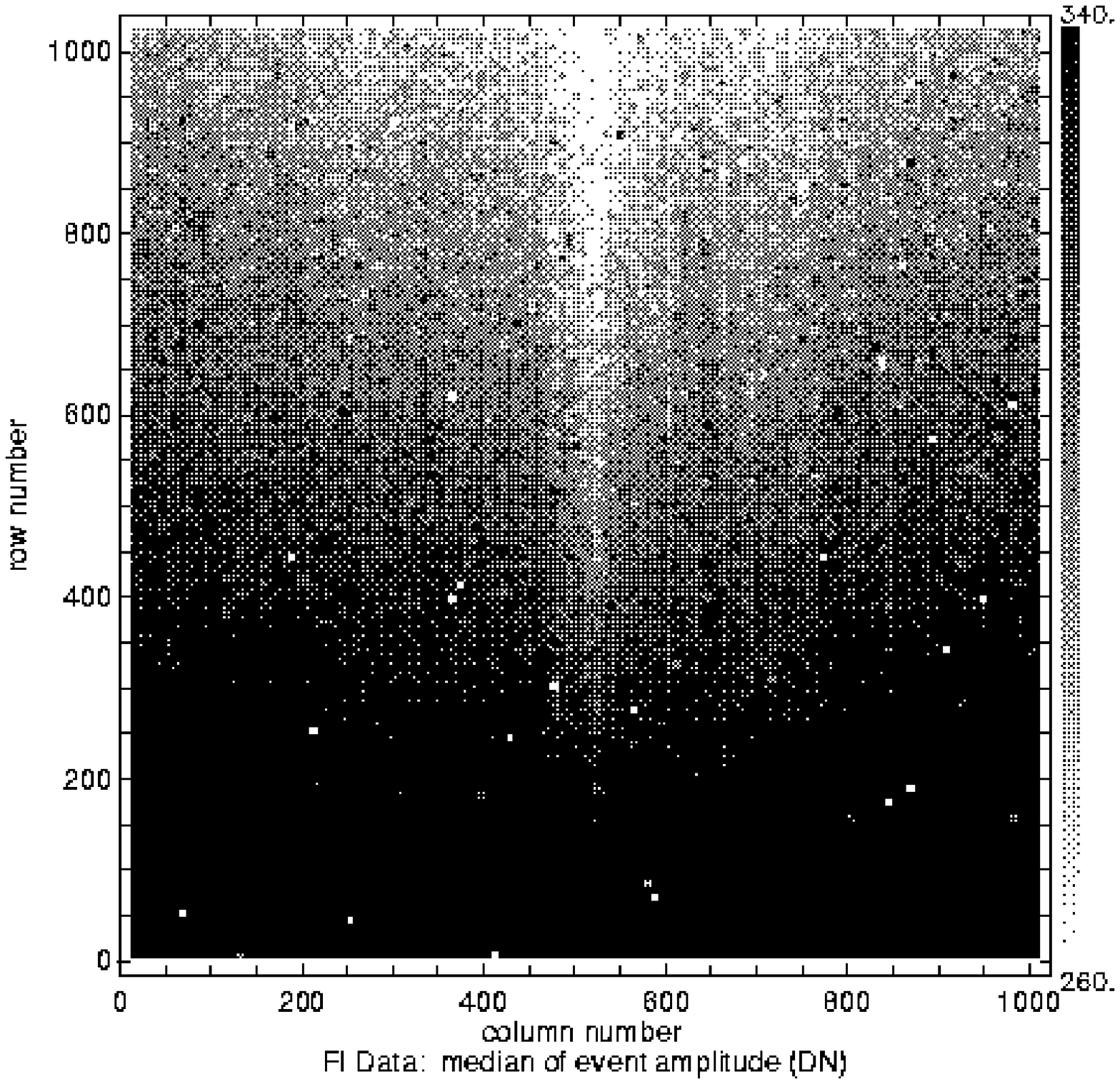,width=3.0in }
         \hspace{0.25in}
         \epsfig{file=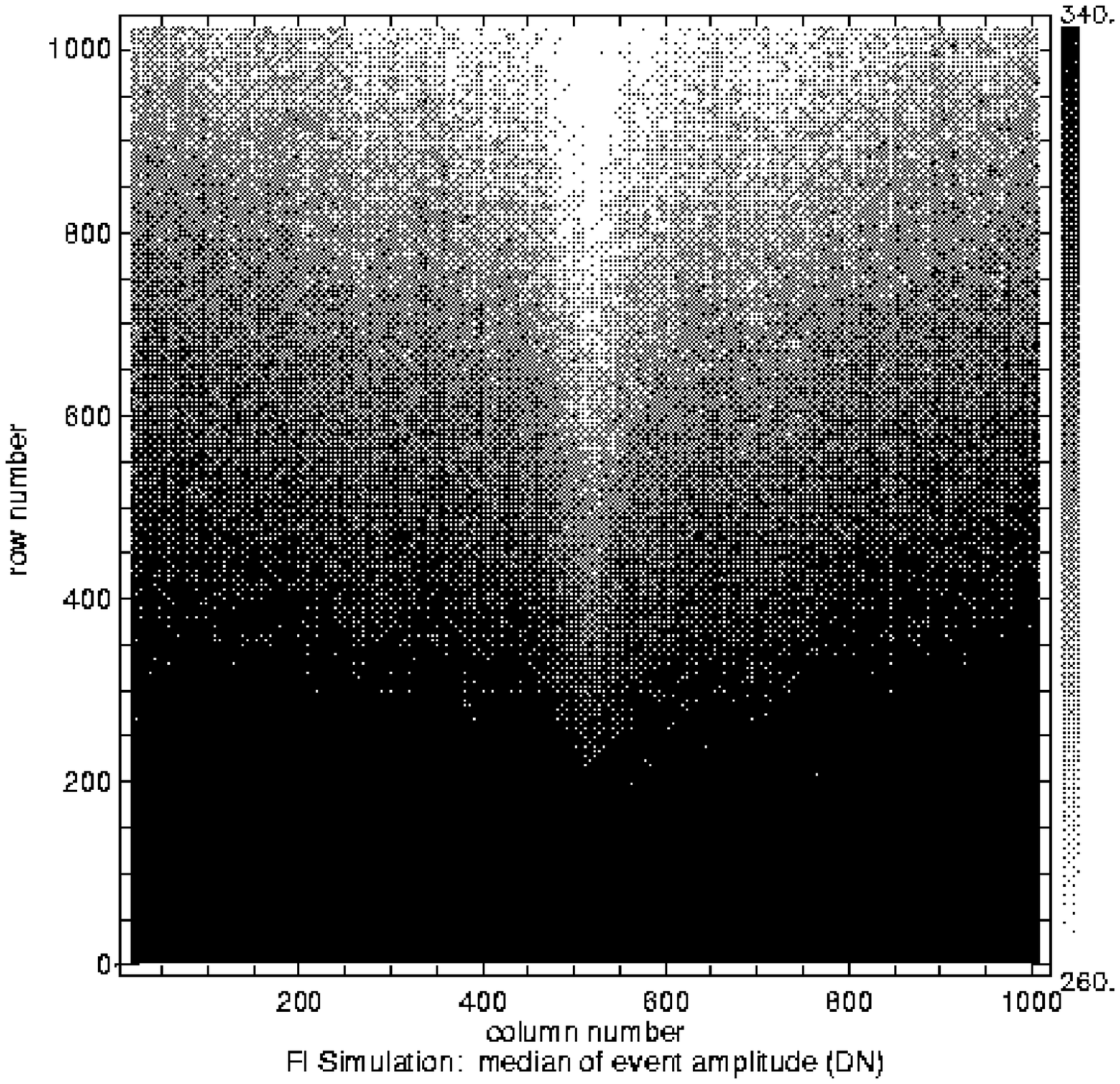,width=3.0in } }}
\centerline{\mbox{
         \epsfig{file=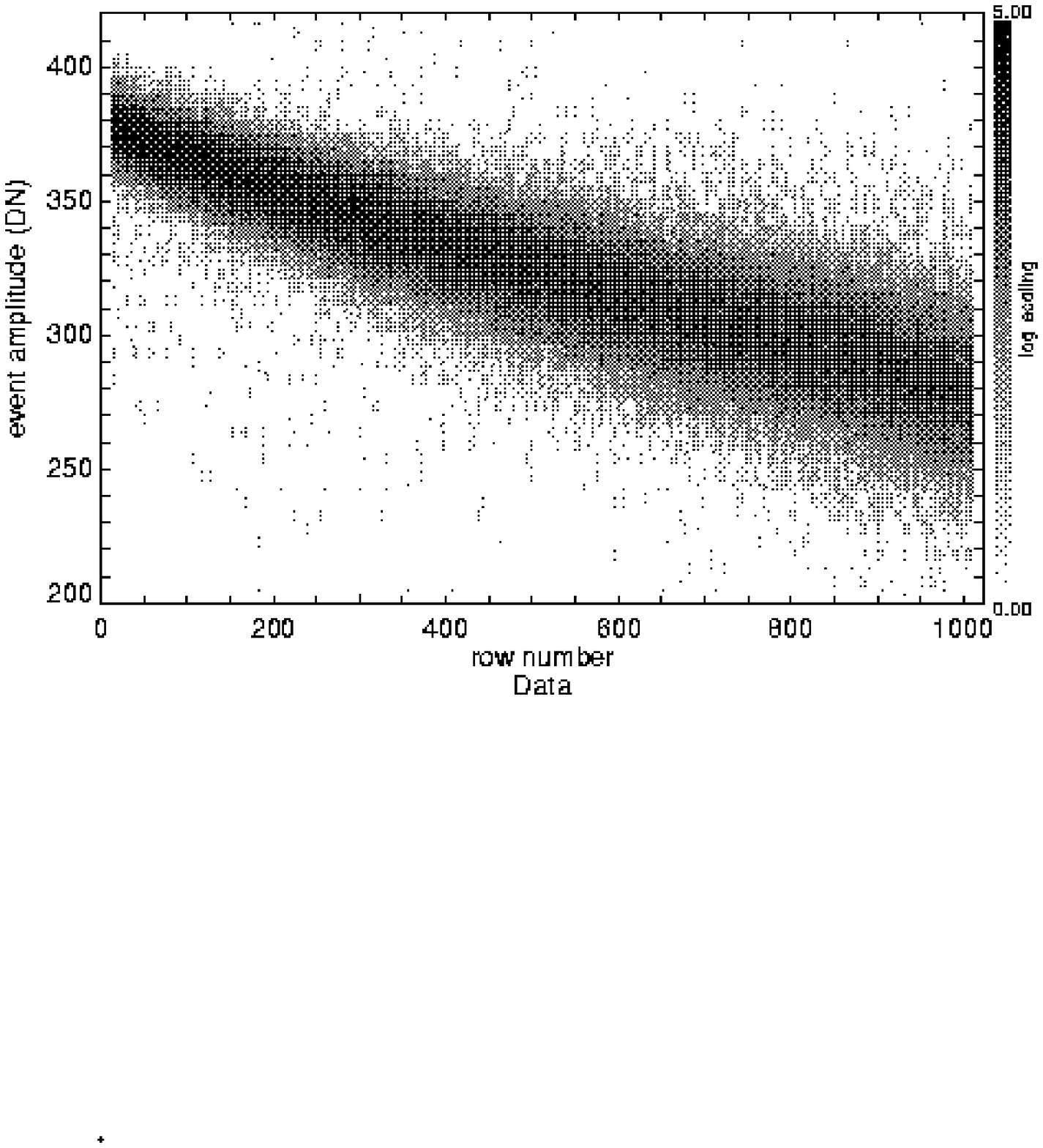,width=3.0in }
         \hspace{0.25in}
         \epsfig{file=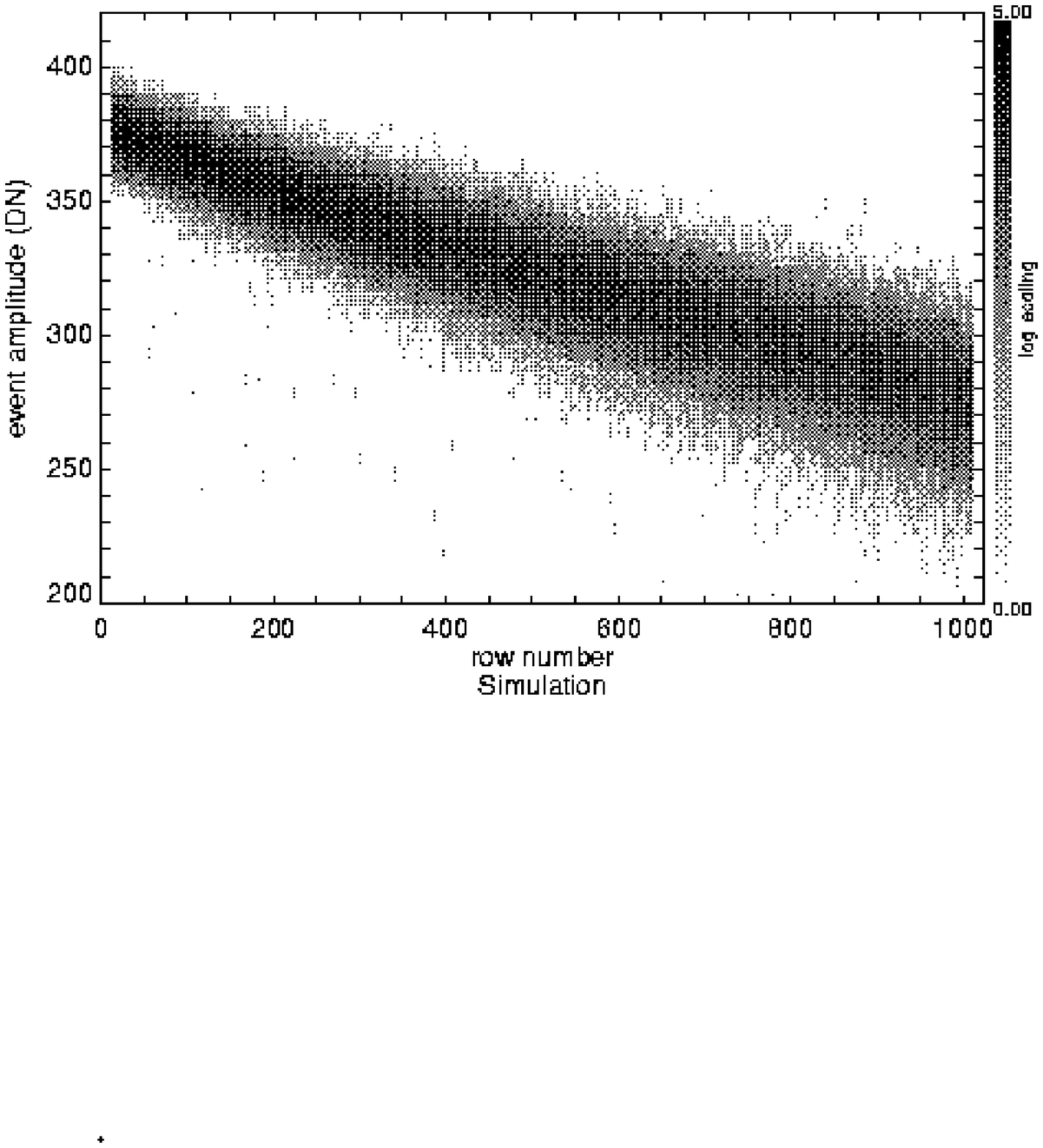,width=3.0in } }}
\caption{\protect \small Data (left) and simulation (right) of Al \Ka on the FI
device I3 at $-$120C.  Top:  Median images of Al \Ka line in units of DN,
binsize $8 \times 8$ pixels.  Bottom:  Amplitude of Al \Ka line in units of
DN, as a function of row number.}

%Upper panels:
%Aspect ratio 1.  Median image of Al \Ka line in PHA space, binsize 8x8.
%Used "partial_index" for data.

%Lower panels:
%Aspect ratio 0.7.  Al Ka line in PHA space vs. chipx, binsize 4x5.
%Used "orig_small_index" for data.
 
\normalsize
\label{fig:fi-image-comp}
\end{figure}

The upper panels of Figure~\ref{fig:fi-image-comp} show images of the
median event amplitude events on an FI device (I3 at
$-$120C).  Both datasets contain $\sim 10^5$ events and are binned at $8
\times 8$ pixels.  With this binning, column-to-column variations are
not visible, but variations with lower spatial frequency are clear and
are reproduced with good fidelity by the simulator.

%-------------------------------------------------------------------------  
%-------------------------------------------------------------------------

When modeling CTI for the event simulator, we add noise to the pixel
adjustments to model the stochastic nature of traps.  We find that a
simple Poisson noise model on the number of electrons lost is not
sufficient for FI devices, so we introduce an energy-dependent tuning
parameter that is multiplied by the event's row number to account for
the additional line broadening with increasing distance from the
readout node that is seen in the ECS data.  The effects of this are
apparent in the lower panels of Figure~\ref{fig:fi-image-comp}.
Row-dependent non-linearities in the CTI give the data a slight S shape
in this plot; this feature is also seen in the simulation.

The data here show more scatter than the simulation; this is due in
part to the background in ECS data and to Al \Ka events that occurred
in the framestore region of the CCD.  These framestore events do not
suffer CTI and appear as a faint horizontal band in the lower left
panel of Figure~\ref{fig:fi-image-comp}.

\begin{figure}[htb]
\centerline{\mbox{
         \epsfig{file=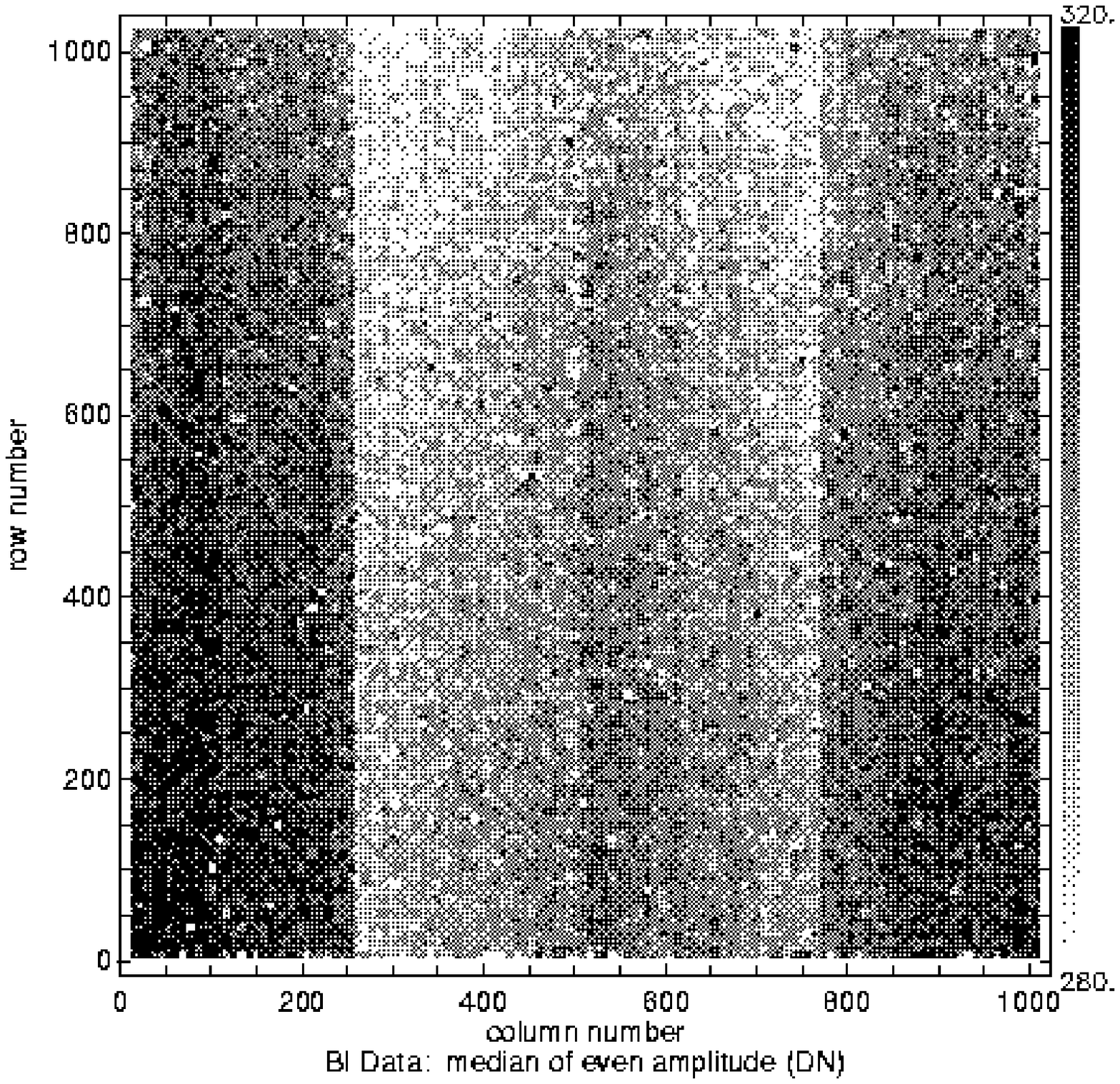,width=3.0in }
         \hspace{0.25in}
         \epsfig{file=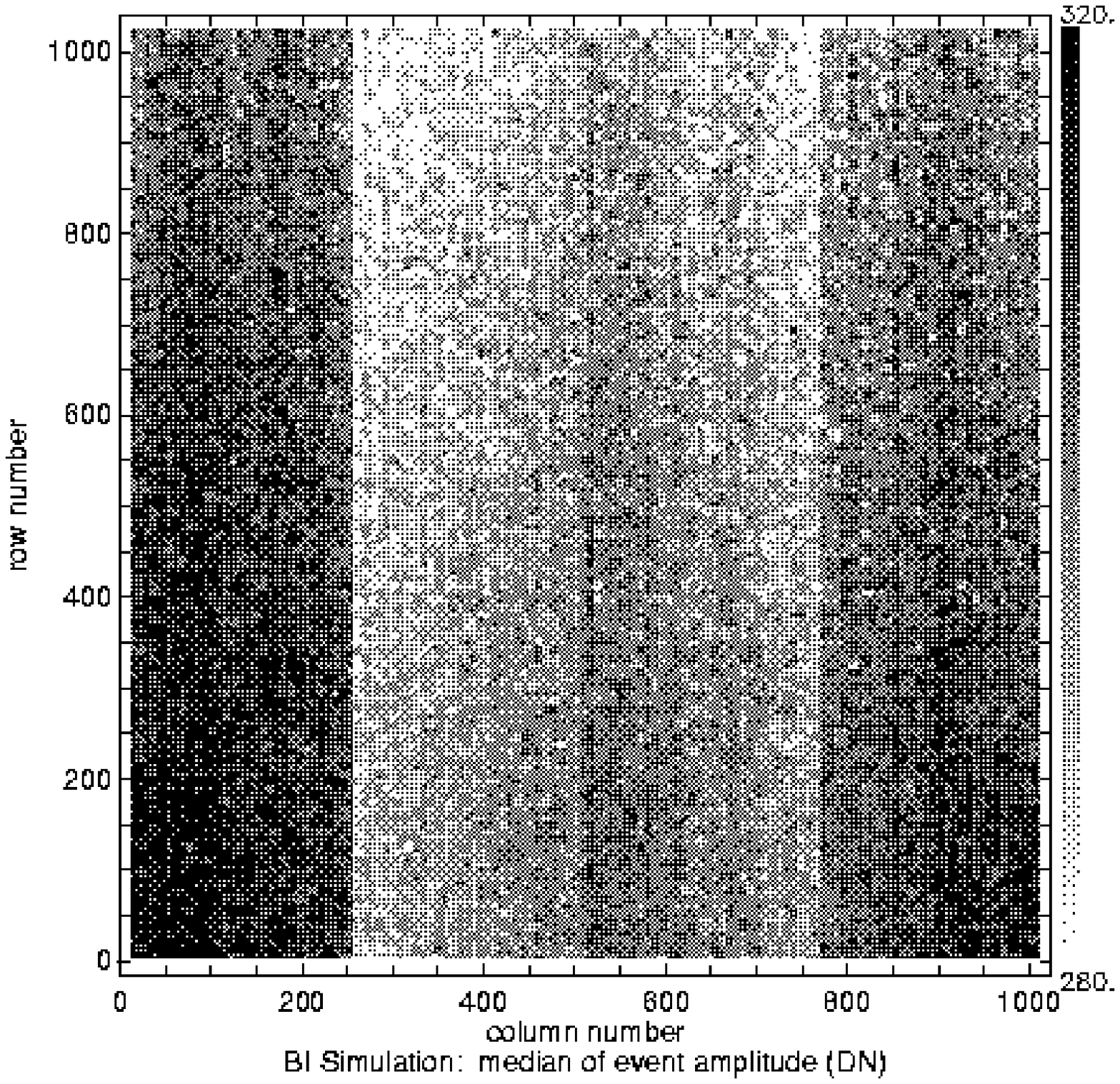,width=3.0in } }}
\centerline{\mbox{
         \epsfig{file=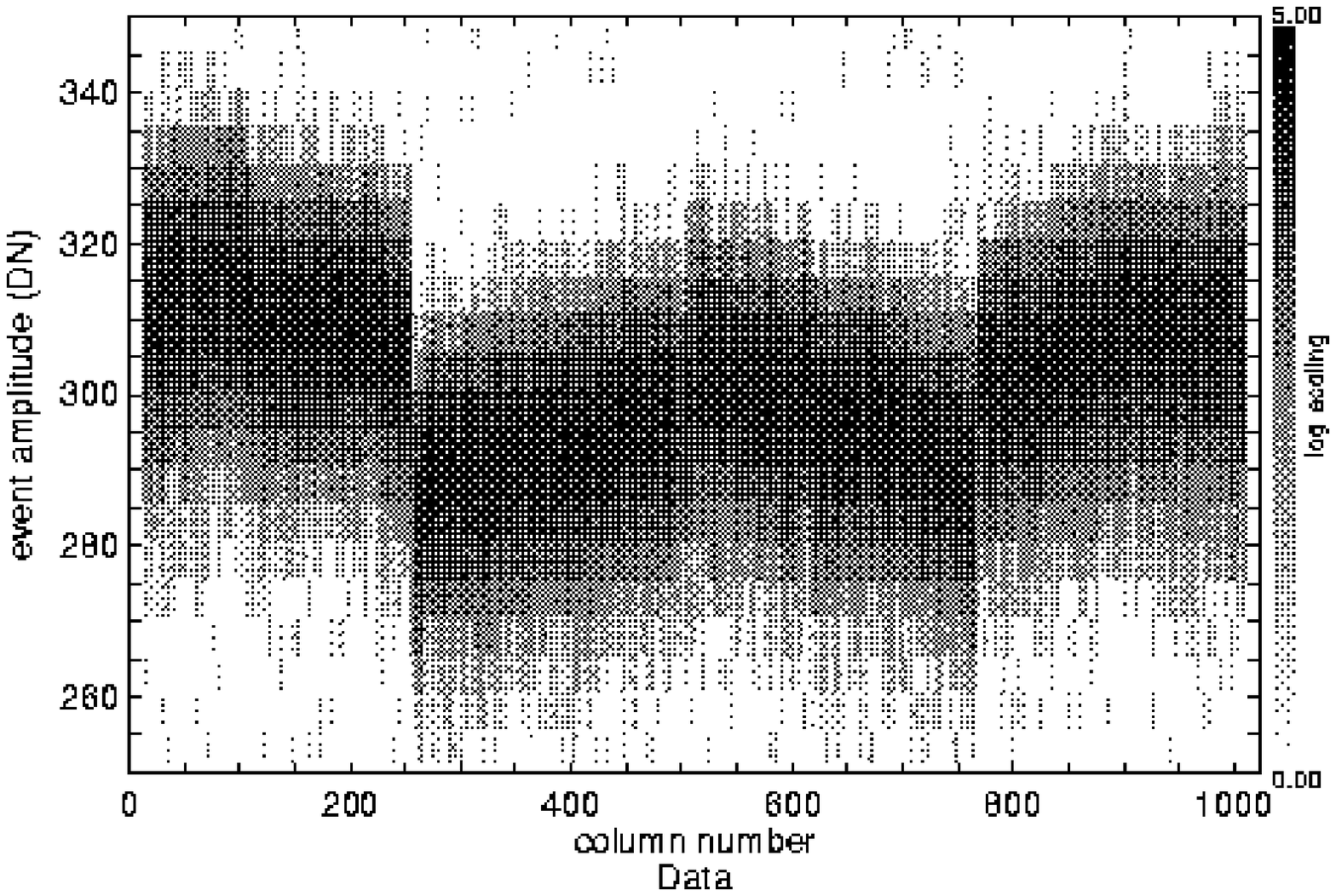,width=3.0in }
         \hspace{0.25in}
         \epsfig{file=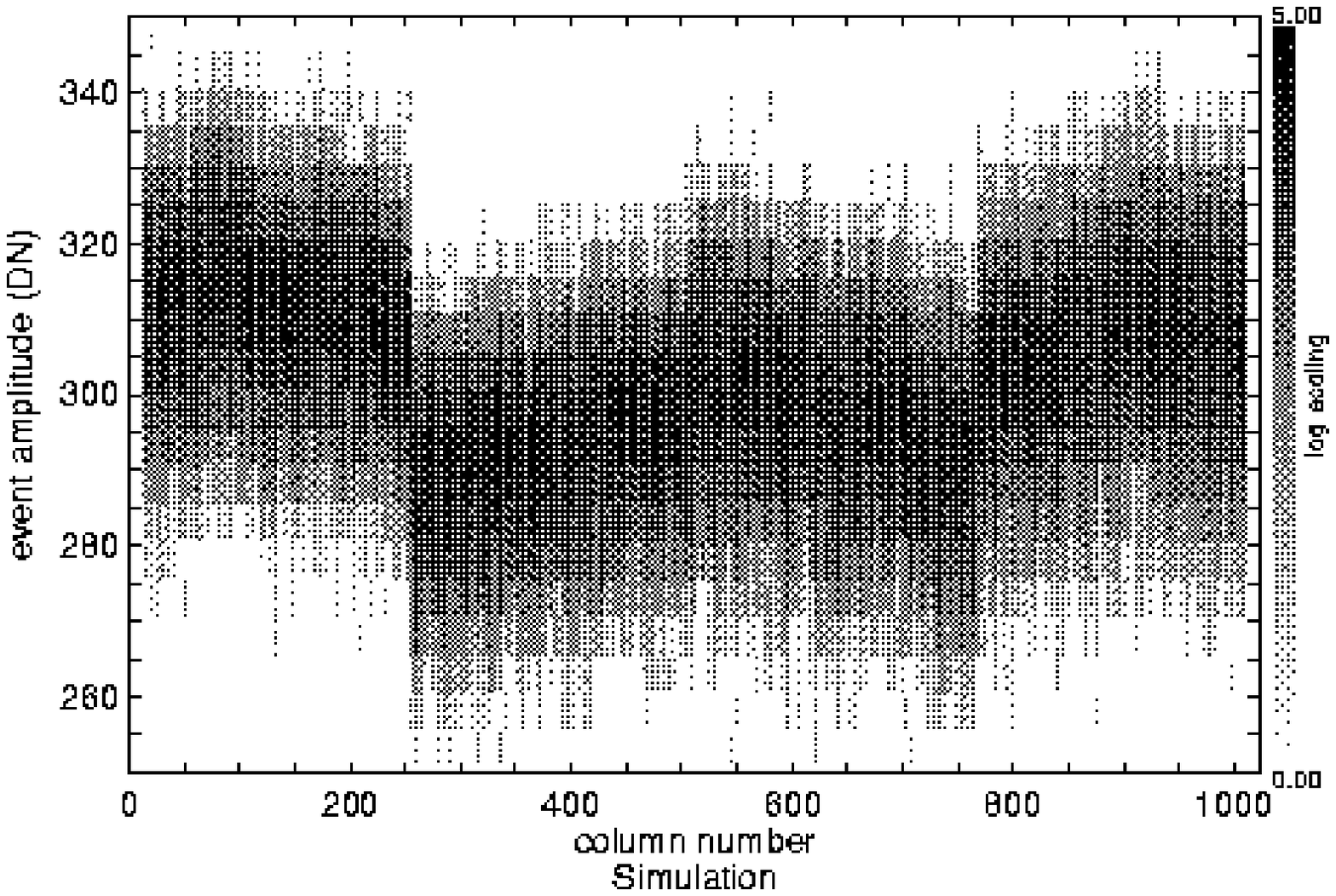,width=3.0in } }}
\caption{\protect \small Data (left) and simulation (right) of Al \Ka on the BI
device S3 at $-$120C.  Top:  Median images of Al \Ka line in units of DN,
binsize $8 \times 8$ pixels.  Bottom:  Al \Ka line in units of DN, as a
function of column number.  All 4 amplifiers are shown.} 

%Upper panels:
%Aspect ratio 1.  Median image of Al Ka line in PHA space, binsize 8x8.
%Used "partial_index" for data.

%Lower panels:
%Aspect ratio 0.7.  Al Ka line in PHA space vs. chipx, binsize 4x5.
%Used "partial_index" for data.
 
\normalsize
\label{fig:bi-image-comp}
\end{figure}

The upper panels of Figure~\ref{fig:bi-image-comp} show the same Al \Ka
map as Figure~\ref{fig:fi-image-comp}, now for the BI device S3.  As
the figure clearly shows, the four amplifiers in the S3 device have
slightly different responses.  This is reproduced well by the
simulation, as are the CTI-induced spatial variations.  The column-wise
variations in the Al \Ka event amplitudes are shown in the lower panels
of Figure~\ref{fig:bi-image-comp}.  Again the simulator does a
reasonable job of reproducing these variations.  Both the imaging and
the framestore sections of BI CCDs suffer CTI, so no horizontal band
of CTI-free events is seen in the lower left panel of Figure~\ref{fig:bi-image-comp}.

%-------------------------------------------------------------------------  
%-------------------------------------------------------------------------

\clearpage

%==========================================================================
\section{The Corrector  
\label{sec:corrector}}

As mentioned above, we have used our Monte Carlo modeling of CTI
effects to build a CTI corrector.  This program attempts to remove
those effects from individual ACIS events, taking a forward-modeling
approach.  For each observed event $O$ we first hypothesize the nine
pixel values of a corresponding clean event $C$, {\em i.e.} the event
that would have been obtained if no CTI effects were present.  This
clean event is passed through the CTI model, producing a model event
$M$ which estimates what we should have observed if our clean event
hypothesis is correct.  The hypothesis (the nine pixel values of $C$)
is then adjusted using the differences between the observed event and
the model event in a simple way:
\begin{equation}
C_{new} = C_{old} + (O - M).
\end{equation}
This iterative process continues until the model and observed events
differ by less than 0.1 DN (for each of the 9 pixels).  When
convergence is obtained, C is returned as the CTI-corrected event.

Tests show that when a limit of 15 iterations is imposed within the
corrector a trivial number of events fail to meet the 0.1 DN
convergence criterion; this limit is imposed as a safety in the code
and the number of iterations needed for convergence (typically 3--5) is
recorded for every simulated event.  The pixel values and thus the
event amplitudes produced by the corrector are real numbers.

Because the corrector operates on event islands (3$\times$3 pixel arrays)
rather than on event amplitudes, it has the potential to improve the
grades in an event list, leading to a different grade distribution.
Thus the CCD quantum efficiency appropriate for grade-filtered
corrected data will differ from that appropriate for similarly
grade-filtered uncorrected data.  A technique for determining this QE
adjustment is presented in Section~\ref{sec:qeu}.

The CTI corrector consists of a set of IDL \cite{idl} programs and
device-dependent parameter files that instantiate the model for each
amplifier on each CCD, for a given epoch of the {\em Chandra} mission.  
Obviously observations must be configured to telemeter the $3 \times 3$
pixel event island in order for the corrector to modify that island.
The corrector operates on the ``Level 1'' event list, one of the
standard data products, and requires no additional information regarding
the observation.  The parameter tuning presumes that the CCDs were
operated in their standard full-frame mode, with a frame readout time of
3.24 seconds.  Since the trap time constants interact with this readout
time to affect the average number of filled traps, observations made
in any of the ``sub-array'' modes (including Continuous Clocking Mode)
will exhibit different CTI and cannot be corrected using this full-frame
tuning.  

The code is available and may be of
interest to other groups concerned with the results of radiation damage
on X-ray and visual CCD detector systems and to {\em Chandra}/ACIS
users.  We have recently completed tuning our CTI corrector for all
chips in the ACIS-I array and chips S2 and S3 in the ACIS-S array, for
focal plane temperatures of $-$110C and $-$120C.  We provide the parameter
files used by the corrector code on our webpage, so any {\em
Chandra}/ACIS user with data on one or more of the allowed CCDs may
experiment with the PSU CTI corrector.

Response matrices have been generated for these device/temperature
combinations as well, using the CTI model as part of the Monte Carlo
CCD simulator to generate simulated events at a large number of
monochromatic energies.  The simulator applies CTI to the events, then
the corrector is used to remove the non-random effects of CTI.  The
resulting events are combined into a (now polychromatic) instrumental
spectrum, which represents a single row of the response matrix.  An
example was shown in T01.  These response matrices are also available to
{\em Chandra}/ACIS users.

%==========================================================================
\section{Corrector Results} \label{sec:results}
%==========================================================================

%====================================
\subsection{Spectral Resolution}

\begin{figure}[htb]
\centerline{\mbox{
	 \epsfig{file=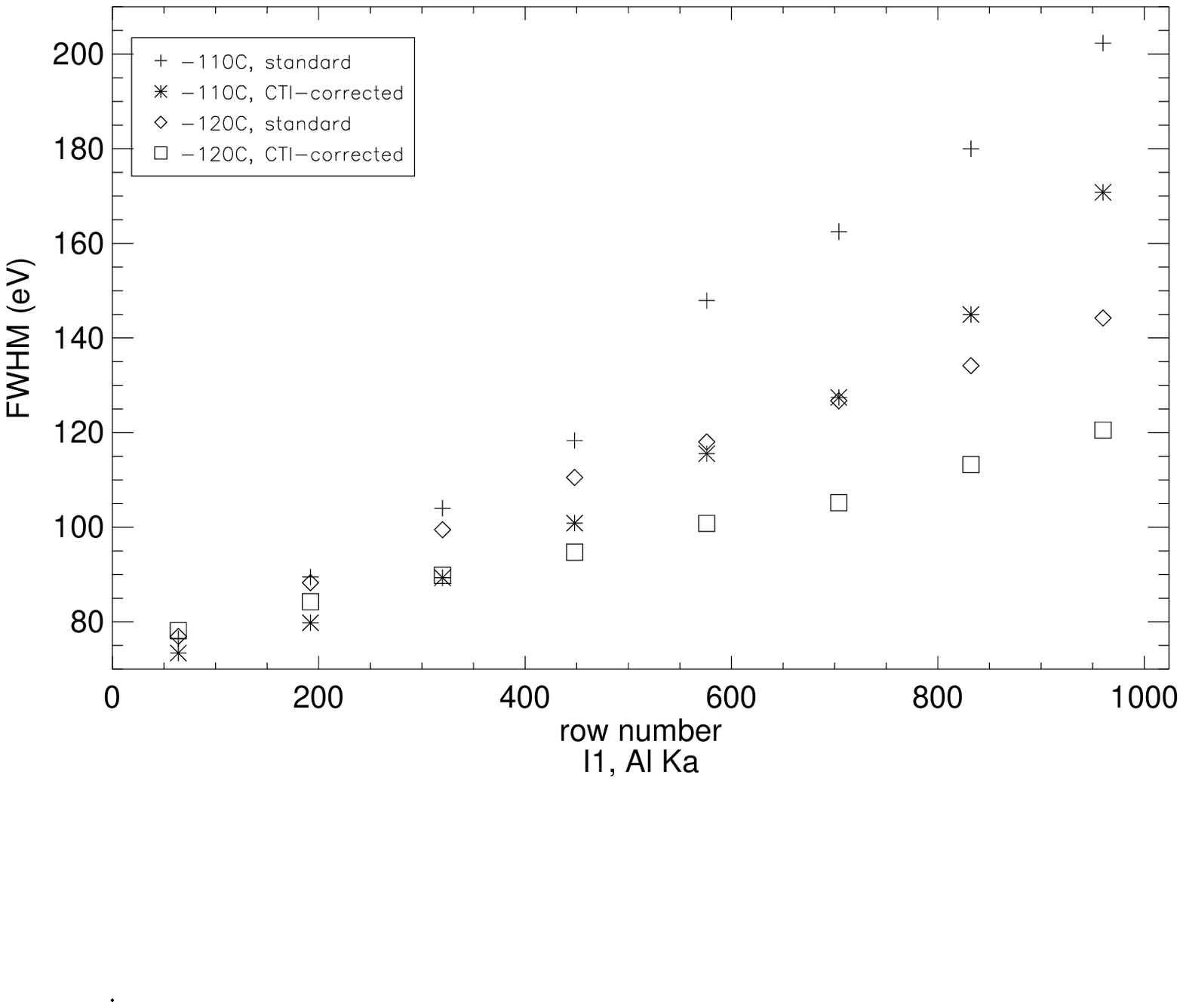,width=3.0in }
	 \hspace{0.25in}
	 \epsfig{file=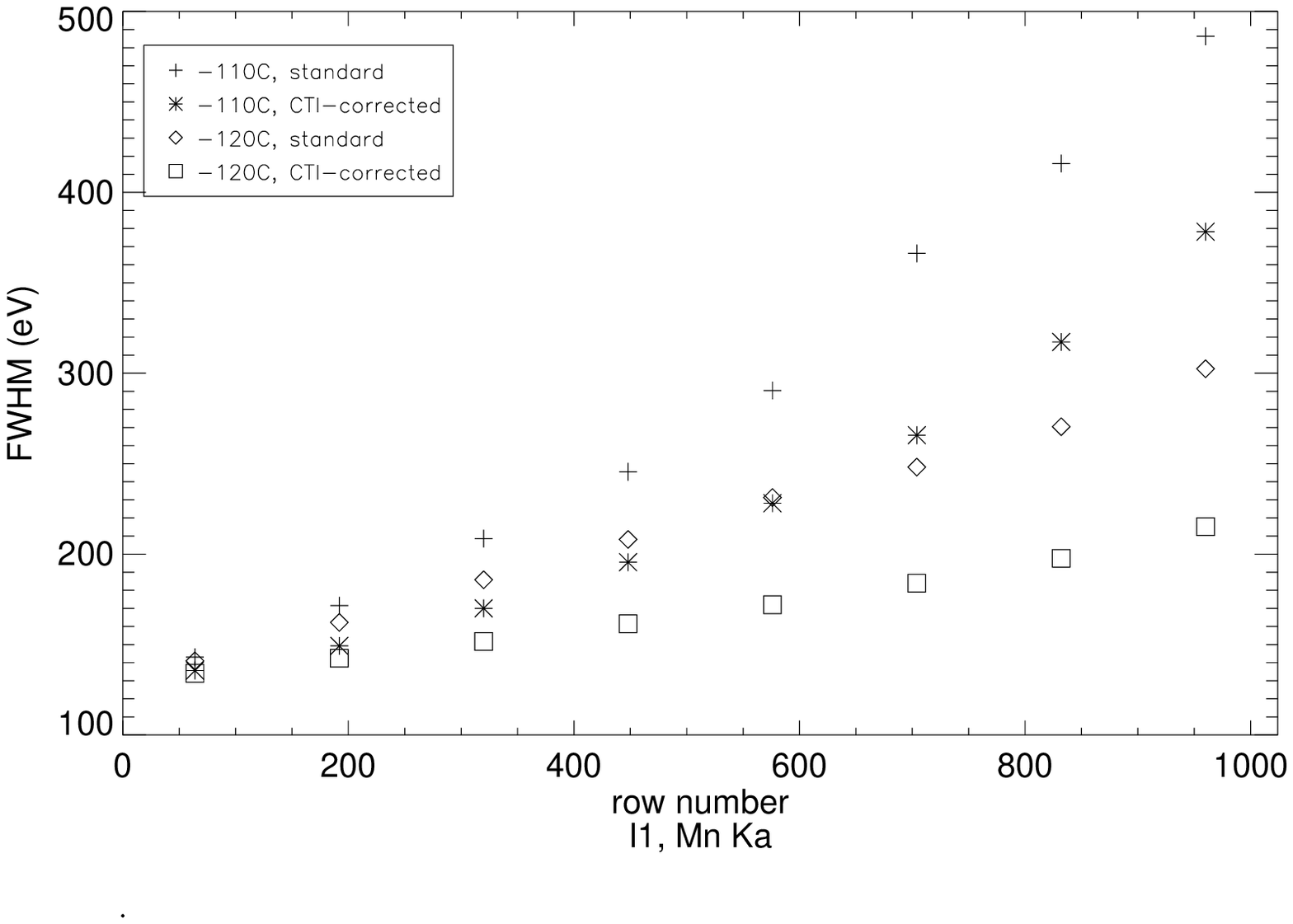,width=3.0in } }}
\caption{\protect \small Comparing linewidths for standard and CTI-corrected
ECS data at 1.486 keV (left) and 5.895 keV (right), for the FI chip I1, using standard (g02346) grades
and all amplifiers.  Note the different scales on the vertical axes.}

\normalsize
\label{fig:fi-linewidthcomp}
\end{figure}

The most fundamental result of CTI correction is the improvement in
spectral resolution.  This is illustrated in Figure~\ref{fig:fi-linewidthcomp}
for an FI device at two ECS energies.  This figure also illustrates the major
improvement in FI spectral resolution made by reducing the ACIS focal
plane temperature from $-$110C to $-$120C.  The CTI corrector improves the
resolution at both focal plane temperatures and all energies for the
uniform illumination of the ECS.  Celestial sources, which tend to be
more pointlike and thus suffer less from the variations across the CCD
than ECS data, will show less dramatic improvement. 

%-------------------------------------------------------------------------
%-------------------------------------------------------------------------

\begin{figure}[htb]
\centerline{\epsfig{file=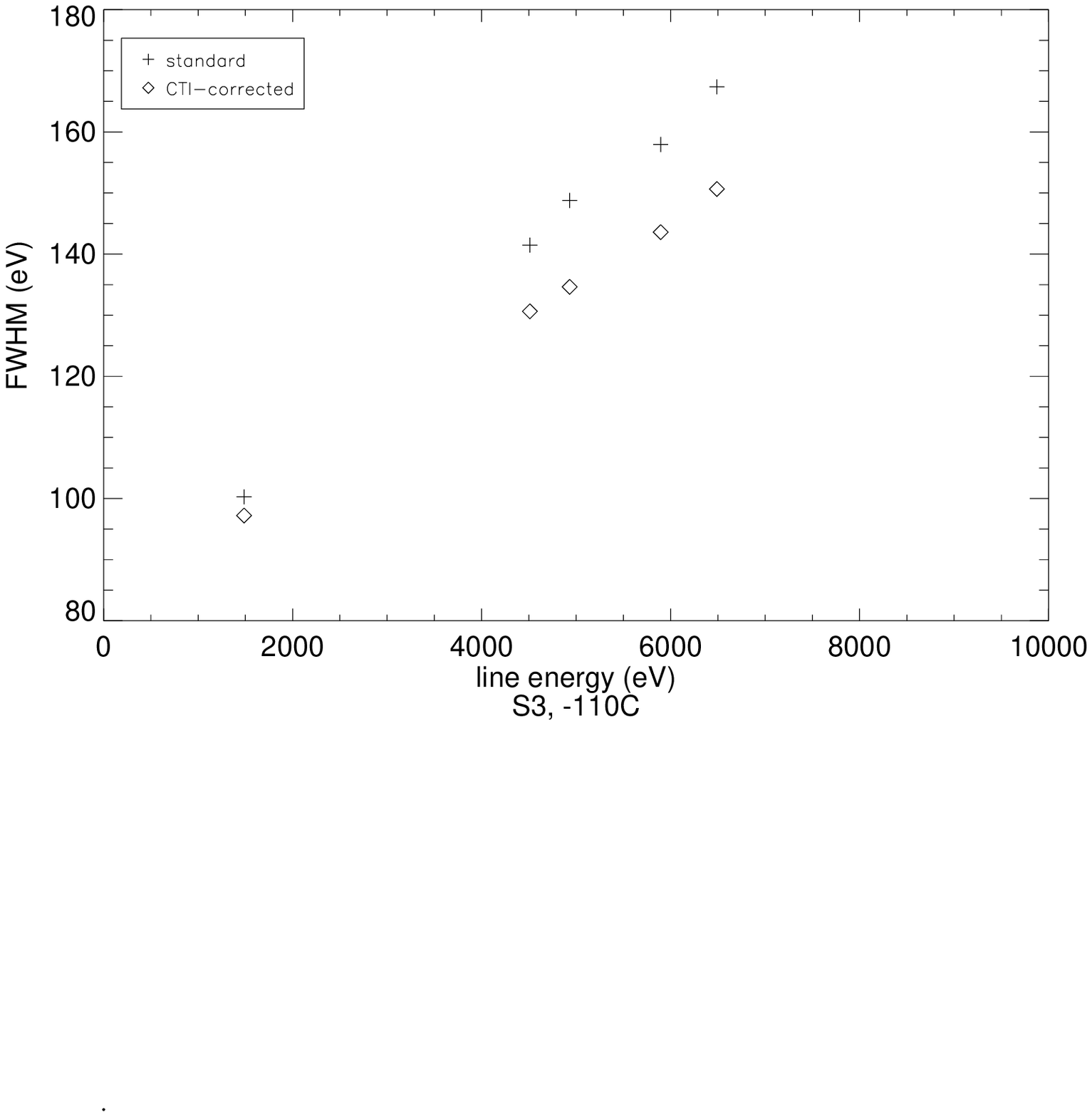,width=3.0in } }
\caption{\protect \small Comparing linewidths for standard and CTI-corrected
ECS data for the BI chip S3, using standard (g02346) grades and all amplifiers. }

\normalsize
\label{fig:bi-linewidthcomp}
\end{figure}

Figure~\ref{fig:bi-linewidthcomp} shows the ECS spectral resolution improvement
with CTI correction on a BI device, now as a function of energy rather
than position, as the BI devices do not exhibit position-dependent
resolution.  Only $-$110C data are shown here; $-$120C results are similar.
Again the spectral resolution is improved at all ECS energies.  Below 1.5~keV,
though, the device's intrinsic resolving ability dominates the CTI effects.
Here, CTI correction will not markedly improve the spectral resolution.
As for FI devices, the spectral resolution improvement for more pointlike
celestial sources will be less dramatic.

%-------------------------------------------------------------------------
%-------------------------------------------------------------------------

Comparing Figures~\ref{fig:fi-linewidthcomp} and
\ref{fig:bi-linewidthcomp} shows that the best ACIS spectral resolution
is achieved by placing targets on the lower half of ACIS FI devices.
In these regions, the FI resolution surpasses the BI resolution.
Without CTI correction, only the lowest quarter of the FI devices
surpassed the BI resolution.  This improvement is particularly
important for targets with extended emission or point sources
distributed across the field, such as star formation regions
\cite{garmire00} and the {\em Chandra} Deep Field observations of the
X-ray background \cite{hornschemeier00}.

%====================================
\subsection{Spectral and Spatial Comparisons}

To give more detail on the effects that the CTI corrector has on ECS
data, figures identical to those in Section~\ref{sec:data} are
presented below.

Figure~\ref{fig:corrspec} shows CTI-corrected ECS spectra.  For ease of
comparison, in this one case we reproduce Figure~\ref{fig:stanspec} in the
top two panels and show the CTI-corrected spectra below.  The spectral 
resolution is improved for both FI (left) and BI (right) spectra.  

%-------------------------------------------------------------------------  
\begin{figure}[htb]
\centerline{\mbox{
         \epsfig{file=spectrum_standard_I3_amp3_120.ps,width=3.0in }
         \hspace{0.25in}
         \epsfig{file=spectrum_standard_S3_amp0_120.ps,width=3.0in } }}
\centerline{\mbox{
         \epsfig{file=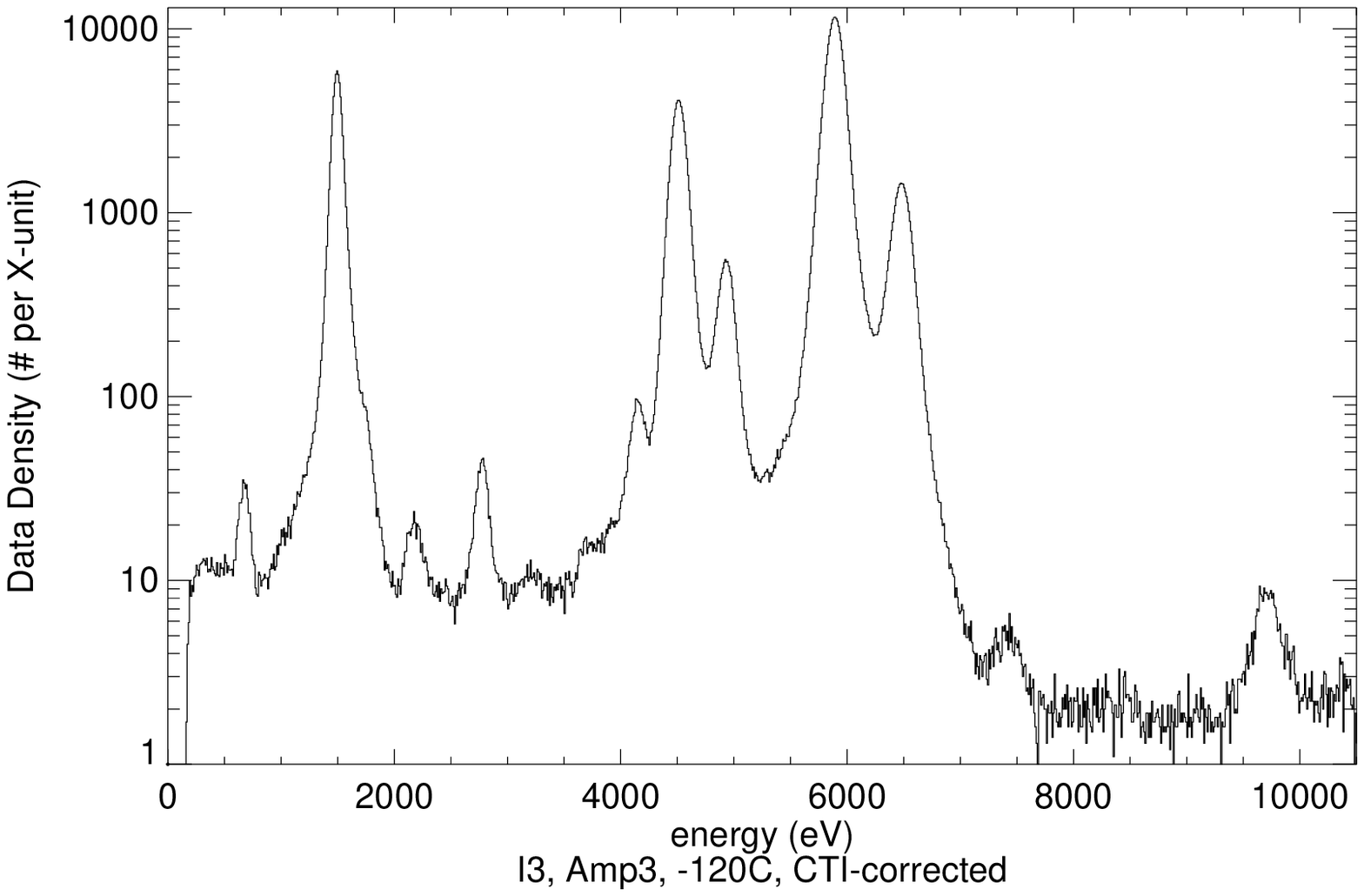,width=3.0in }
         \hspace{0.25in}
         \epsfig{file=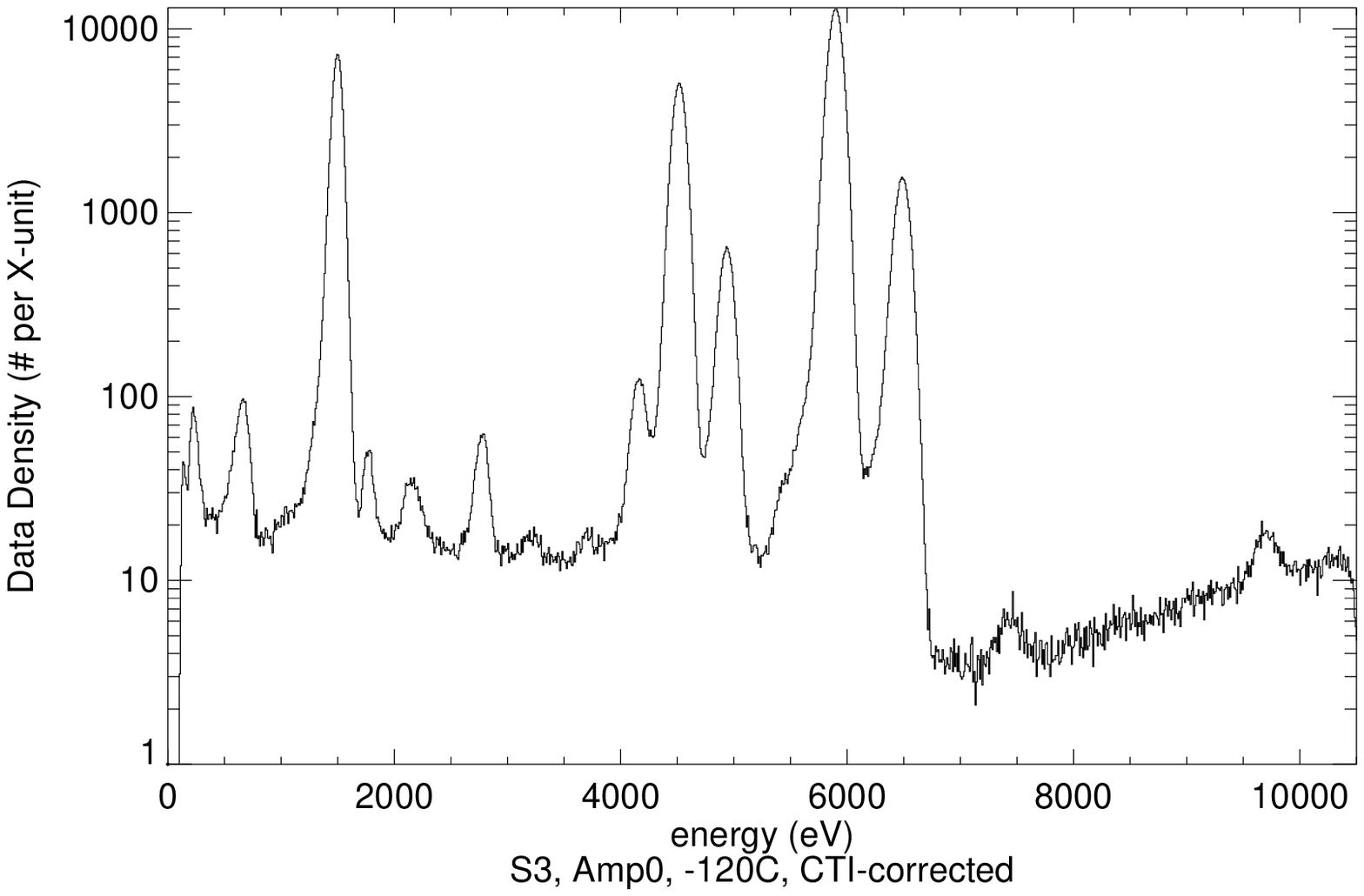,width=3.0in }}}
\caption{\protect \small ECS spectra, made by combining over 100
observations to give $\sim 4 \times 10^6$ events.  The top panels are
reproduced from Figure~\ref{fig:stanspec}; the bottom panels show
the CTI-corrected data.  These spectra use standard ASCA g02346
grade selection (applied after CTI correction for the bottom panels).  
Left:  all rows of I3, Amplifier
3, at $-$120C.  Right: all rows of S3, Amplifier 0, at $-$120C.}

%yrange (log) 1:13000, xrange 0:10500.  Aspect ratio 0.7.  Left margin 9.
%Used "short_index" in cti_correction subdirectory.
 
\normalsize
\label{fig:corrspec}
\end{figure}
%-------------------------------------------------------------------------

\begin{figure}[htb]
\centerline{\mbox{
         \epsfig{file=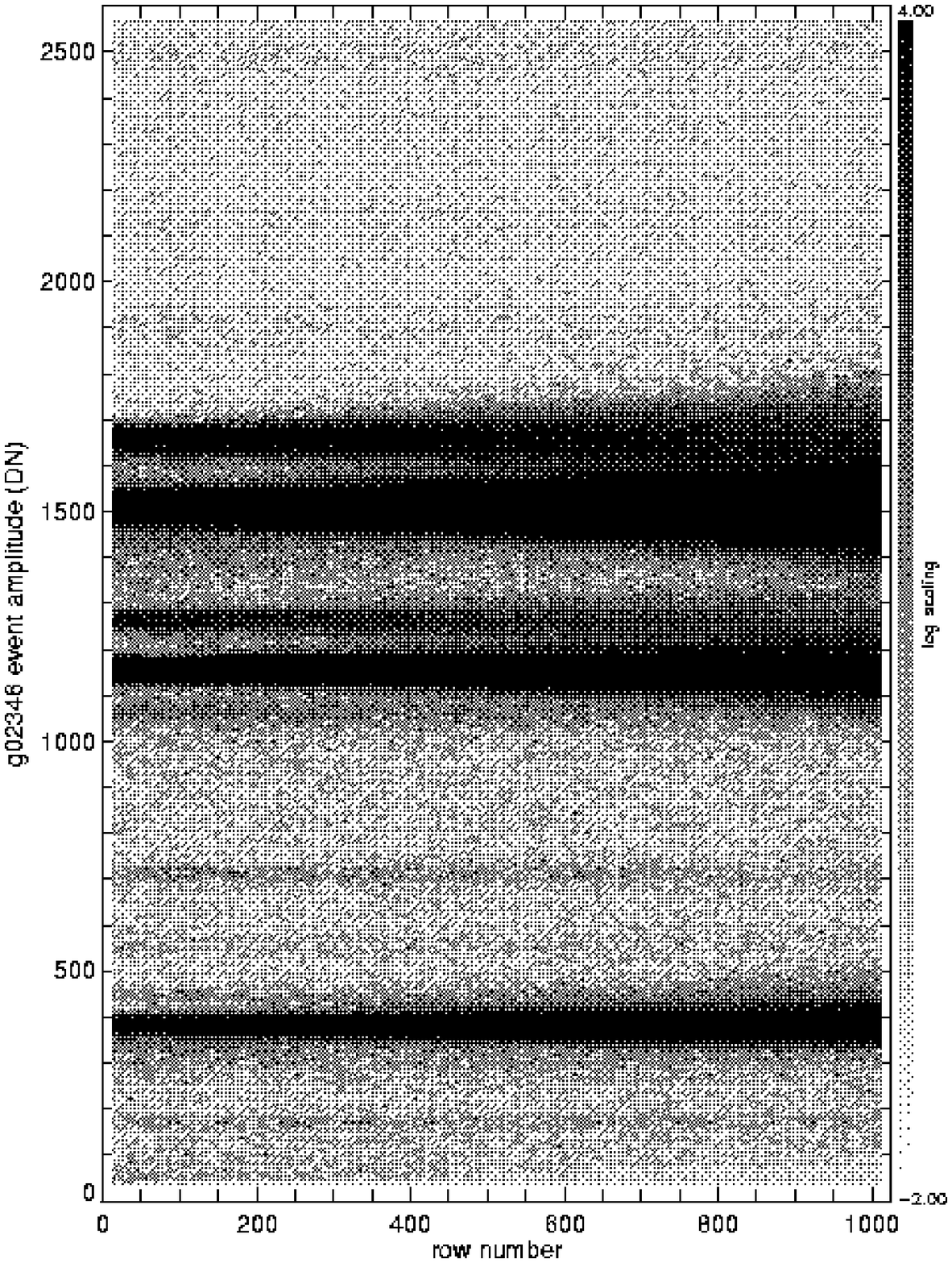,height=4.0in }
         \hspace{0.5in}
         \epsfig{file=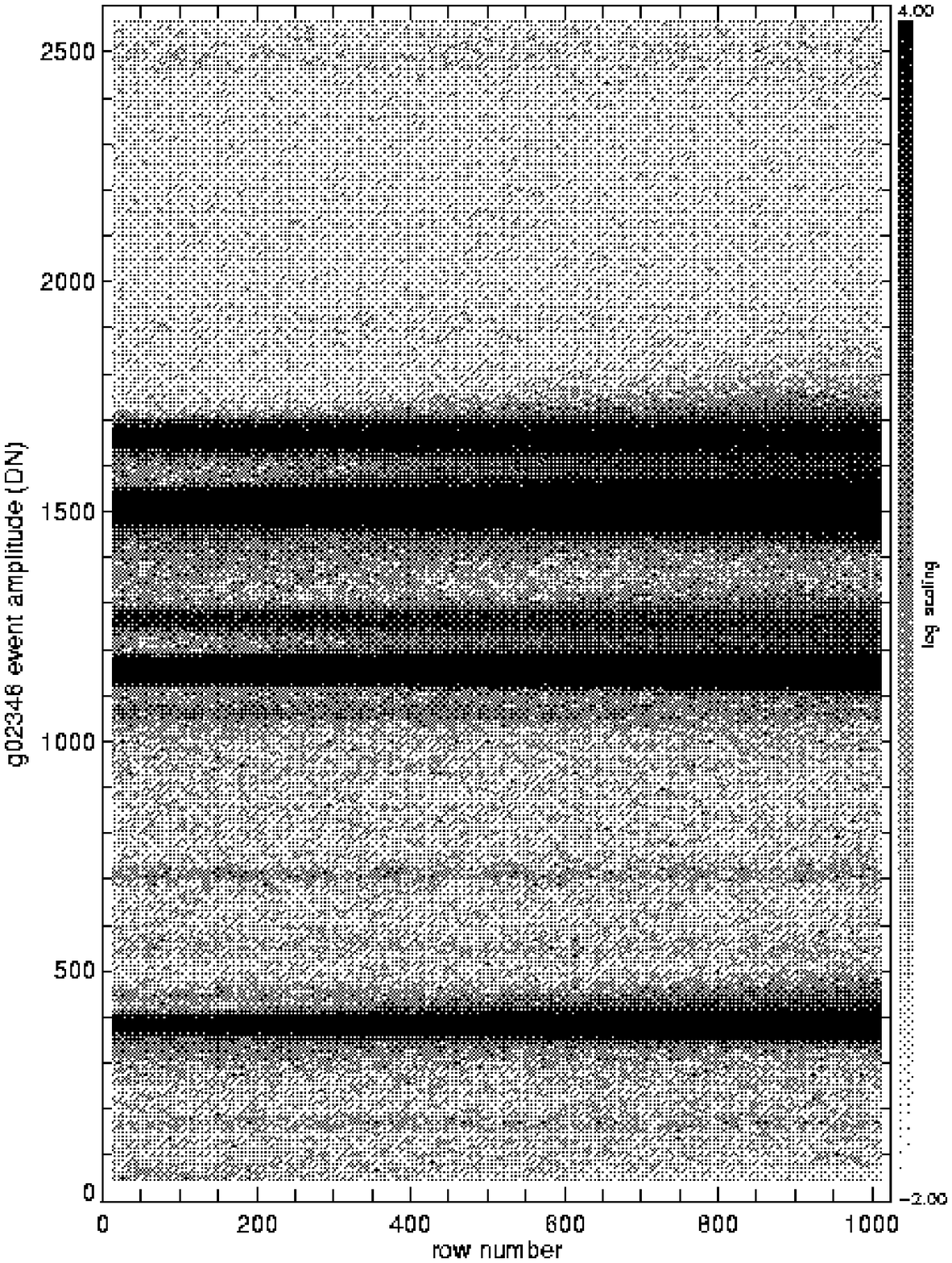,height=4.0in }}}
\centerline{\mbox{
         \epsfig{file=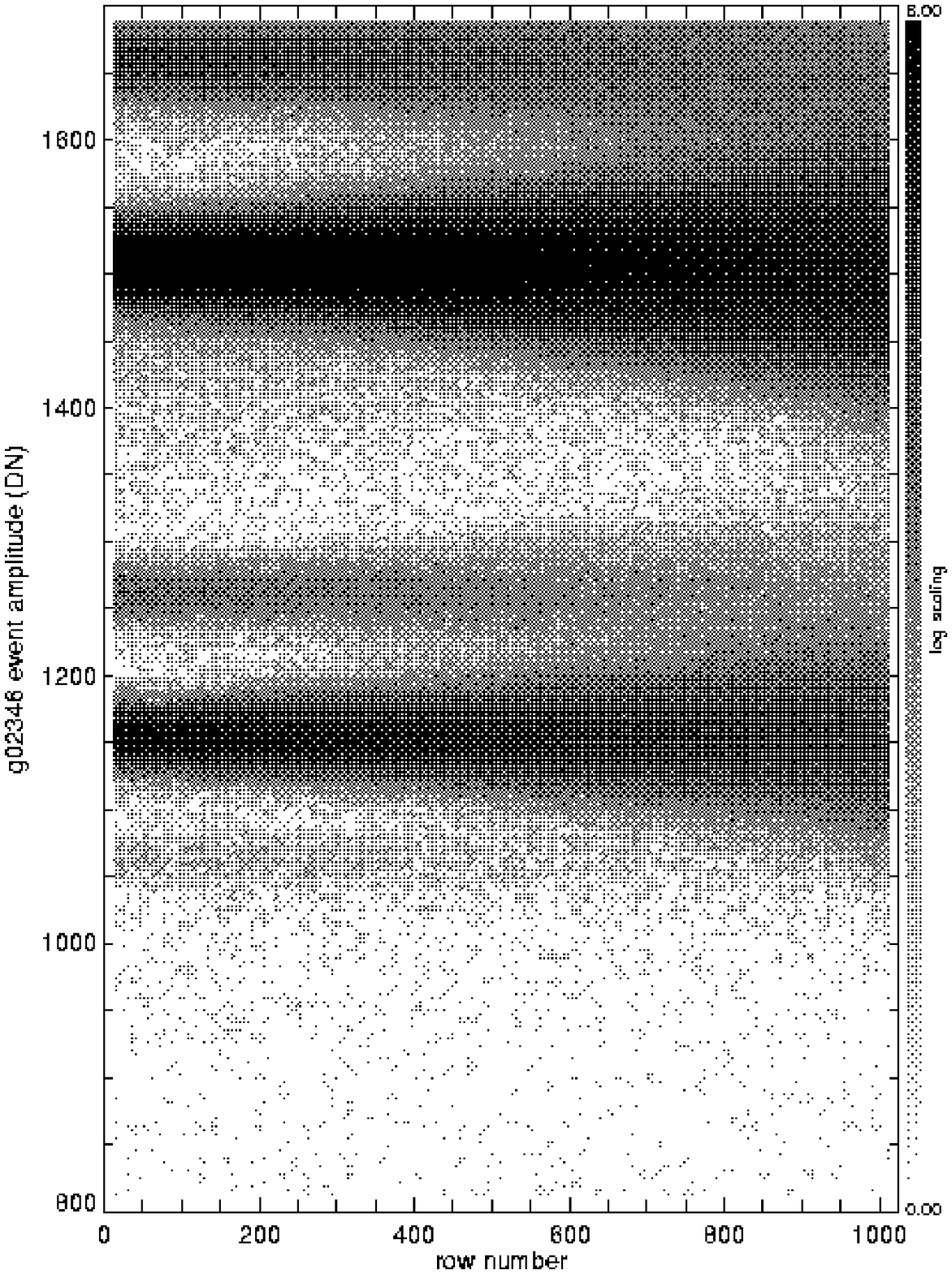,height=4.0in }
         \hspace{0.5in}
         \epsfig{file=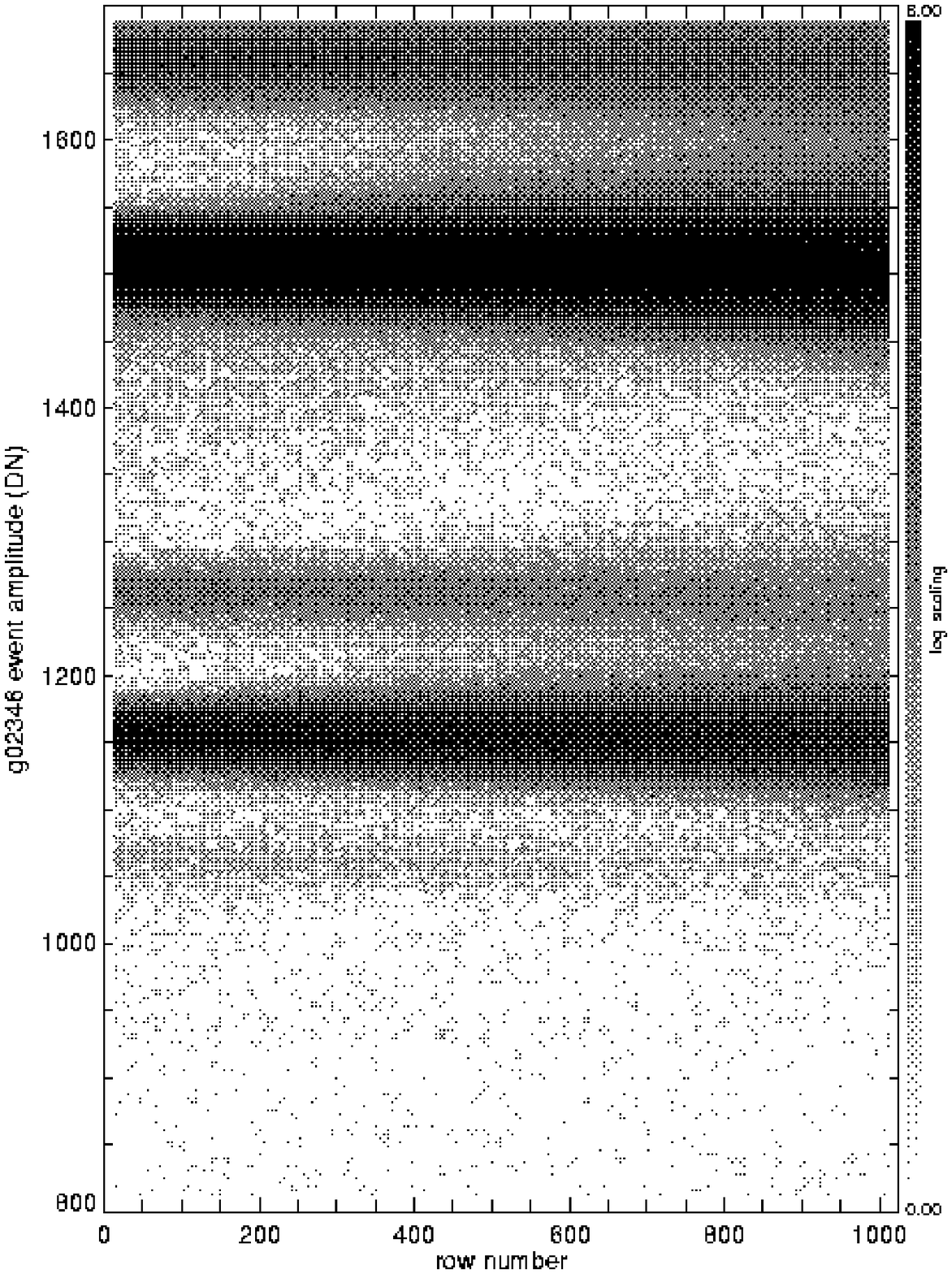,height=4.0in }}}
\caption{\protect \small Parallel CTI removed from the FI chip I3,
Amplifier 3, illustrated with data from the ECS, to be
compared with Figure~\ref{fig:allenergy-cti-fi}.  Standard grades
(``g02346'') are included.  Note the remaining row-dependent energy
resolution.  Left panels:  Corrected data at a focal plane temperature of
$-$110C.  The upper panel shows the full energy range; note the faint
but useful Au L$\alpha$ line at $\sim$2500 DN.  The lower panel is an
expanded view of the Ti and Mn calibration lines.  Right panels:  the
same plots for a focal plane temperature of $-$120C.  Both $-$110C and
$-$120C datasets contain about $1.8 \times 10^6$ events to facilitate
comparison. }

% Aspect ratio for these figures is 1.4.
% Binsize used for the image was 8x4.  Y-range 0-2600 for all, 800-1700 
% for zoom.

\normalsize
\label{fig:allenergy-no-cti-fi}
\end{figure}

The performance of the corrector in removing the observed spatial
variation of event amplitude in FI ECS data is shown in
Figure~\ref{fig:allenergy-no-cti-fi}.  The corrector has largely
removed the gain changes across the device; compare to
Figure~\ref{fig:allenergy-cti-fi}.  Since the spectral broadening with
row number is primarily the result of a random process involving one of
the trap species \cite{antunes93}, it cannot be completely suppressed
in our reconstructed events.

%-------------------------------------------------------------------------
%-------------------------------------------------------------------------

\clearpage

\begin{figure}[htb]
\centerline{\epsfig{file=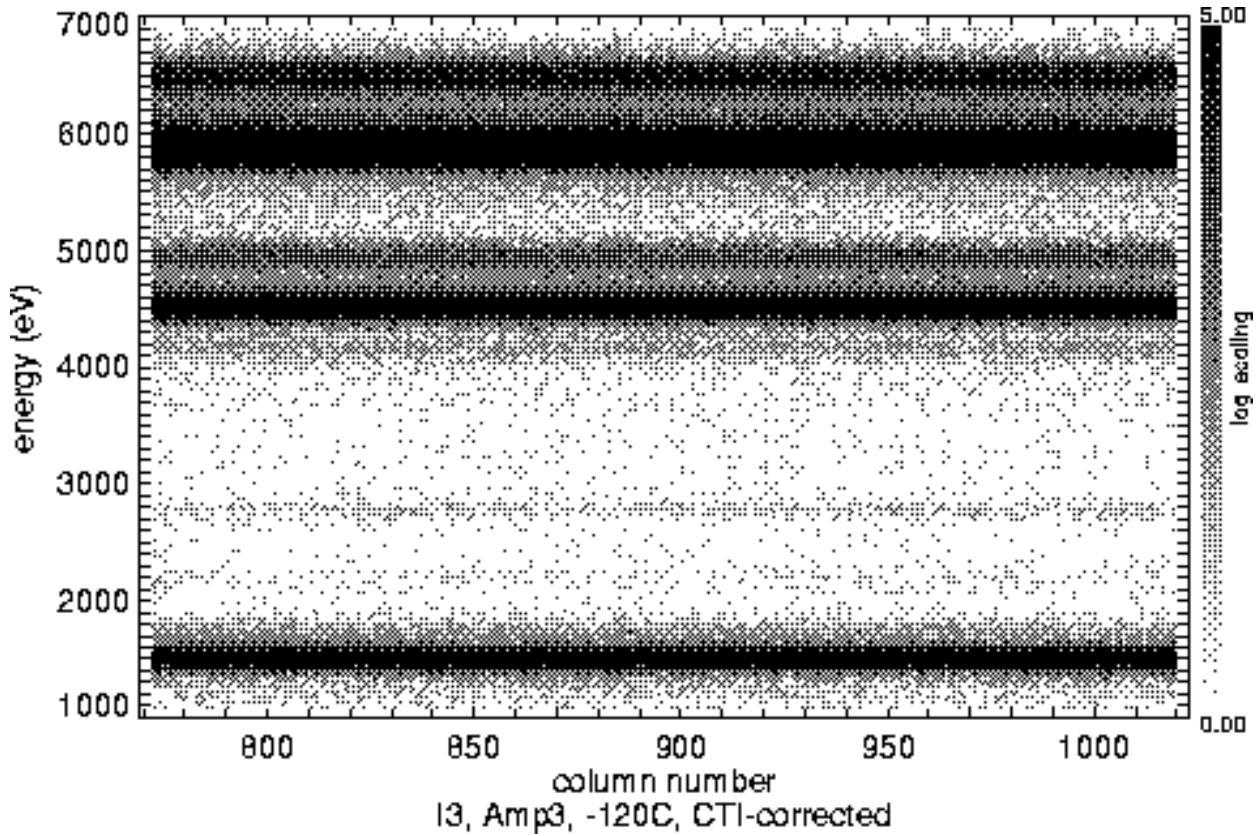, width=6.5in }}
\caption{\protect \small Image showing energy vs.\ column number at the
main calibration energies in the ECS, averaged over all rows on
I3, Amplifier 3 at $-$120C, after CTI correction. Compare to Figure~\ref{fig:xgainvars}.}

% Aspect ratio for this figure is 0.7.  Binsize was 1x10.  Used all g02346
% events, ~4e6.
 
\normalsize
\label{fig:xgainvars_nocti}
\end{figure}

Figure~\ref{fig:xgainvars_nocti} illustrates the remaining
non-uniformities after the CTI corrector, with its deviation map, is used
to remove the column-dependent gain variations caused by CTI.  Comparing this to
Figure~\ref{fig:xgainvars}, it is clear that accounting for column-to-column
gain variations improves the quality of the data.

%-------------------------------------------------------------------------  
%-------------------------------------------------------------------------
%\clearpage

\begin{figure}[htb]
\centerline{\mbox{
         \epsfig{file=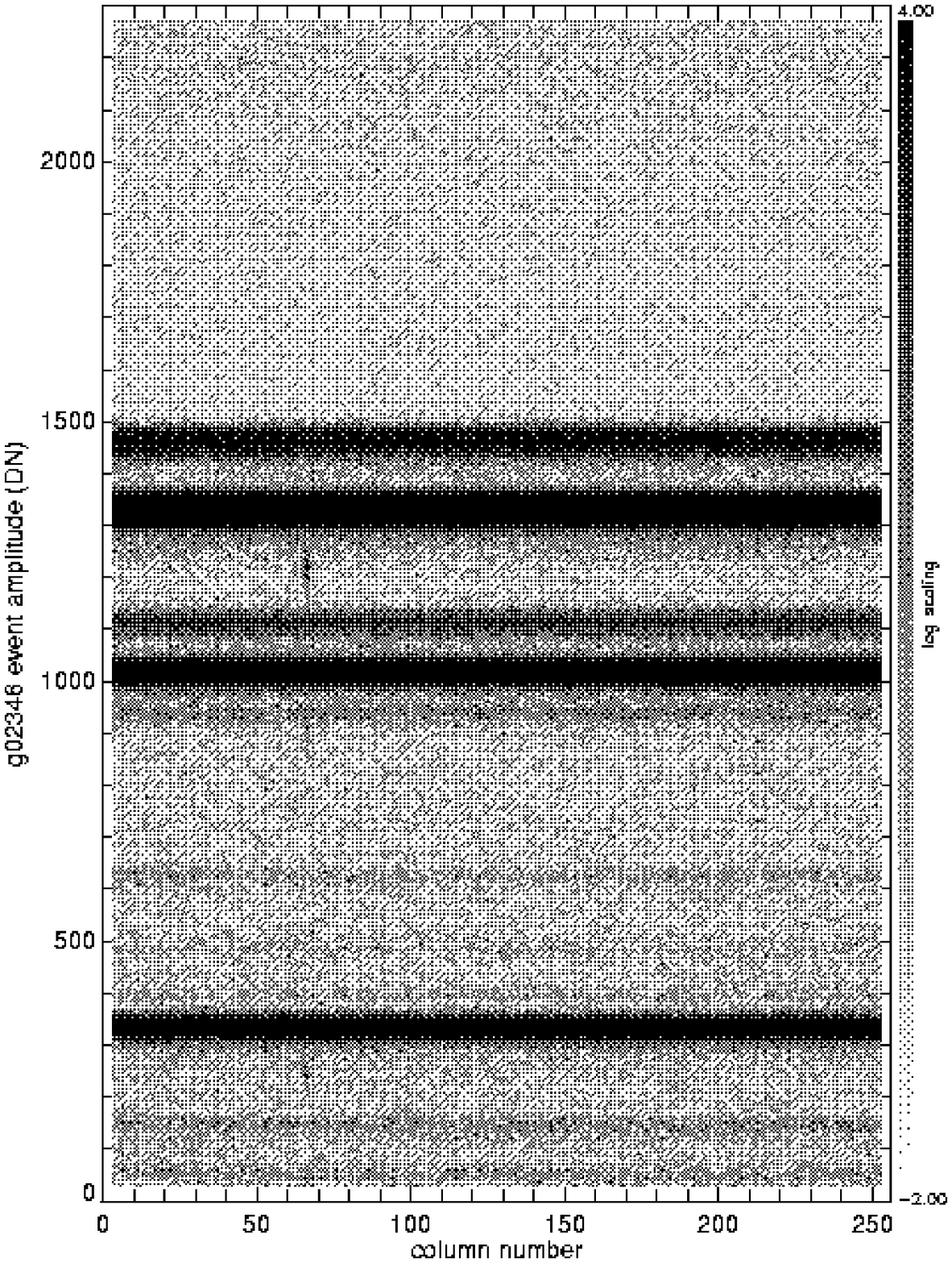,height=4.0in }
         \hspace{0.5in}
         \epsfig{file=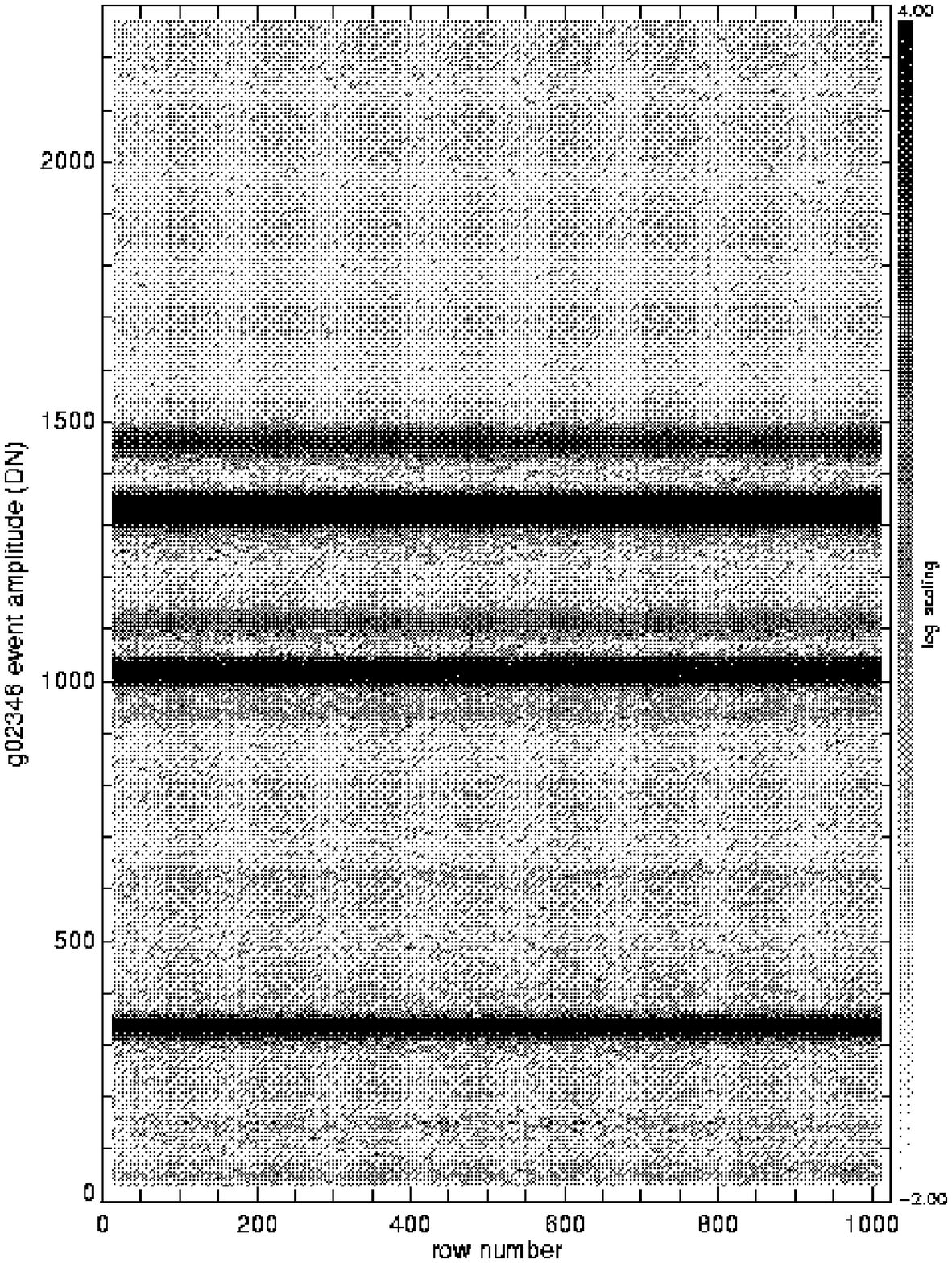,height=4.0in }}}
\centerline{\mbox{
         \epsfig{file=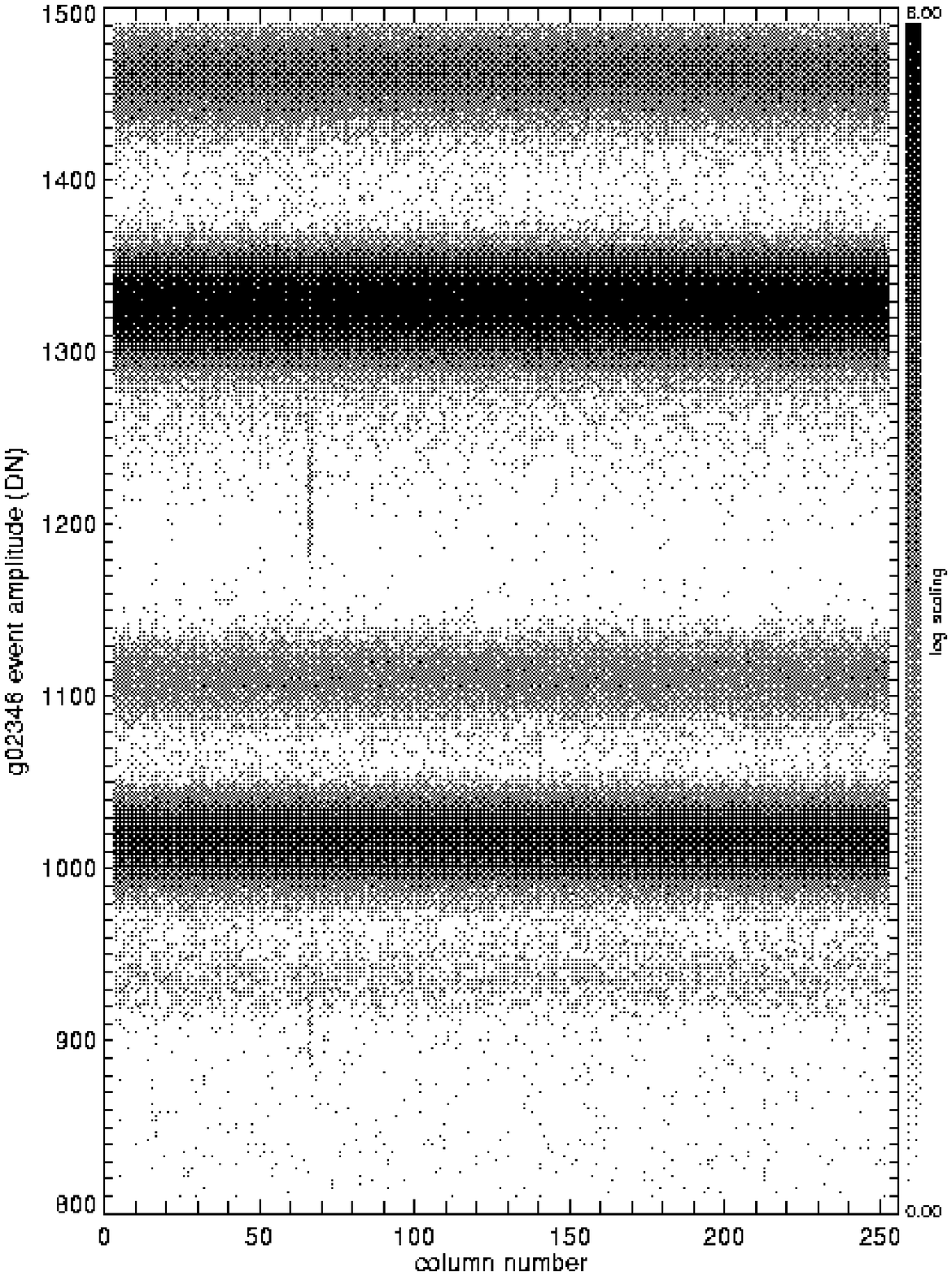,height=4.0in }
         \hspace{0.5in}
         \epsfig{file=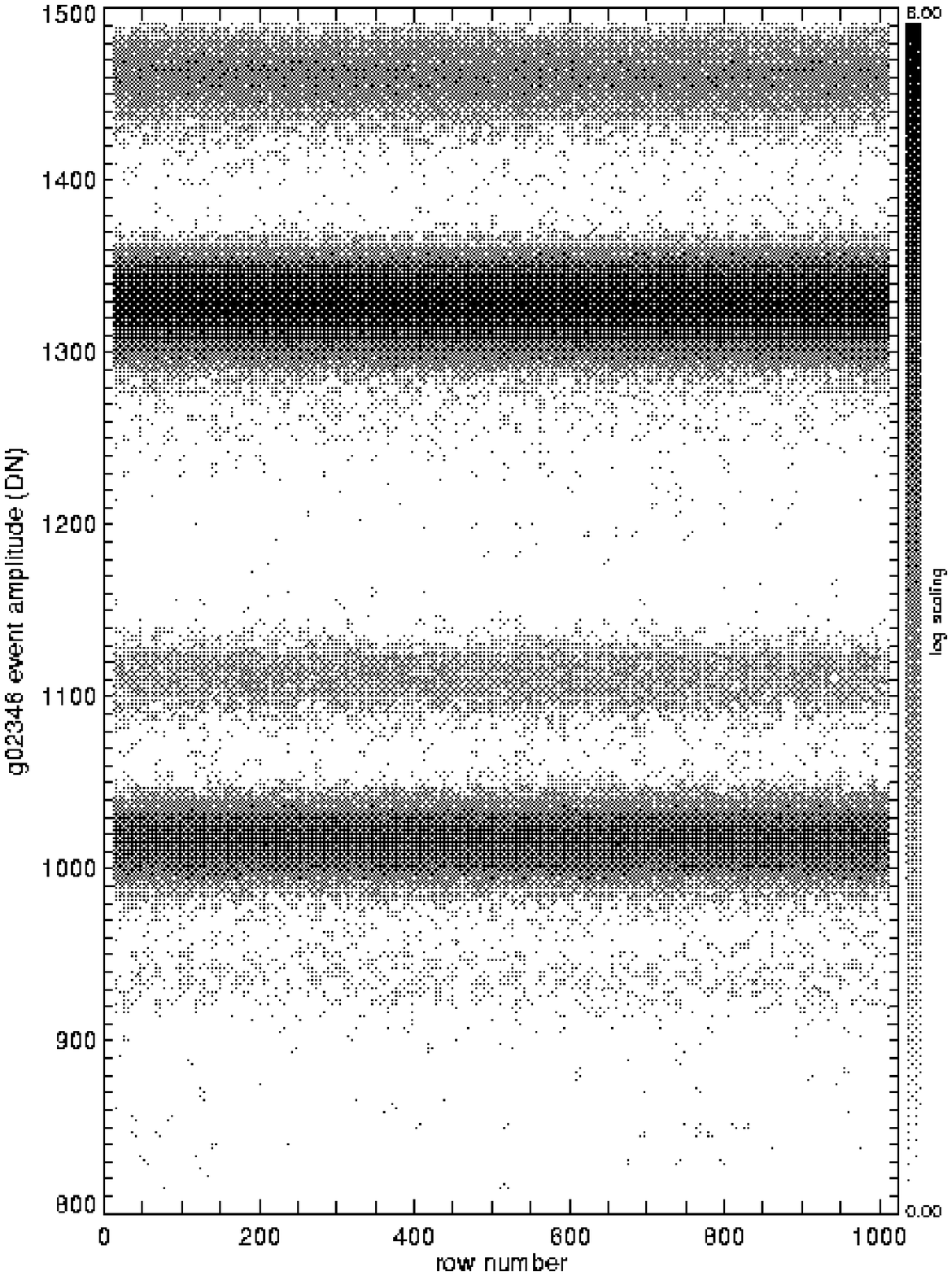,height=4.0in }}}
\caption{\protect \small Serial and parallel CTI removed from the BI
chip S3, Amplifier 0 (at $-$120C), illustrated with data from the ECS.
Standard grades (``g02346'') are included.  The row-dependent gain and
energy resolution and the energy-dependent charge loss are all largely
suppressed.  Left panels:  serial CTI.  The upper panel shows the full
energy range the lower panel is an expanded view of the Ti and Mn calibration
lines.  Right panels:  similar plots for parallel CTI.  As in
Figure~\ref{fig:allenergy-cti-bi}, only events in the 64 columns
closest to the readout node are shown.  Both datasets contain about
$1.8 \times 10^6$ events to facilitate comparison with each other and
with the FI data in Figure~\ref{fig:allenergy-no-cti-fi}. }

% Aspect ratio for these figures is 1.4.
% Binsize used was 8x4 for parallel, 1x4 for serial.  Y-range 0-2300 for all,
% 800-1500 for zoom.

\normalsize
\label{fig:allenergy-no-cti-bi}
\end{figure}

Now turning to a BI device, Figure~\ref{fig:allenergy-no-cti-bi} shows
the ECS data from S3, Amplifier 0, at $-$120C, after CTI correction.  The
spatial effects of CTI are much more subtle on BI devices than on FI devices;
the most obvious improvement compared to Figure~\ref{fig:allenergy-cti-bi}
is in the lower left panel, where the column-to-column gain variations
have been largely suppressed by the CTI corrector.

%-------------------------------------------------------------------------
%-------------------------------------------------------------------------

%\clearpage

\begin{figure}[htb]
\centerline{\mbox{
         \epsfig{file=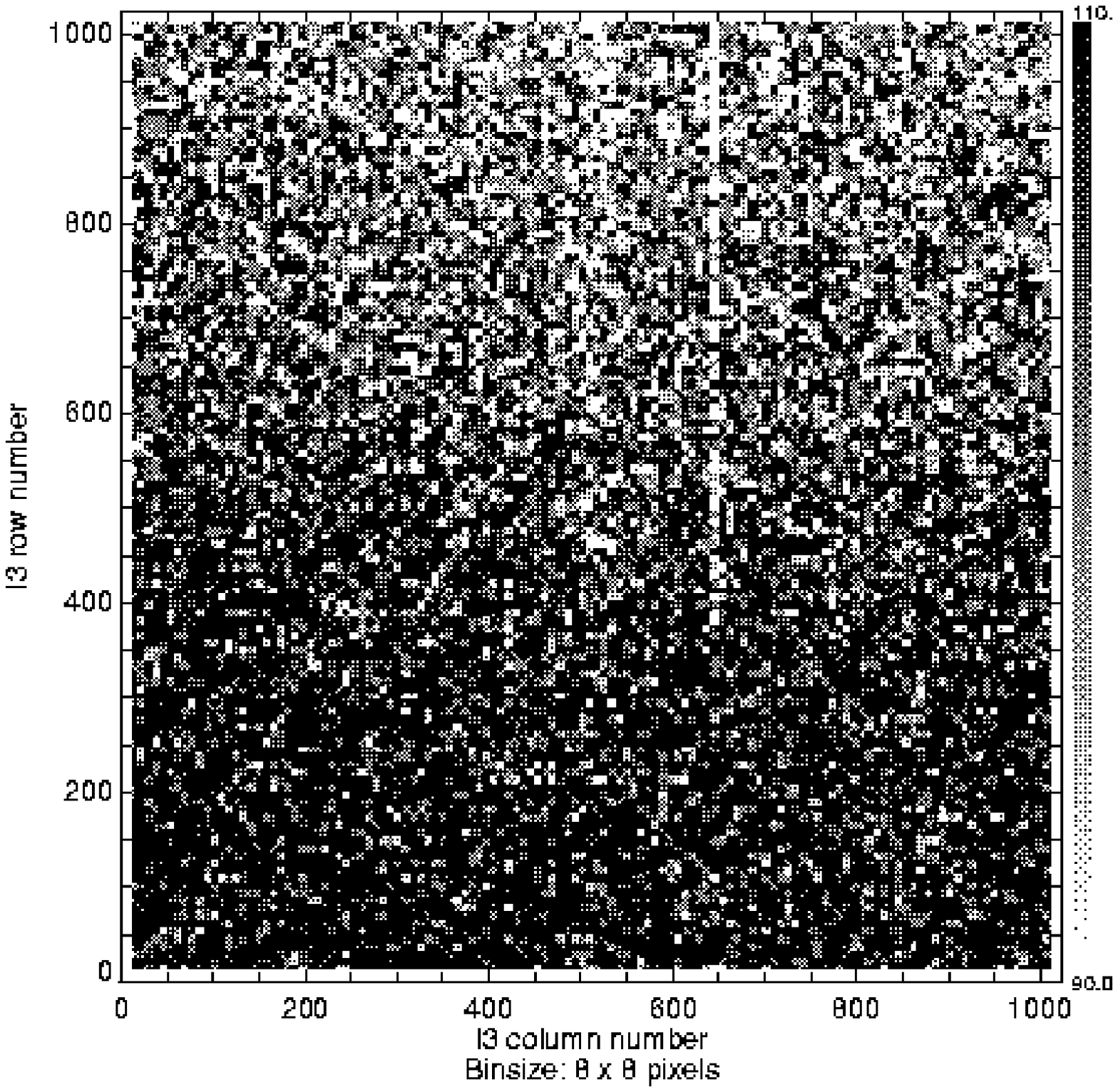,width=3.0in }
         \hspace{0.5in}
         \epsfig{file=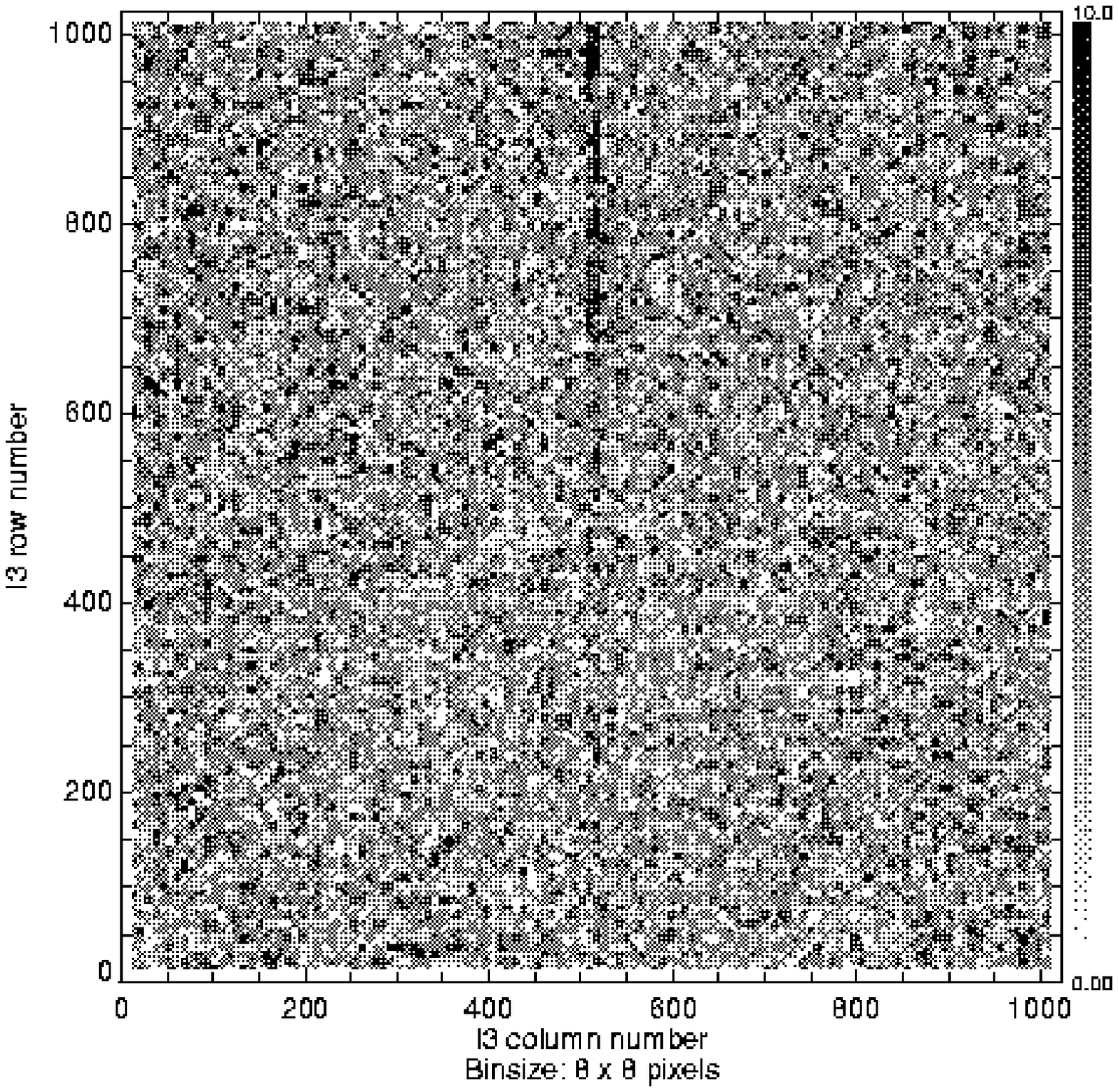,width=3.0in }}}
\centerline{\mbox{
         \epsfig{file=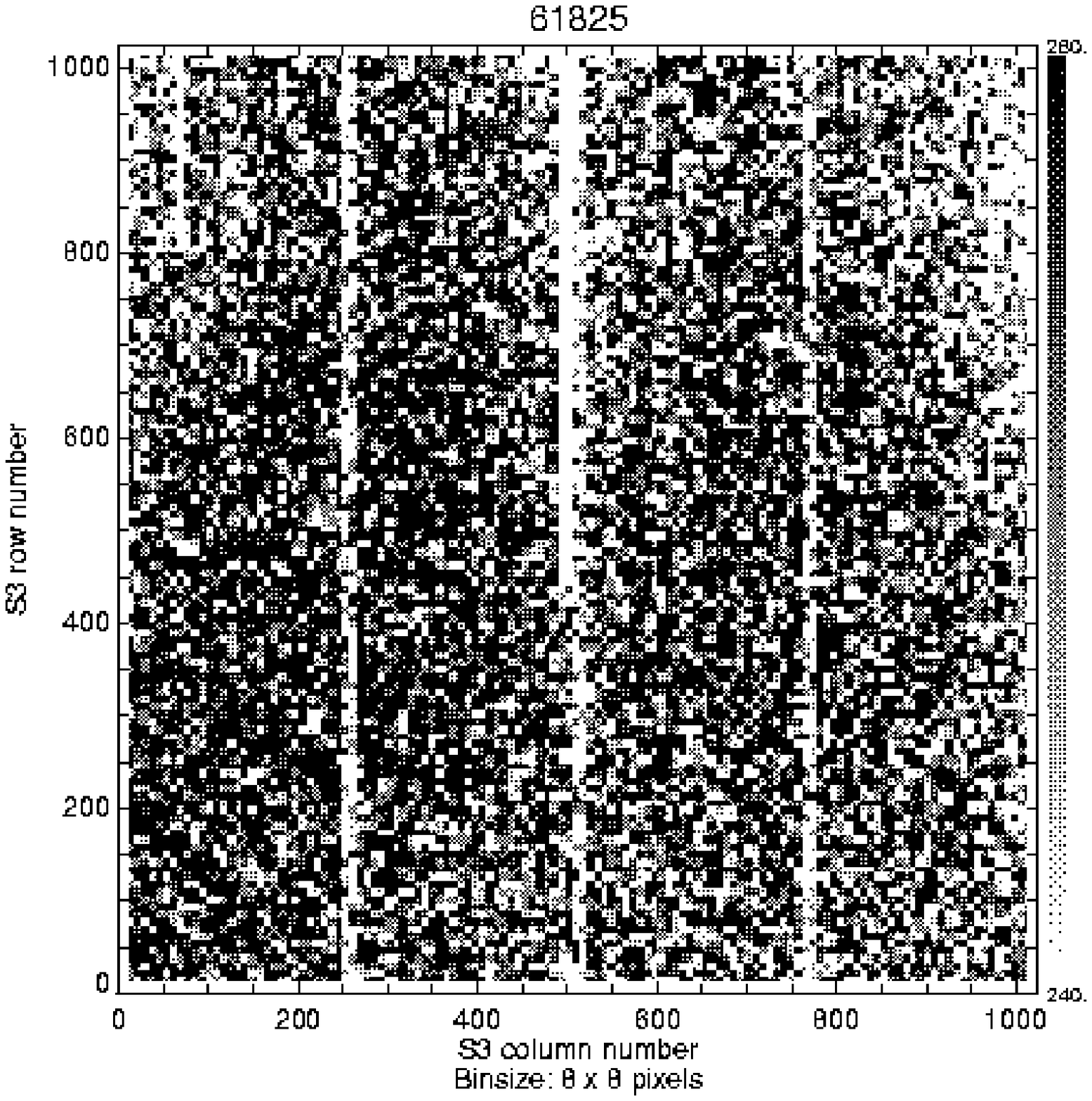,width=3.0in }
         \hspace{0.5in}
         \epsfig{file=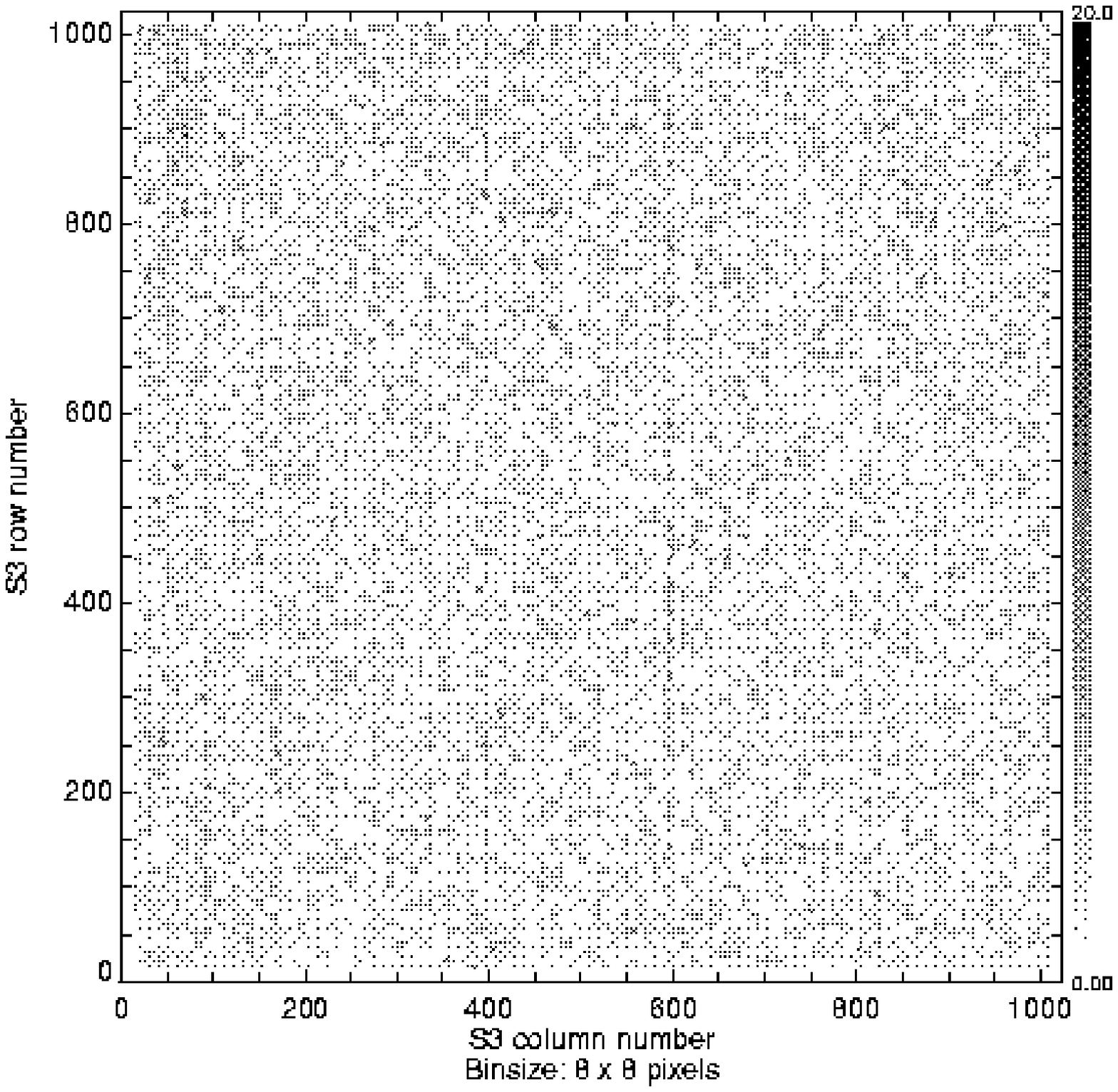,width=3.0in }}}

\caption{\protect \small Grade morphing reduced by CTI correction:  I3
at $-$110C (top); S3 at $-$120C (bottom).  Compare to
Figure~\ref{fig:qeumap}.  These images give the number of ASCA g02346
events (left) and the complementary grades 1, 5, and 7 (right) as a
function of chip position for the Mn K lines ($\sim 6$keV) in ECS data,
after CTI correction. }

\normalsize
\label{fig:qeumap-nocti}
\end{figure}

Finally, Figure~\ref{fig:qeumap-nocti} shows the grade-specific 6~keV
images from Figure~\ref{fig:qeumap} after CTI correction.  The position
dependence of undesirable grades (1, 5, and 7) is much reduced.  
%-------------------------------------------------------------------------
%-------------------------------------------------------------------------
\clearpage

%====================================
\subsection{Event Grades}

As mentioned above, grade morphing due to CTI is a substantial concern
for FI observations early in the mission, due to the higher focal plane
temperature of $-$110C.  This problem was largely solved by lowering the
focal plane temperature, but it is worth noting here the effects at
$-$110C for archival researchers.

\begin{figure}[htb]
\centerline{\epsfig{file=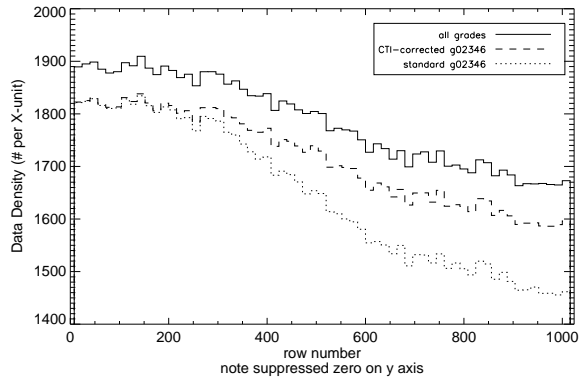, width=3.0in}}
\caption{\protect \small The number of events as a function
of row number for the Mn \Ka + Mn \Kb lines, on I3 at $-$110C.  All event
grades are compared to ASCA g02346 grade filtering obtained via 
standard analysis and from the CTI corrector.}

%Aspect ratio 0.7.  Binsize 16, bin edge 8.  X-range 0-1024, Y-range 
%1400-2000.

%There is a color version of this:  I3_qe_6keV_comparison_color.ps.
 
\normalsize
\label{fig:qecomp}
\end{figure}

Figure~\ref{fig:qecomp} shows the effect of grade-morphing on the $-$110C
FI detector quantum efficiency, for the Mn \Ka and Mn \Kb lines
combined.  At this energy, many events are lost forever because they
morphed into grades that were not telemetered.  For low-energy events
({\em e.g.}\ Al K) even at $-$110C, the charge is more spatially
concentrated, leading to more single-pixel events.  CTI causes these to
morph into grades that are still captured in the standard grade
filtering, so the QE at the top of the device is not affected by CTI.

For this high-energy dataset, keeping all grades, there are 12\% fewer
events at the top of the chip compared to the bottom due to grade
morphing and subsequent on-orbit grade rejection.  For standard grade
filtering with standard processing, this loss is augmented to 20\% at
the top of the chip.  This is because grade morphing caused events to
migrate out of acceptable grades.  By applying the standard grade
filter after applying the CTI corrector, though, there are only 13\%
fewer events at the top of the chip compared to the bottom.  Since this
is similar to the curve for all grades, we can conclude that we have
recovered virtually all useful events available on the ground.

%-------------------------------------------------------------------------  
%-------------------------------------------------------------------------

For the BI devices, grade morphing is an issue even at $-$120C.  The
success of the CTI corrector in regularizing the grade distribution for
the top half of a BI device (where grade morphing is most pronounced),
using on-orbit ECS data on the S3 chip at $-$120C, is shown in
Table~\ref{table:gradesbi}, where original and corrected ASCA-like
grade distributions are listed for the Al, Ti, and Mn calibration
lines.

Some events are not recovered into their original grades because the
corrector is unable to recover charge that has been eroded by CTI down
below the split threshold.  At that level, the charge is
indistinguishable from noise and the corrector is purposely not allowed
to include such pixels in its reconstruction.

A good test of grade recovery is to compare the number of Grade 2
events to the sum of Grade 3 and Grade 4 events -- for devices with no
CTI, these quantities should be roughly the same.  Applying this
diagnostic to Table~\ref{table:gradesbi} shows that the corrector is
quite useful in recovering appropriate grades.

\protect \footnotesize
%-------------------------------------------------------------------------
\begin{table}[htb] \centering
\begin{tabular}{||c|c|c|c|c|c|c||} \hline

ASCA	& \multicolumn{2}{c|}{1.486~keV} & \multicolumn{2}{c|}{4.511~keV}  & \multicolumn{2}{c||}{5.895~keV}\\    
grade   & Standard (\%) & Corrected (\%) & Standard (\%) & Corrected (\%) & Standard (\%) & Corrected (\%)\\ \hline \hline
 
0  &  26.7 $\pm$0.1  &  30.9 $\pm$0.1 &  13.5 $\pm$0.0  &  22.0 $\pm$0.1 &  14.9 $\pm$0.0  &  28.7 $\pm$0.0 \\ \hline
1  &   0.0 $\pm$0.0  &   0.0 $\pm$0.0 &   0.0 $\pm$0.0  &   0.0 $\pm$0.0 &   0.0 $\pm$0.0  &   0.0 $\pm$0.0 \\ \hline
2  &  30.8 $\pm$0.1  &  25.3 $\pm$0.1 &  30.8 $\pm$0.1  &  20.3 $\pm$0.1 &  33.2 $\pm$0.0  &  18.4 $\pm$0.0 \\ \hline
3  &  10.1 $\pm$0.0  &  12.1 $\pm$0.0 &   6.1 $\pm$0.0  &  11.2 $\pm$0.0 &   5.0 $\pm$0.0  &  10.2 $\pm$0.0 \\ \hline
4  &   9.9 $\pm$0.0  &  12.3 $\pm$0.0 &   6.0 $\pm$0.0  &  11.4 $\pm$0.0 &   4.9 $\pm$0.0  &  10.4 $\pm$0.0 \\ \hline
5  &   0.3 $\pm$0.0  &   0.4 $\pm$0.0 &   0.7 $\pm$0.0  &   0.5 $\pm$0.0 &   0.8 $\pm$0.0  &   0.4 $\pm$0.0 \\ \hline
6  &  21.8 $\pm$0.1  &  18.9 $\pm$0.0 &  39.6 $\pm$0.1  &  34.4 $\pm$0.1 &  37.1 $\pm$0.0  &  31.6 $\pm$0.0 \\ \hline
7  &   0.3 $\pm$0.0  &   0.0 $\pm$0.0 &   3.3 $\pm$0.0  &   0.2 $\pm$0.0 &   4.2 $\pm$0.0  &   0.2 $\pm$0.0 \\ \hline
 
\end{tabular}
\caption{\protect \footnotesize A comparison of BI grade distributions (branching ratios) between standard processing and after CTI correction.  This uses ECS data for the S3 chip at $-$120C.  Only the top half of the CCD is included.}
\label{table:gradesbi}
\end{table}
%-------------------------------------------------------------------------
\normalsize

\begin{figure}[htb]
\centerline{\mbox{
         \epsfig{file=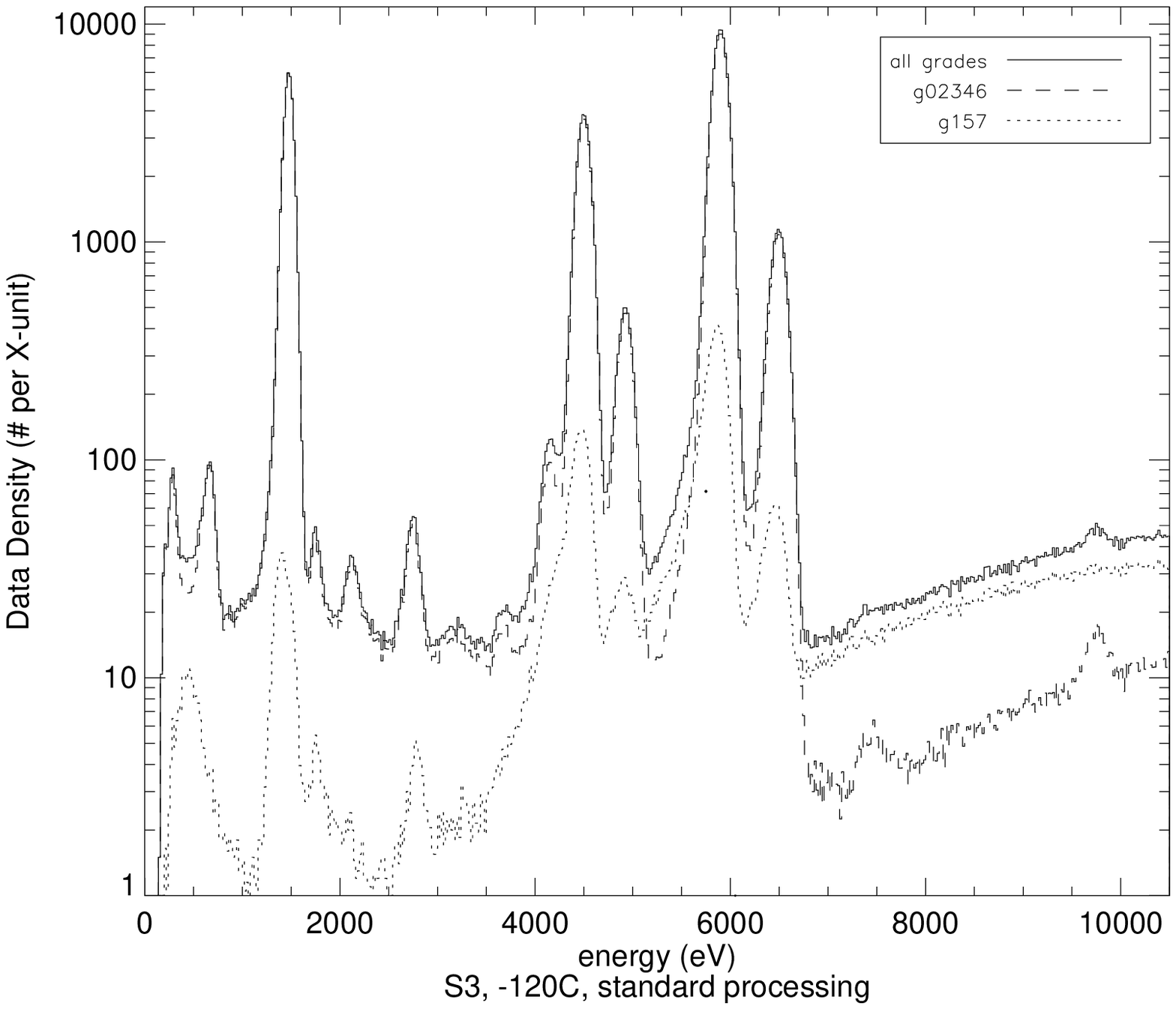,width=3.0in }
         \hspace{0.25in}
         \epsfig{file=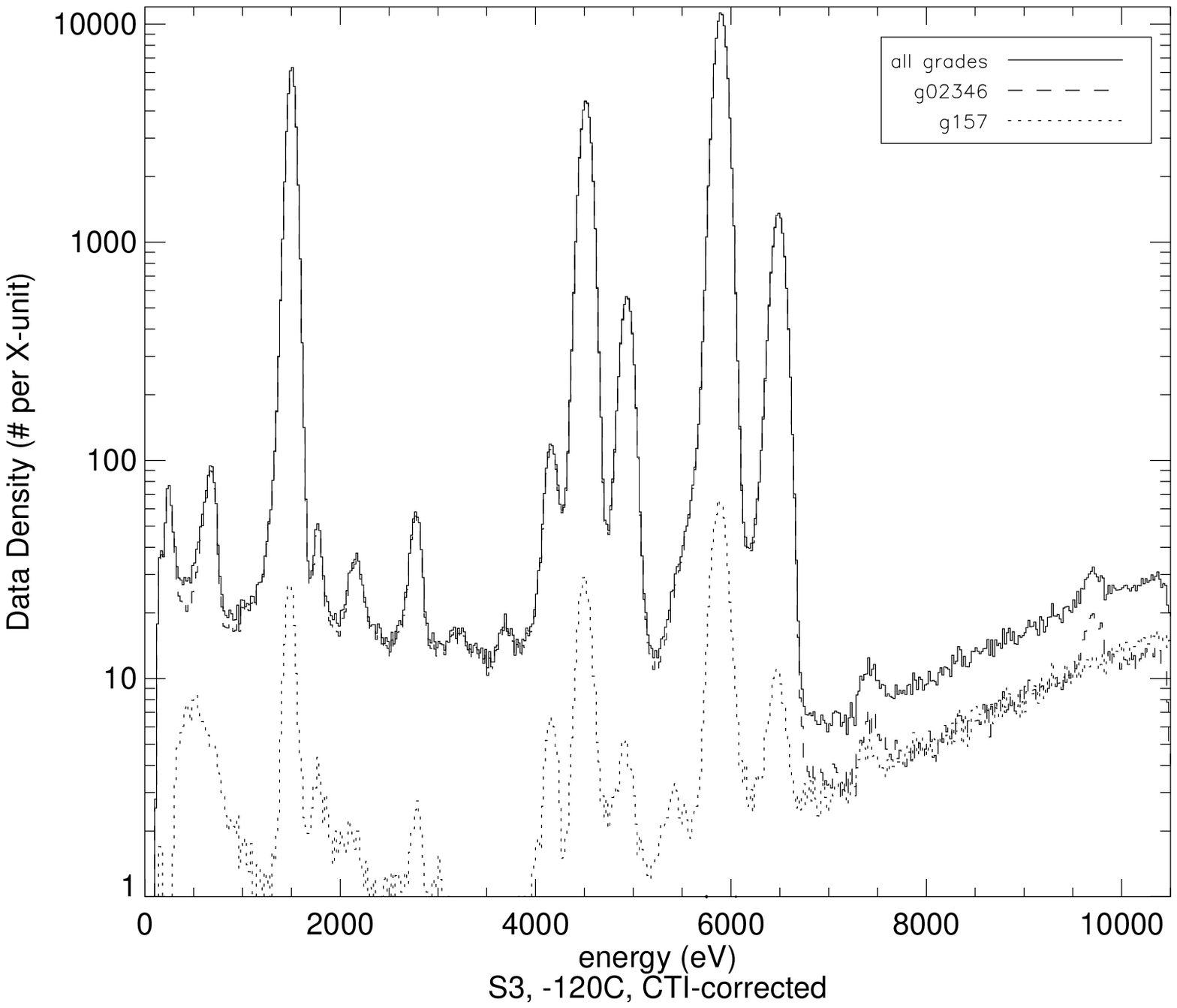,width=3.0in }}}
\caption{\protect \small ECS spectra separated by ASCA grade, for the
top half of S3 at $-$120C (all columns).  The left panel shows the standard
data, while the right shows the same data after CTI-correction.} 

%yrange (log) 1:12000, xrange 0:10500.  Aspect ratio 0.9.  Left margin 9.
%Used "all_index" and "orig_all_index" in 120C/ccd7/ subdirectory.  Binsize 20 eV.

%There are color versions of these two figures if Elsevier can handle them:
%S3_graded_spectra_color.ps and S3_graded_spectra_nocti_color.ps.
 
\normalsize
\label{fig:graded-spec-bi}
\end{figure}

Grade-dependent BI spectra are shown in
Figure~\ref{fig:graded-spec-bi}:  the left-hand panel gives the results
for standard processing and the right-hand panel those for the CTI
corrector.  Between 1 and 7~keV, it appears that including ASCA grades
1, 5, and 7 might improve the signal-to-noise ratio in CTI-corrected
data, as the ECS spectral lines dominate over the background spectrum.
In general, though, CTI correction has restored the original intent of
ASCA g02346 grade filtering, to suppress the particle background and
improve signal-to-noise across the full spatial and spectral range of the device.

%-------------------------------------------------------------------------  
%-------------------------------------------------------------------------
%\clearpage

%====================================
\subsection{Gain linearity}

\begin{figure}[htb]
\centerline{\mbox{
	\epsfig{file=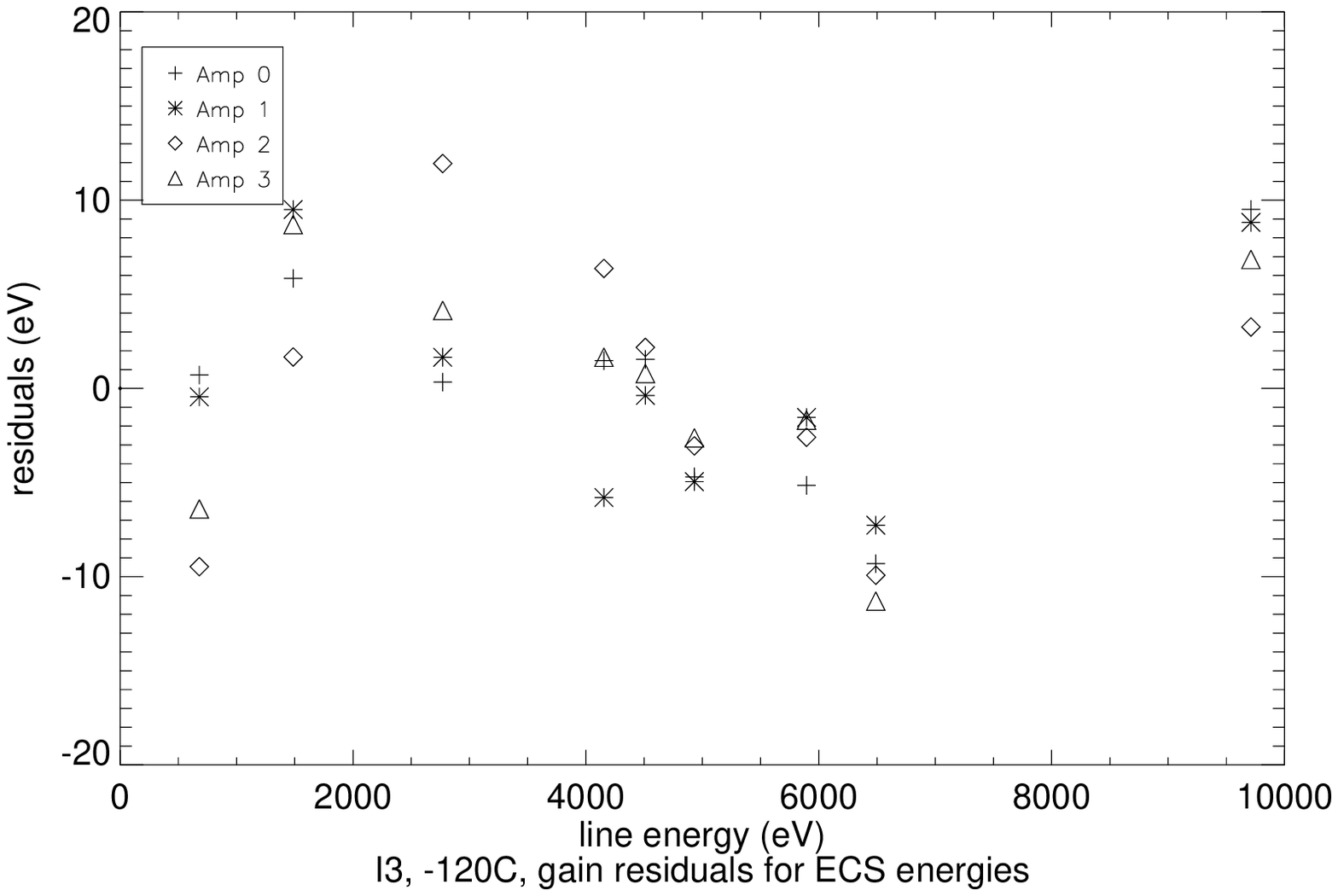,width=3.25in }
	\epsfig{file=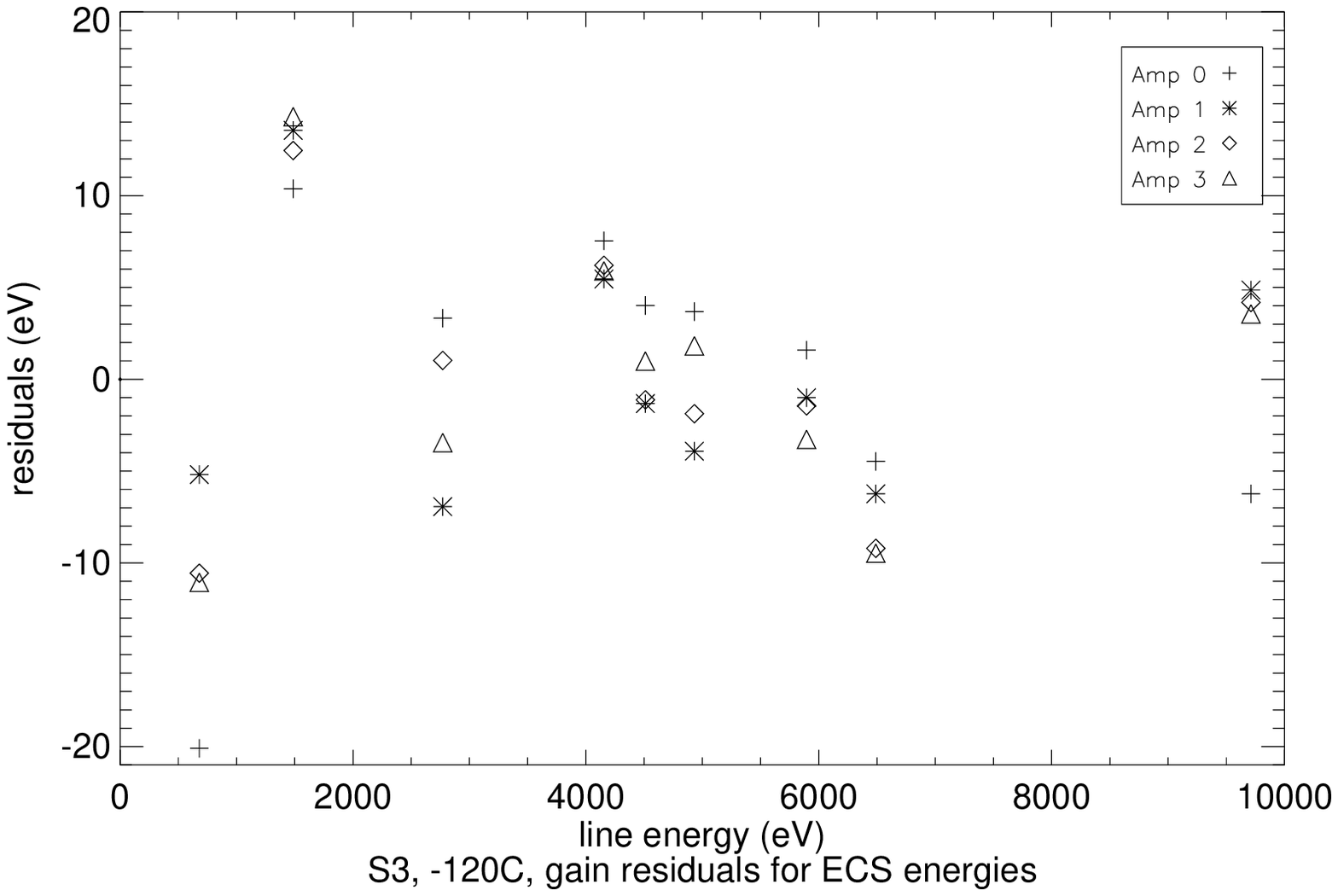,width=3.25in } }}
\caption{\protect \small Gain residuals for all amplifiers on I3, $-$120C (left)
and S3, $-$120C (right).  Note that the Y axis is scaled slightly differently
in the two plots.}
 
\normalsize
\label{fig:gainresid}
\end{figure}

Once the CTI corrector is applied to adjust the amplitude and grade of
each ACIS event, the DN-to-eV conversion is recomputed using the ECS
lines.  A linear fit is made to obtain the gain (slope) and offset
(y-intercept) for each amplifier.  We can then compute the difference
between the measured ECS line energies and the values predicted by this
linear model; these residuals are a measure of the non-linearity of the
amplifier after correcting for CTI in the data.
Figure~\ref{fig:gainresid} shows the ECS line energy residuals for all
amps on I3 (left) and S3 (right) at $-$120C.

%-------------------------------------------------------------------------  
%-------------------------------------------------------------------------

These residuals show that a linear gain model is reasonable for CTI-corrected
ECS data on FI devices at all relevant energies; deviations are seldom larger
than $\pm 10$~eV.  For the BI device S3, the deviations are larger at
energies below 2~keV.  The ECS line complex at $\sim$680~eV is faint
and more information is needed to search for gain non-linearities in
the important BI spectral range 0.2-0.5~keV.  Unfortunately virtually
no calibration information at such low energies is available at this time.

\begin{figure}[htb]
\centerline{\mbox{
	\epsfig{file=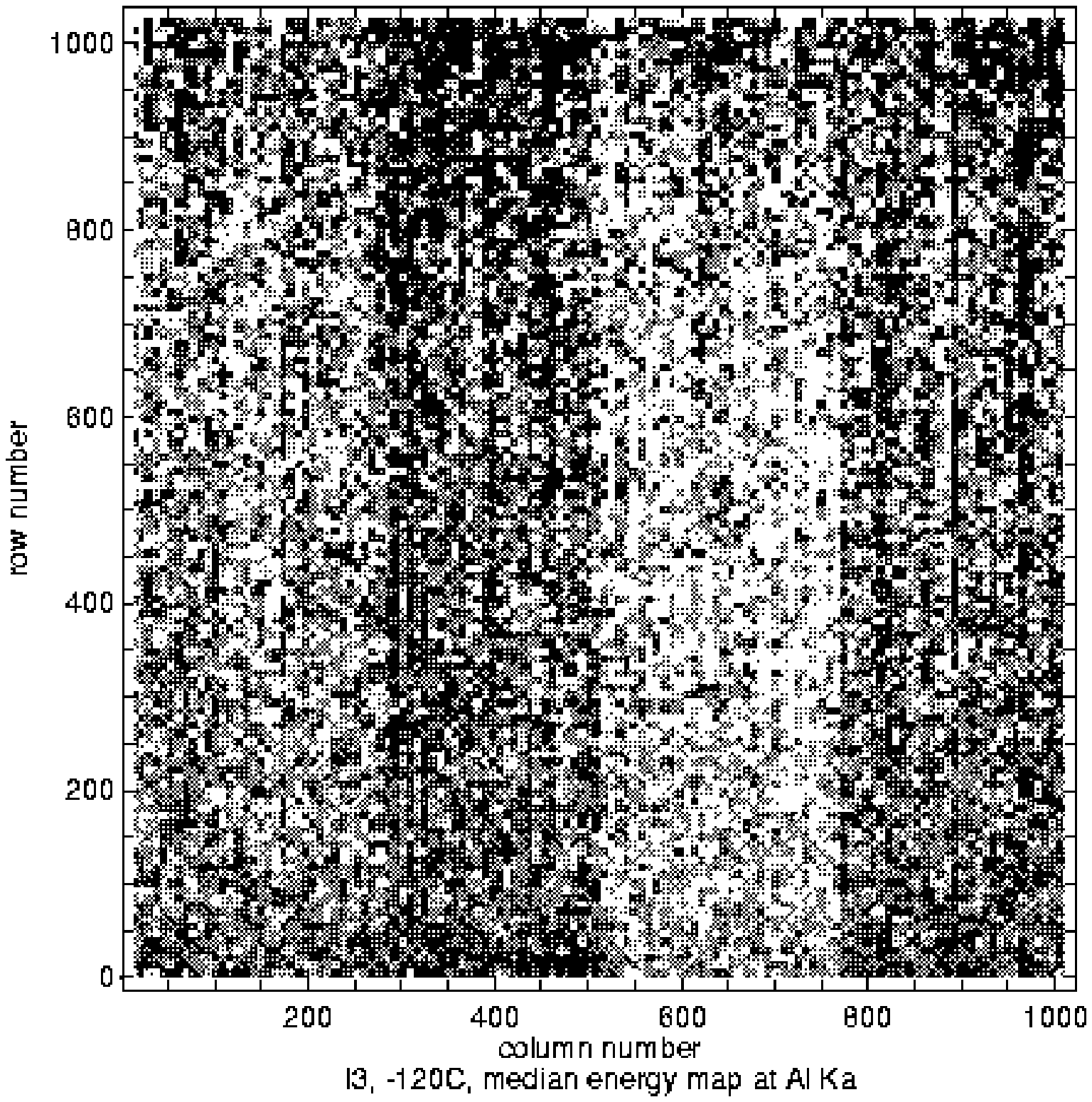,width=3.0in }
	\epsfig{file=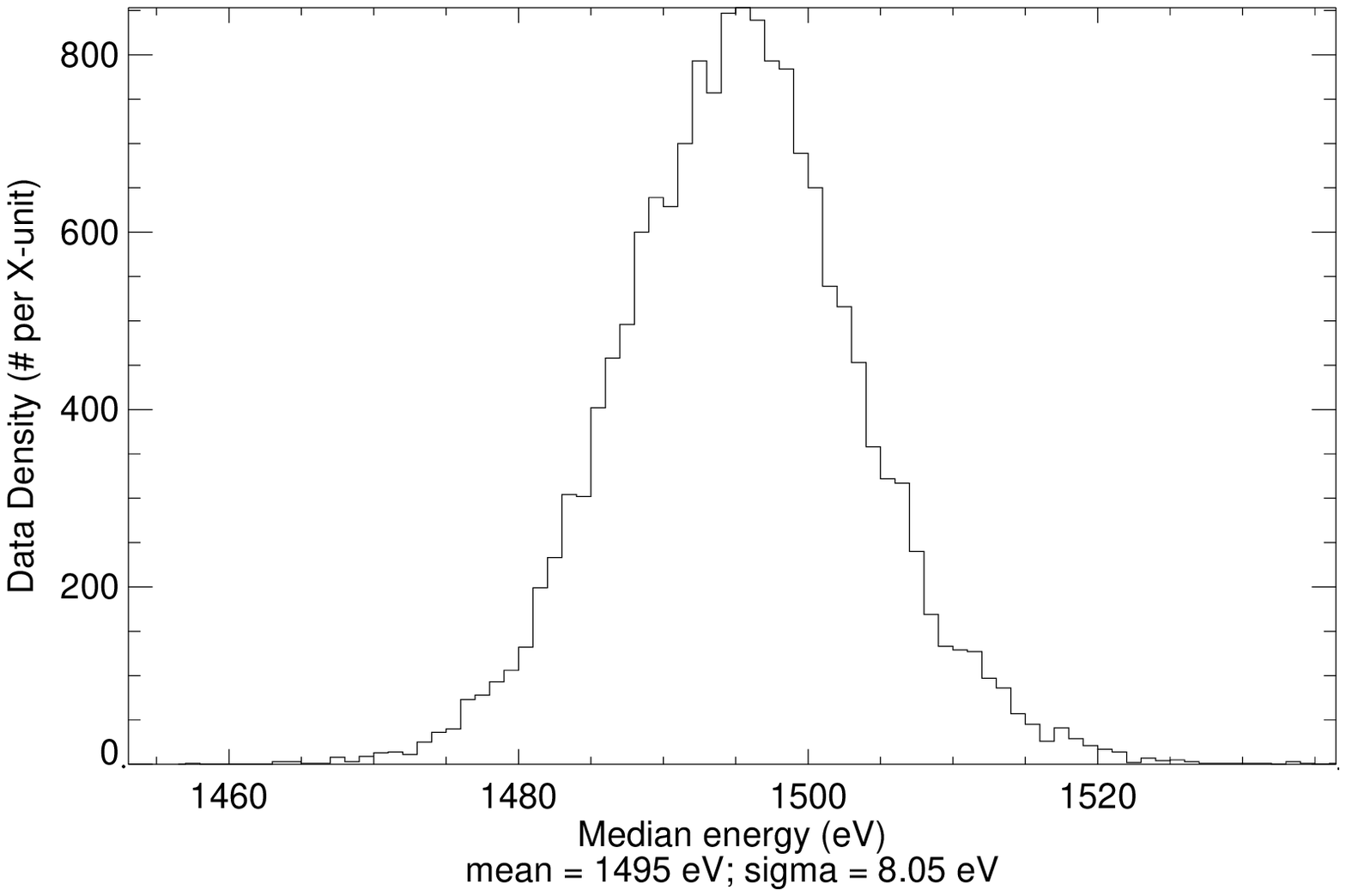,width=3.0in } }}
\centerline{\mbox{
	\epsfig{file=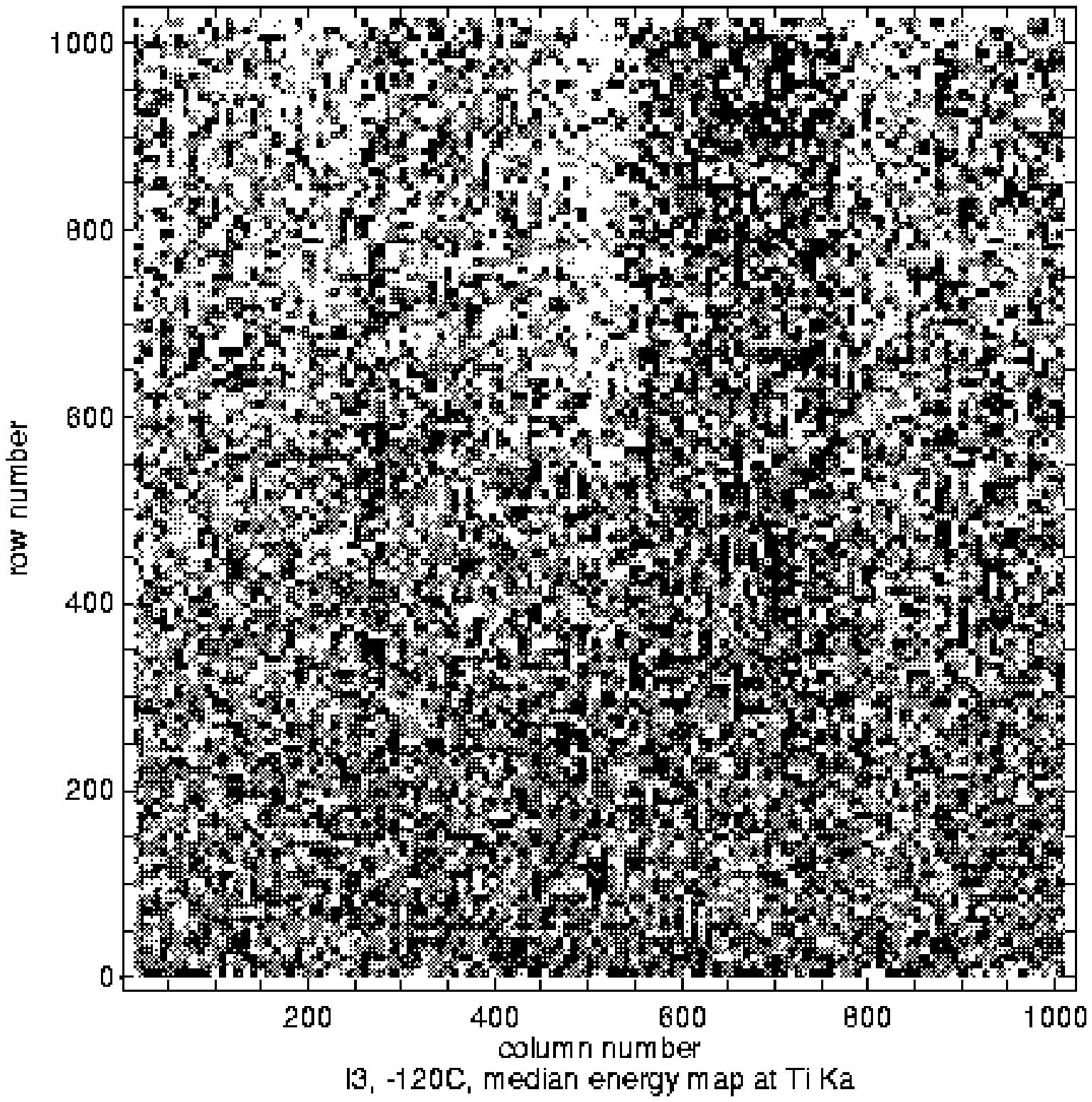,width=3.0in }
	\epsfig{file=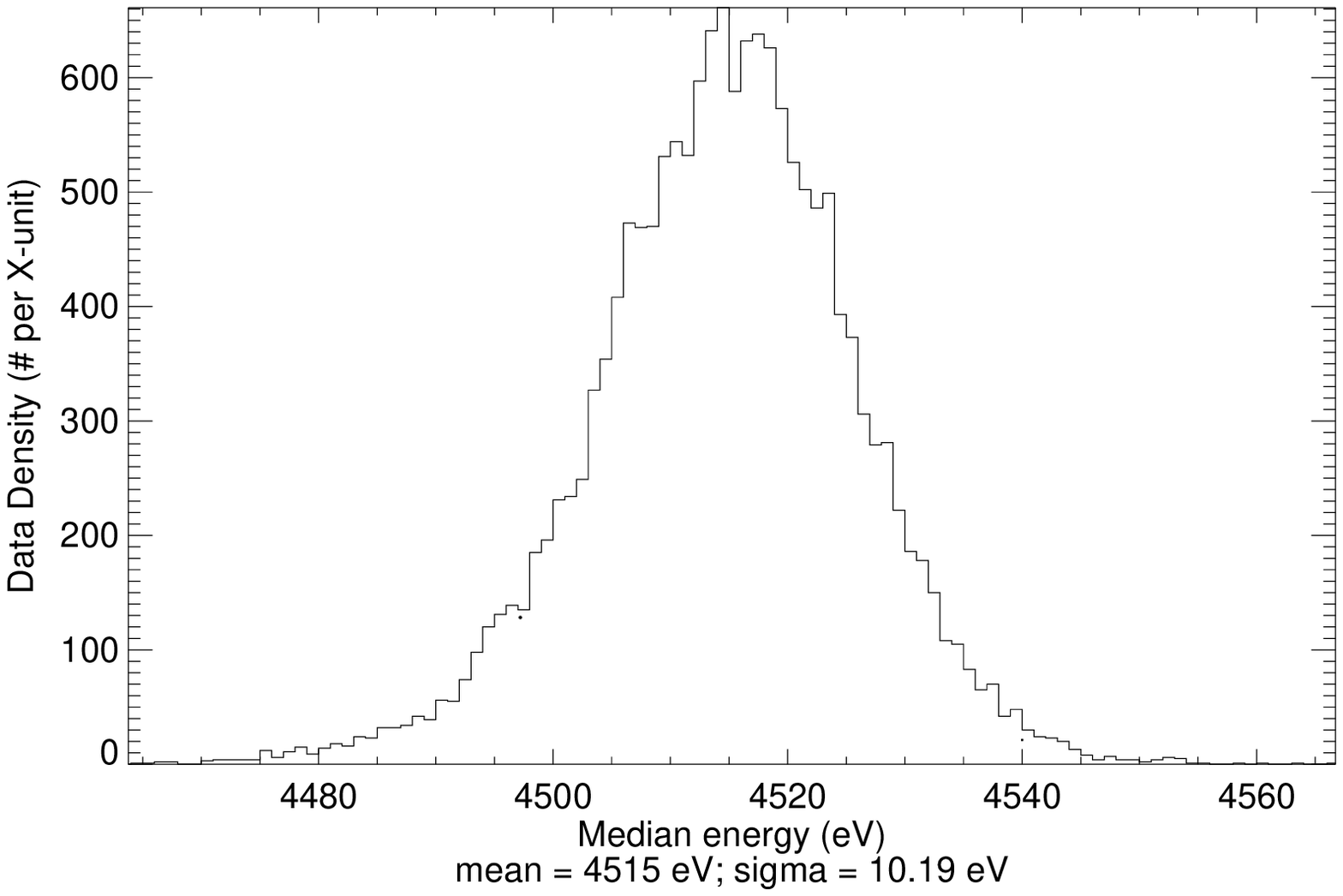,width=3.0in } }}
\centerline{\mbox{
	\epsfig{file=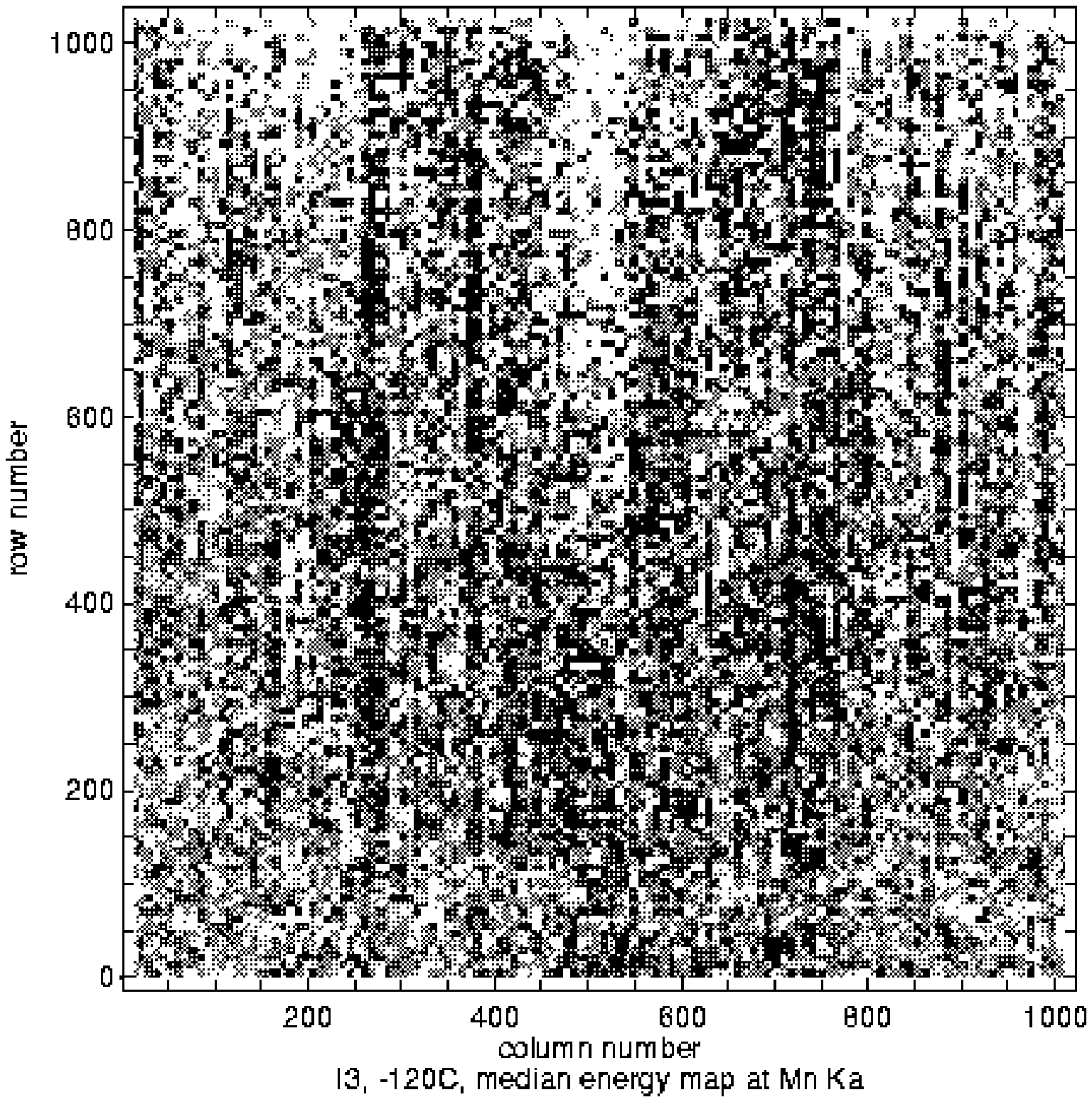,width=3.0in }
	\epsfig{file=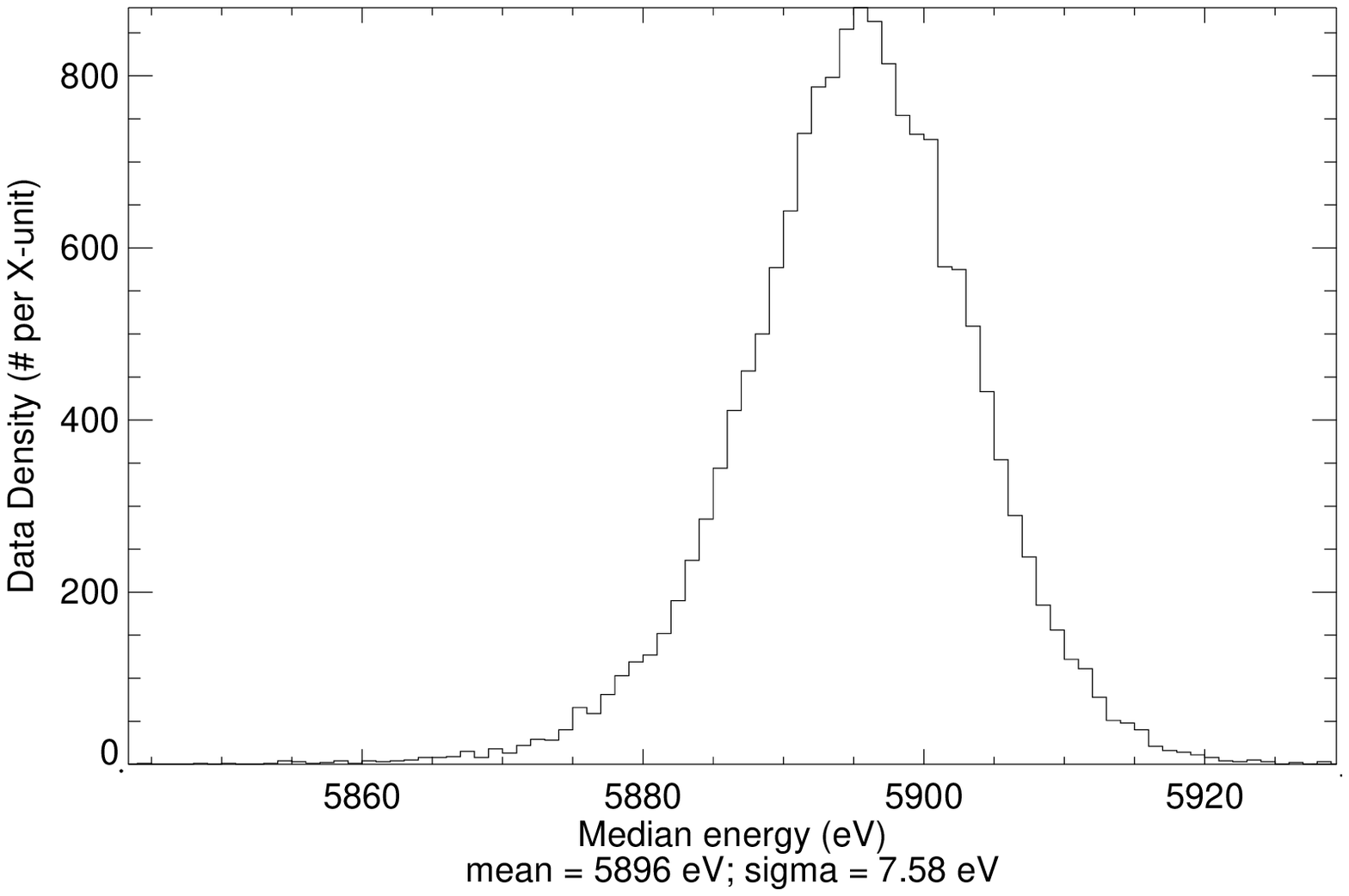,width=3.0in } }}
\caption{\protect \small Median energy maps (left) and distributions (right)
for the three bright ECS emission lines, on I3 at $-$120C, after CTI correction.
The images are displayed in the negative, with a linear greyscale range of
$\pm 1 \sigma$ and a binsize of $8 \times 8$ pixels.}
 
\normalsize
\label{fig:I3-median-maps}
\end{figure}

\begin{figure}[htb]
\centerline{\mbox{
	\epsfig{file=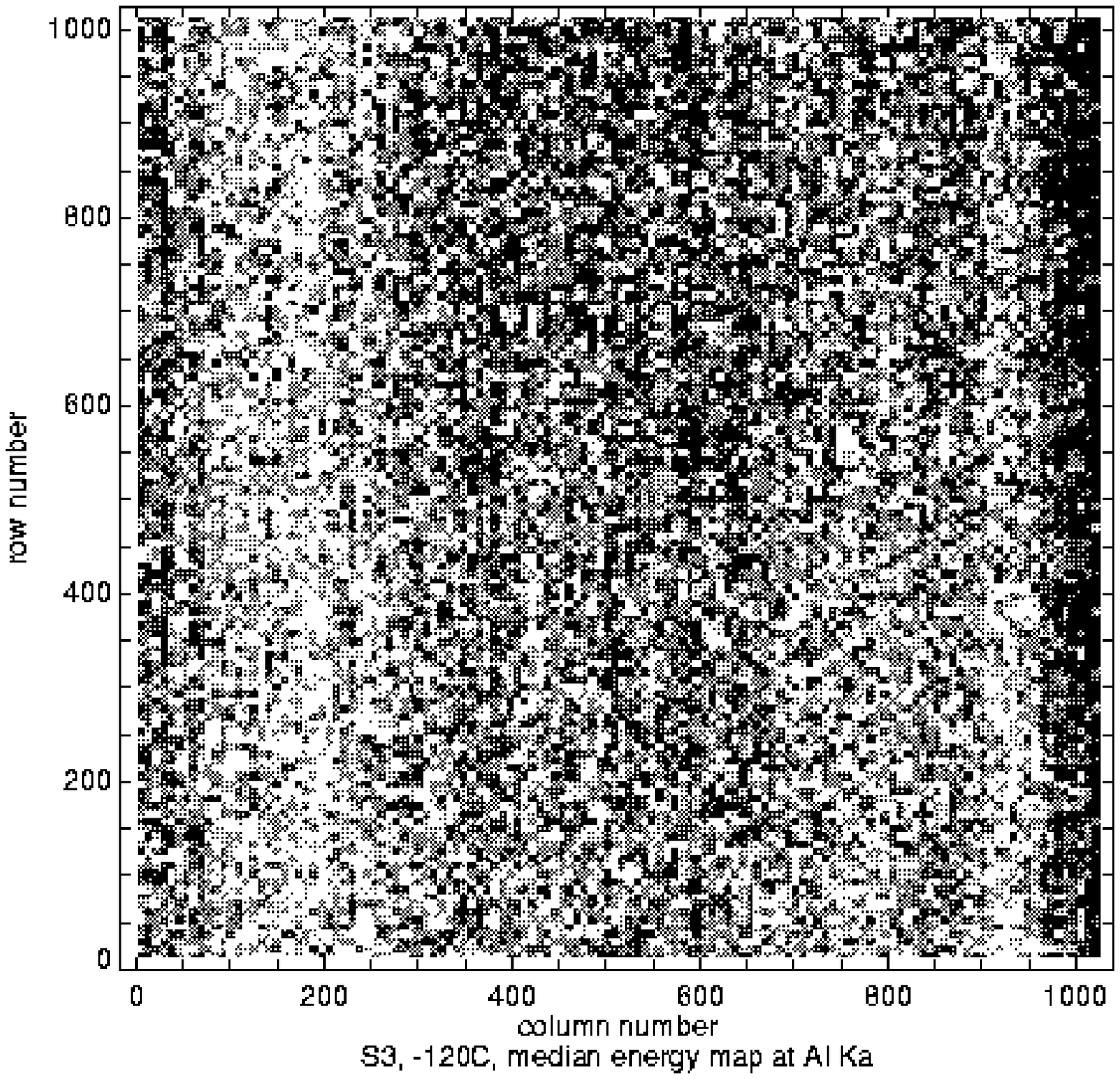,width=3.0in }
	\epsfig{file=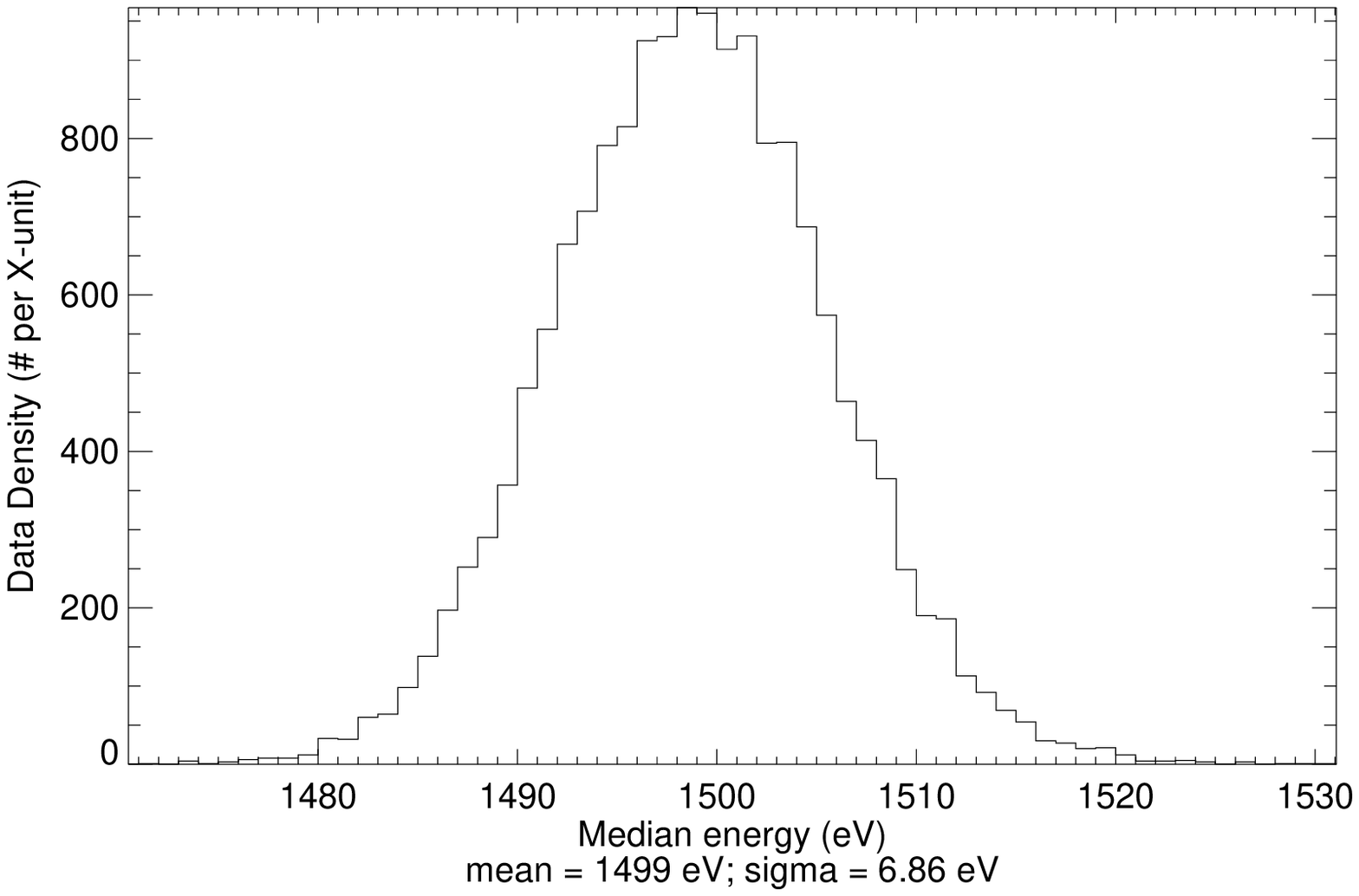,width=3.0in } }}
\centerline{\mbox{
	\epsfig{file=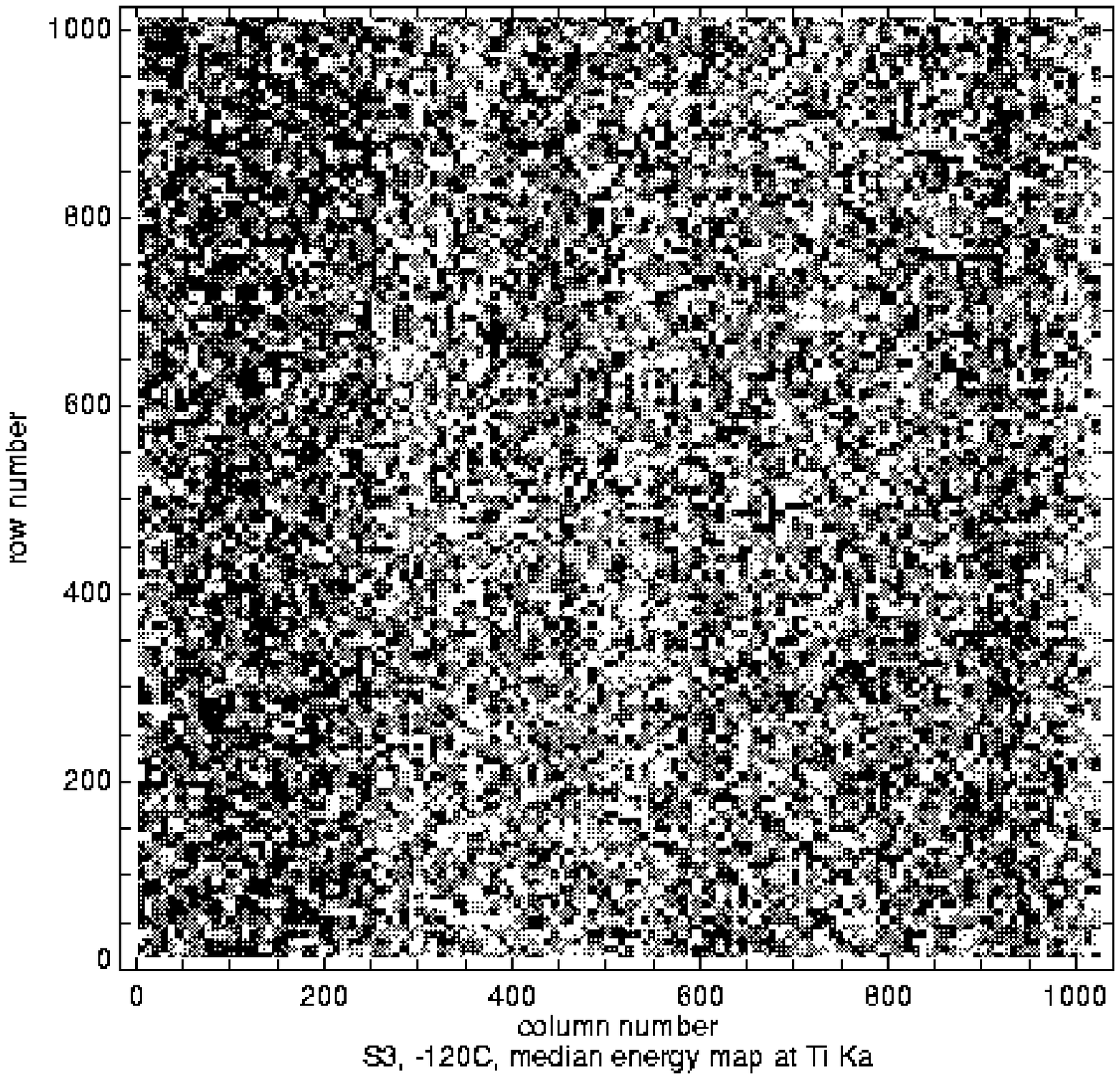,width=3.0in }
	\epsfig{file=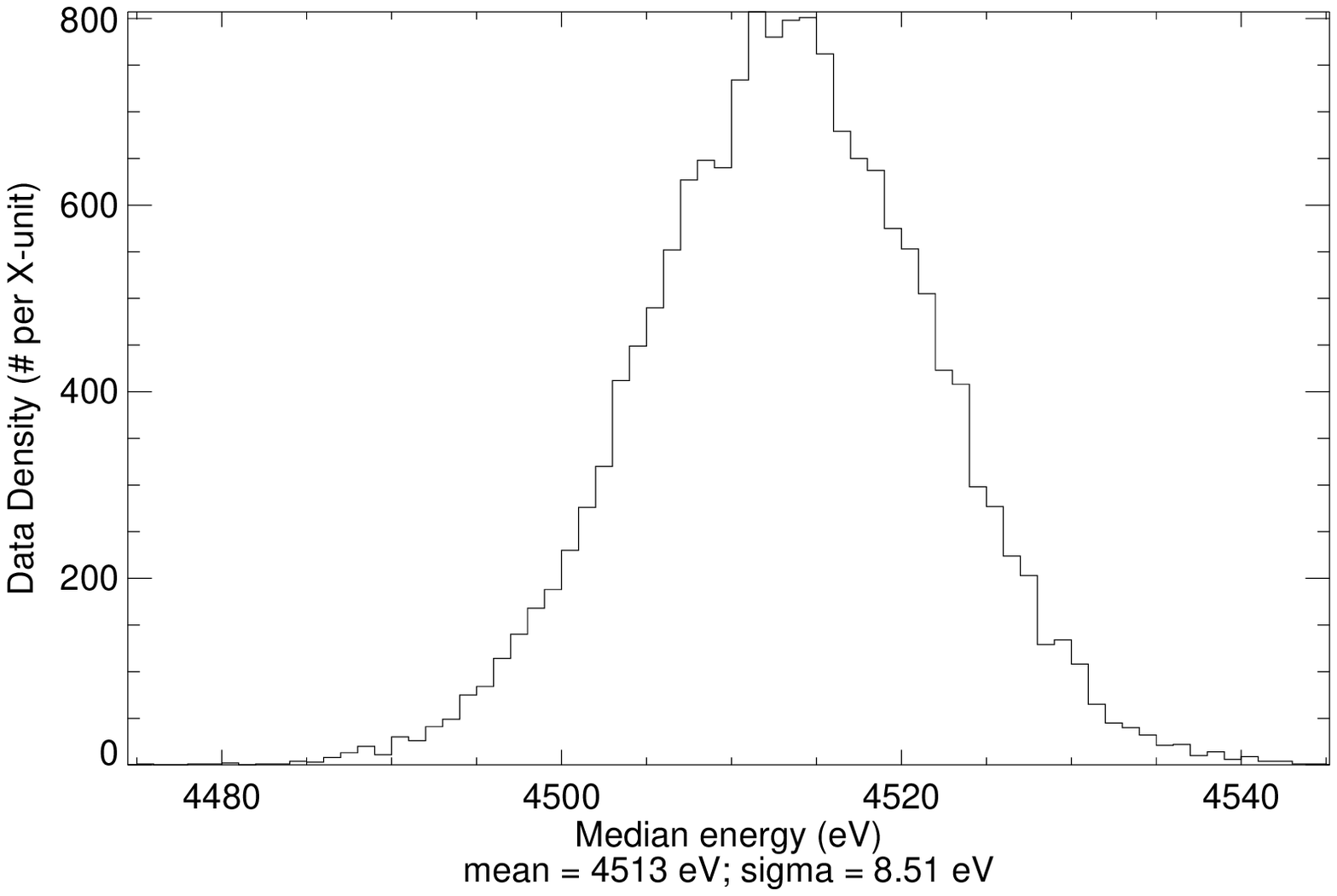,width=3.0in } }}
\centerline{\mbox{
	\epsfig{file=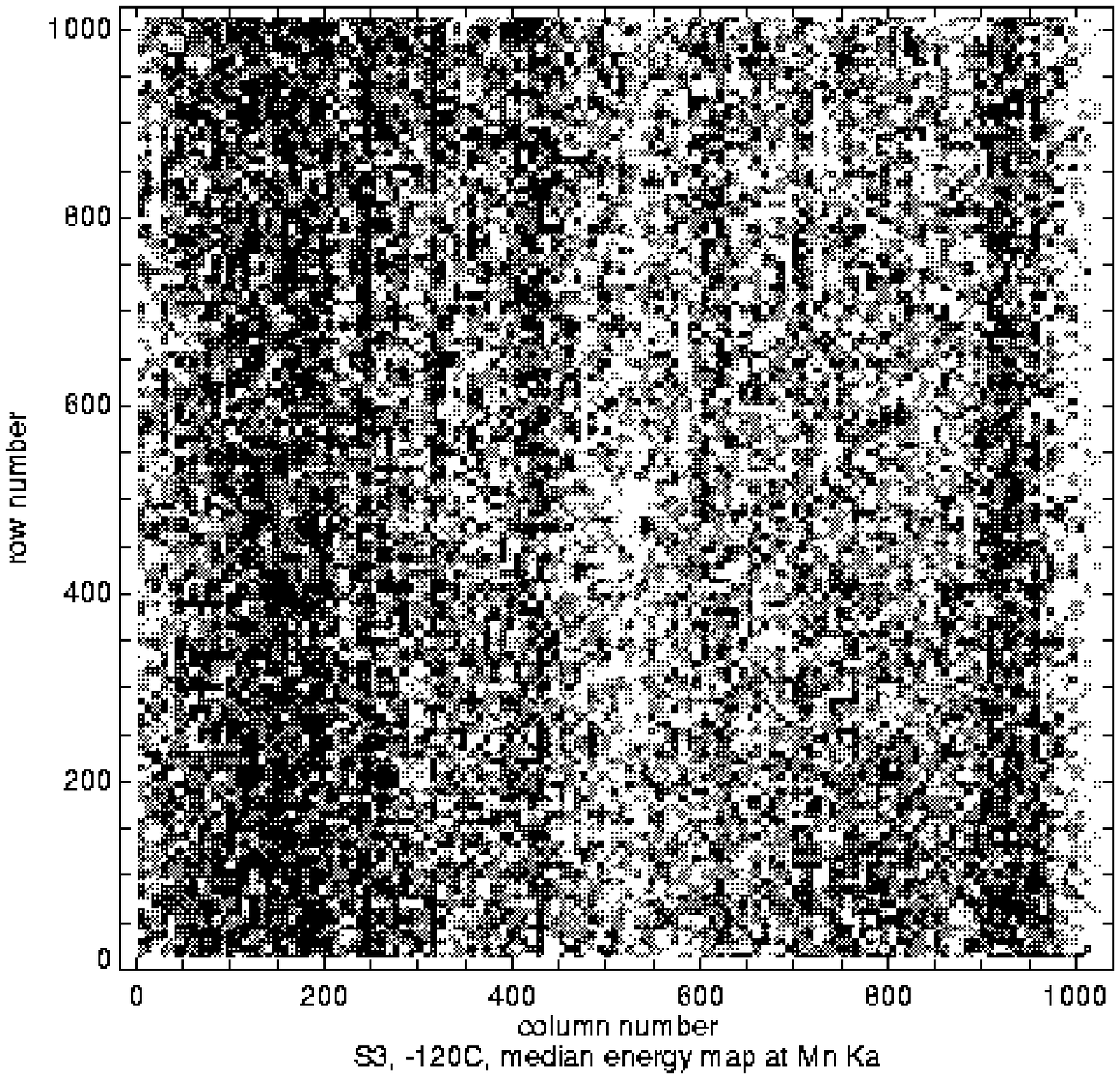,width=3.0in }
	\epsfig{file=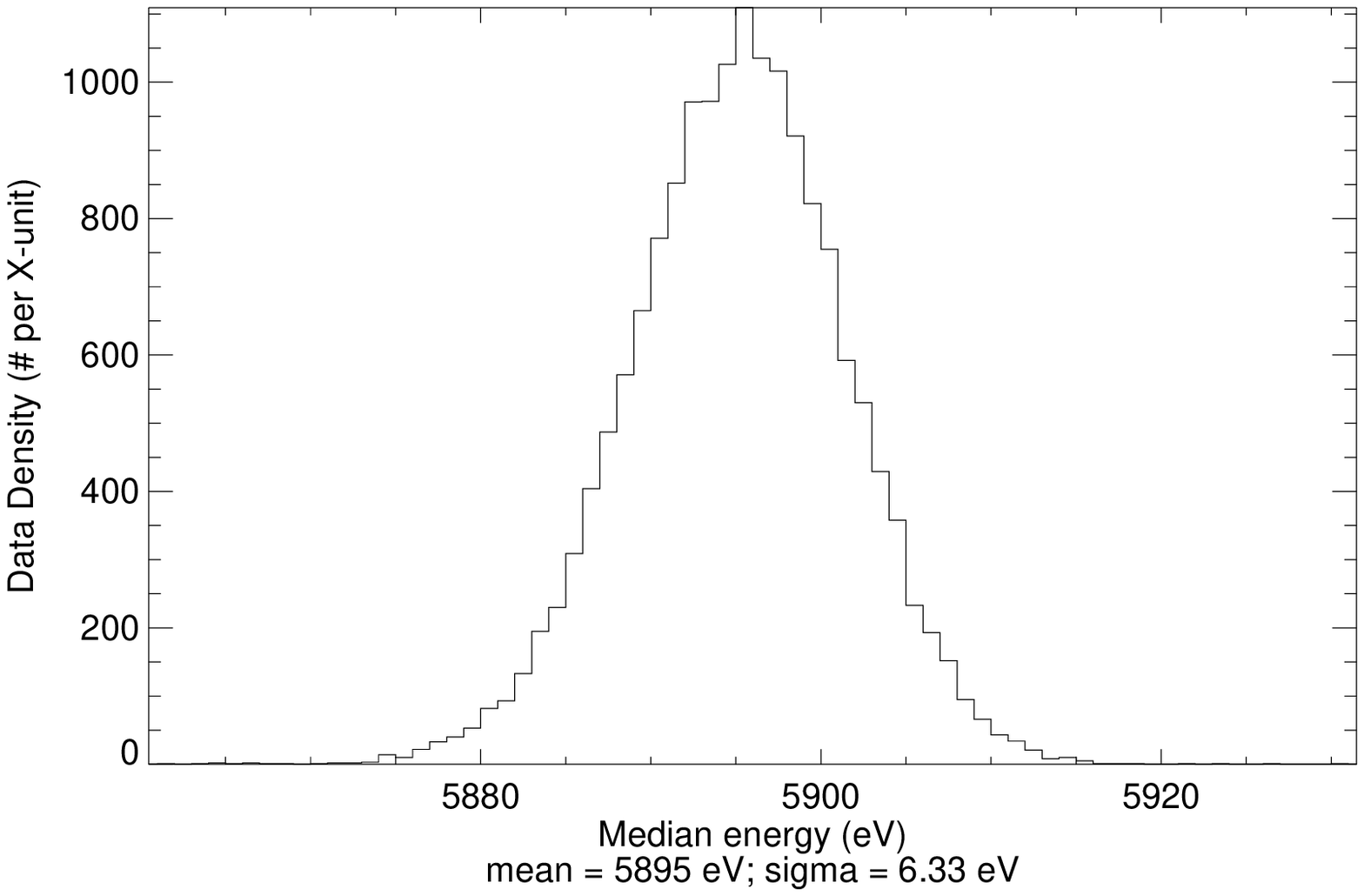,width=3.0in } }}
\caption{\protect \small Median energy maps (left) and distributions (right)
for the three bright ECS emission lines, on S3 at $-$120C, after CTI correction.
The images are displayed in the negative, with a linear greyscale range of
$\pm 1 \sigma$ and a binsize of $8 \times 8$ pixels.}
 
\normalsize
\label{fig:S3-median-maps}
\end{figure}

Figures~\ref{fig:I3-median-maps} and \ref{fig:S3-median-maps} show the
position dependence of gain variations remaining after CTI correction.
Three median energy maps are given, for the three main ECS lines.  These
greyscale maps are displayed such that black is $+1\sigma$ from the mean energy
and white is $-1\sigma$ from the mean energy.  The spectral lines are
shown in the right panels.

For the FI device I3 at $-$120C, Figure~\ref{fig:I3-median-maps} shows that
there is some remaining structure following amplifier boundaries.  There
is no clear correlation between energies, however; regions that are light
at one energy are neutral or dark at another.  Thus the algorithm,
although imperfect, appears to be removing gain variations in a reasonable way.

%-------------------------------------------------------------------------  
%-------------------------------------------------------------------------

Figure~\ref{fig:S3-median-maps} shows that the same is true for the BI
device S3.  Non-random gain deviations exist, but they are not correlated
across energies.  It might be possible to generate a second-stage deviation
map to try to remove these residual features, but our attempts using
this dataset (roughly one year of ECS observations) have been unsuccessful.  

%-------------------------------------------------------------------------  
%-------------------------------------------------------------------------

\clearpage

%==========================================================================
\section{Quantum Efficiency Adjustments} \label{sec:qeu}
%==========================================================================

As mentioned above, grade adjustments by the CTI corrector combined
with subsequent grade filtering result in CCD quantum efficiency
improvement.  This change in QE must be incorporated into the ACIS
calibration products so that source fluxes can be estimated properly.
Since the CTI-induced grade morphing changes with position on the CCD,
the most direct way to account for these QE changes is via the Quantum
Efficiency Uniformity (QEU) files, one component of the Auxiliary
Response File (ARF) used in spectral fitting.  The ARF incorporates the
vignetting function of the telescope, the positionally-averaged QE of
the CCD, and position-dependent deviations from that average QE via the
QEU files.  

Two approaches to adjusting the QEU files for CTI-corrected
data are necessary:  one for FI devices that were effectively CTI-free
before launch, and a second for BI devices that have always suffered
measurable CTI.  These are described below.  

\subsection{Estimating QE Loss with Respect to an Undamaged Device}

\begin{figure}[htb]
\centerline{\epsfig{file=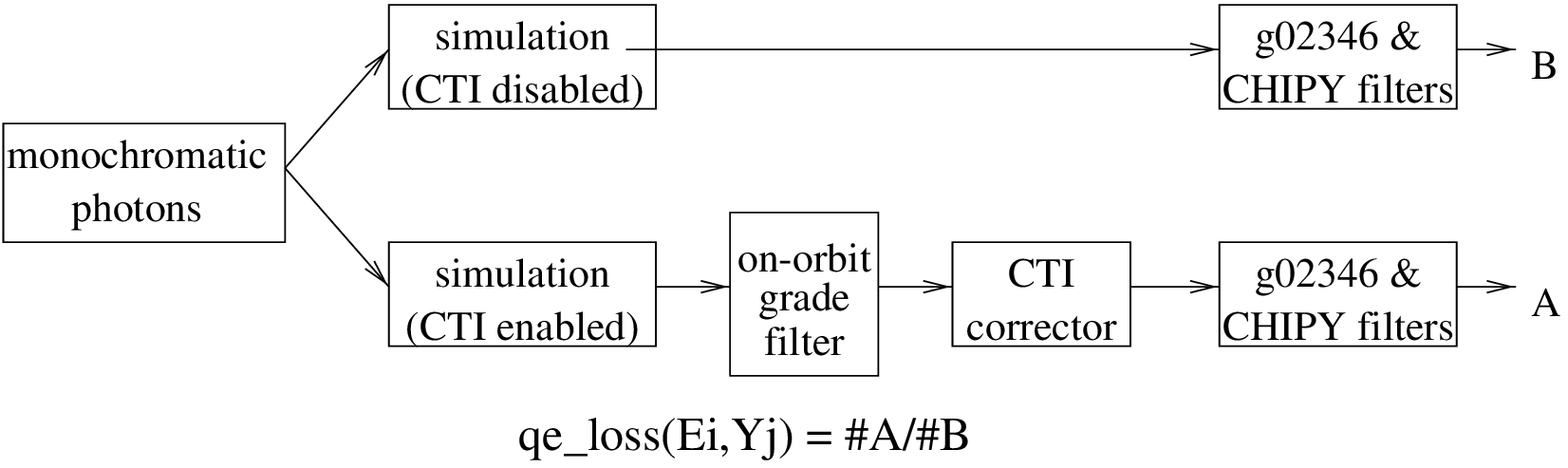,width=6.0in,angle=-90}}
\caption{\protect \small Estimating the QE loss for FI devices.}
\label{fig:fi_qe_loss}
\end{figure}

This approach is used for FI devices.  We estimate how the QE of
undamaged devices changes when the damage is applied and the resulting
events are corrected for CTI (for example by the simulation experiments
shown in Figure~\ref{fig:fi_qe_loss}).  Such a loss is of course a
function of both energy and position on the CCD.  Then the QEU files
that were constructed for undamaged devices are adjusted downward to
account for unrecoverable CTI losses.  For each energy $E_i$,

\[ QE(E_i,X_j,Y_k) = QE_0(E_i) * QEU(E_i,X_j,Y_k) * qe\_loss(E_i,X_j,Y_k) \]

In this formulation, the $QE_0$ \& $QEU$ terms are the mean QE and QE
uniformity products produced by standard processing for undamaged
devices, and the $qe\_loss$ term expresses the QE lost by the damage
plus correction process.  For ACIS FI devices the $qe\_loss$ term is
not actually a function of X (CCD column number).  The pre-launch FI QE is
very spatially uniform below $\sim$8~keV, consistent with unity to within
measurement errors \cite{calreport}.  Thus for ACIS FI devices
the $QEU$ term above is set to unity.

%-------------------------------------------------------------------------
%-------------------------------------------------------------------------

\subsection{Estimating QE Recovery with Respect to a Damaged Device}

\begin{figure}[htb]
\centerline{\epsfig{file=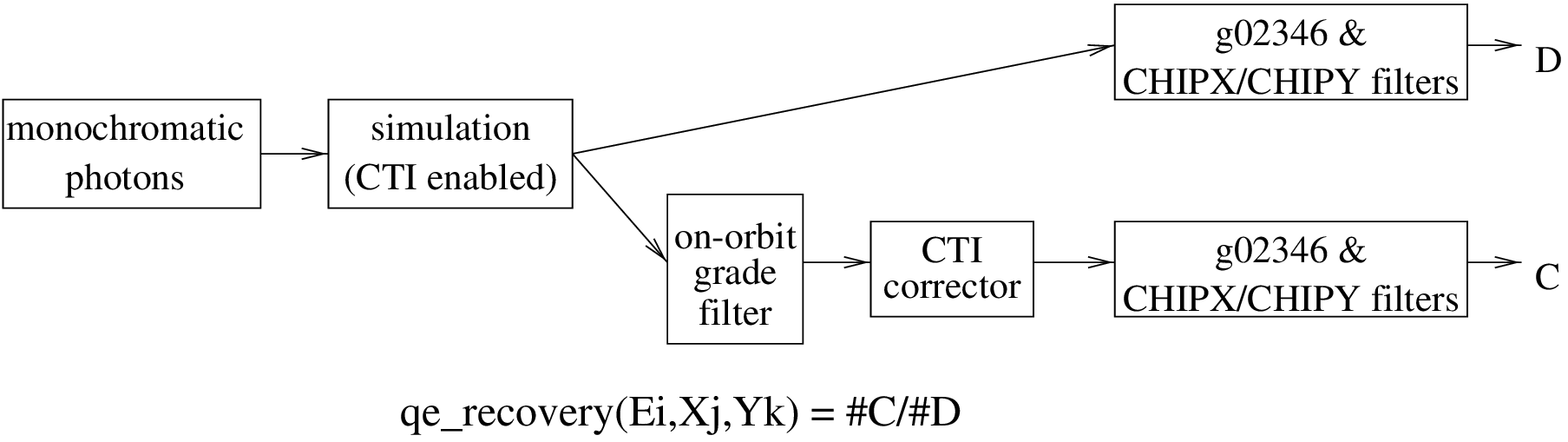,width=7.0in,angle=-90}}
\caption{\protect \small Estimating the QE recovery for BI devices.}
\label{fig:bi_qe_recovery}
\end{figure}

This approach is used for BI devices.  We estimate the QE recovery of
the CTI corrector via simulation/correction experiments, as shown in
Figure~\ref{fig:bi_qe_recovery}.  This recovery is also a function of
both energy and position on the CCD.  Then QEU products that were
constructed for damaged devices are adjusted upward to account for the
increased QE after CTI correction.  So for each energy $E_i$,

\[ QE(E_i,X_j,Y_k) = QE_0(E_i) * QEU(E_i,X_j,Y_k) * qe\_recovery(E_i,X_j,Y_k) \]

In this formulation, the $QE_0$ and $QEU$ terms are the mean QE and QE
uniformity products generated by standard processing for damaged
devices, and the $qe\_recovery$ term expresses the QE recovered by the
correction process.

%-------------------------------------------------------------------------
%-------------------------------------------------------------------------

%\clearpage

%==========================================================================
\section{Model Performance on Celestial Data} \label{sec:performance}
%==========================================================================

There is an uncertainty introduced by using on-orbit calibration data
to tune this or any other model:  we must assume that the CTI measured
by the on-orbit calibration source is representative of that present in
celestial data.  This assumption could be incorrect, even for celestial
data obtained at nearly the same time as the calibration data, due to
differences in the count rates (photons plus particle background)
between these two configurations.  It is important to remember,
however, that {\em most} ACIS events are due to particle interactions
and are not even telemetered, so the event list supplied to the user is
not representative of all the potentially trap-filling interactions
that occurred.  Since most of these particle events are from cosmic
rays that enter from all angles and are not absorbed by the Observatory
structure, it is reasonable to assume that calibration and celestial
data have comparable particle fluxes, thus a comparable trap
population.

We have used the instrumental Au L$\alpha$ (9.7 keV) and Ni K$\alpha$
(7.5 keV) lines as rough diagnostics of the corrector's performance on
celestial data.  These spectral features are due to fluorescence from
hardware components.  The lines are faint, but in long exposures it is
possible to measure their gain changes with position on the chip just
like we measure lines in the calibration data.  The tests show that the
corrector works fairly well even at these high energies, reducing the
charge loss per pixel transfer by a factor of $\sim$10.

Another implication of tuning the corrector to the ECS data is that we
have only minimal calibration information below the Al K$\alpha$ line
at 1.486 keV and no information below the Fe-Mn L complex at
$\sim$680~eV.  We have extrapolated to lower energies using fits of
sensible functional forms, but it is likely that our assumptions break
down for small amounts of charge.  Understanding the behavior of the
CCDs between 0.2 and 1.5 keV is important for obtaining accurate
hydrogen column densities, comparing ACIS results to findings from
other X-ray missions, and inferring valid astrophysics for the large
number of soft sources {\em Chandra} is examining.

We can use celestial data to test and improve our calibration of the
CTI at low energies.  The best available target is the supernova
remnant E0102-72.3 (``E0102'') in the Large Magellanic Cloud, used as a
soft calibration source for {\em Chandra}.  We used this source in
Figure~\ref{fig:lossfitse0102} above to confirm that the
piecewise-linear CTI model was reasonable at low energies.

\begin{figure}[htb]
\centerline{\mbox{
	\epsfig{file=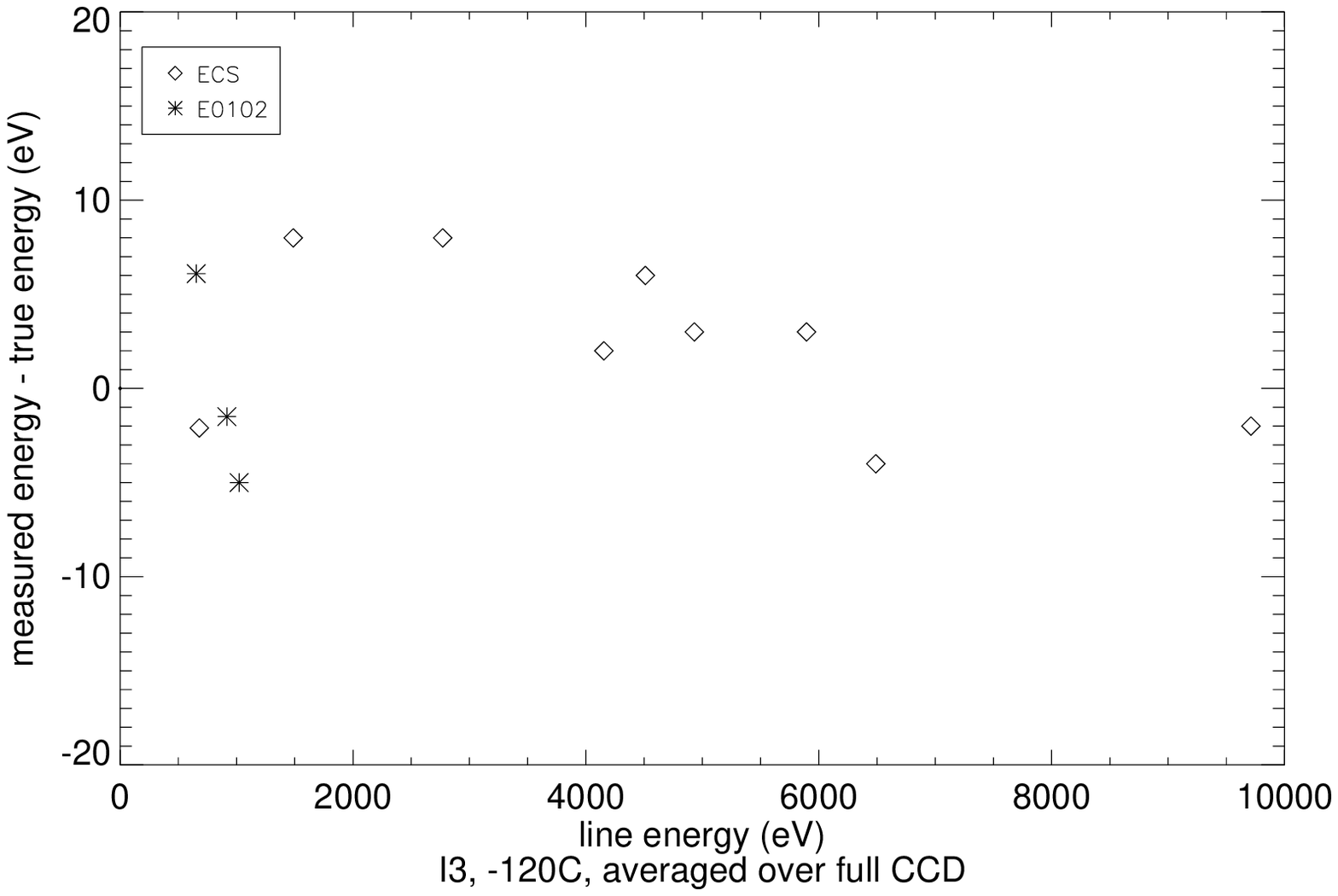,width=3.25in }
	\epsfig{file=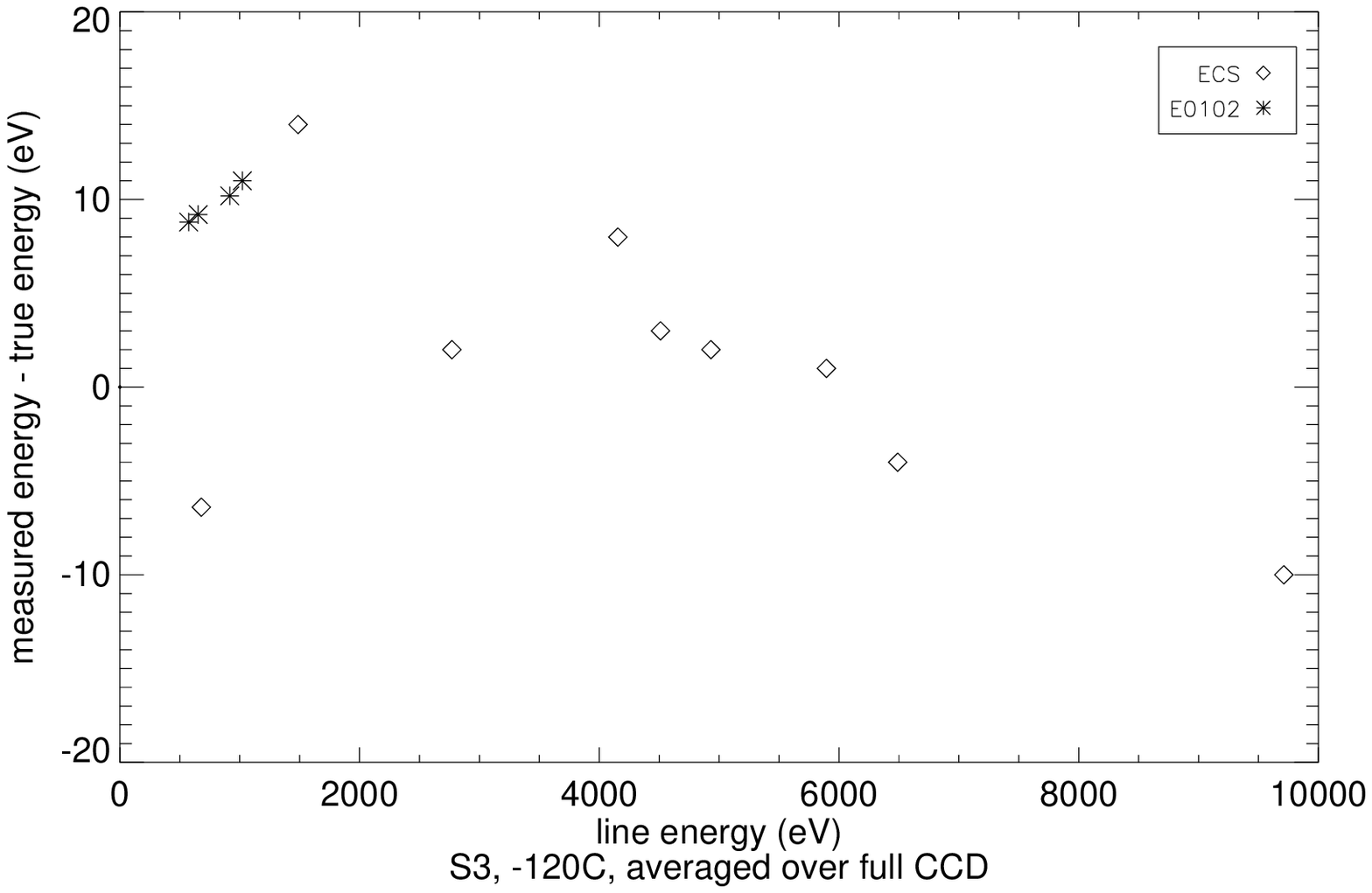,width=3.25in } }}
\caption{\protect \small Difference between measured and true line energies,
averaged over all CCD event positions, for I3, $-$120C (left)
and S3, $-$120C (right), including E0102-72.3 data.}
 
\normalsize
\label{fig:energydiff}
\end{figure}

Figure~\ref{fig:energydiff} shows the difference between measured and
true line energies from ECS data and the E0102 soft spectral lines.
Energies for the E0102 lines were provided by R. Edgar (private communication)
and can be inferred from Rasmussen {et al.} \cite{rasmussen01}.
This shows that the model can recover reasonable line energies for both
FI and BI devices down to $\sim$0.5~keV.  The I3 plot includes E0102
points obtained from combining Observation ID's 136, 140, 439, 440,
1314, 1315, 1316, and 1317.  The S3 plot includes E0102 points obtained
from combining Observation ID's 141, 1308, 1311, 1530, 1531, and 1702.

%-------------------------------------------------------------------------  
%-------------------------------------------------------------------------

\begin{figure}[htb]
\centerline{\mbox{
	\epsfig{file=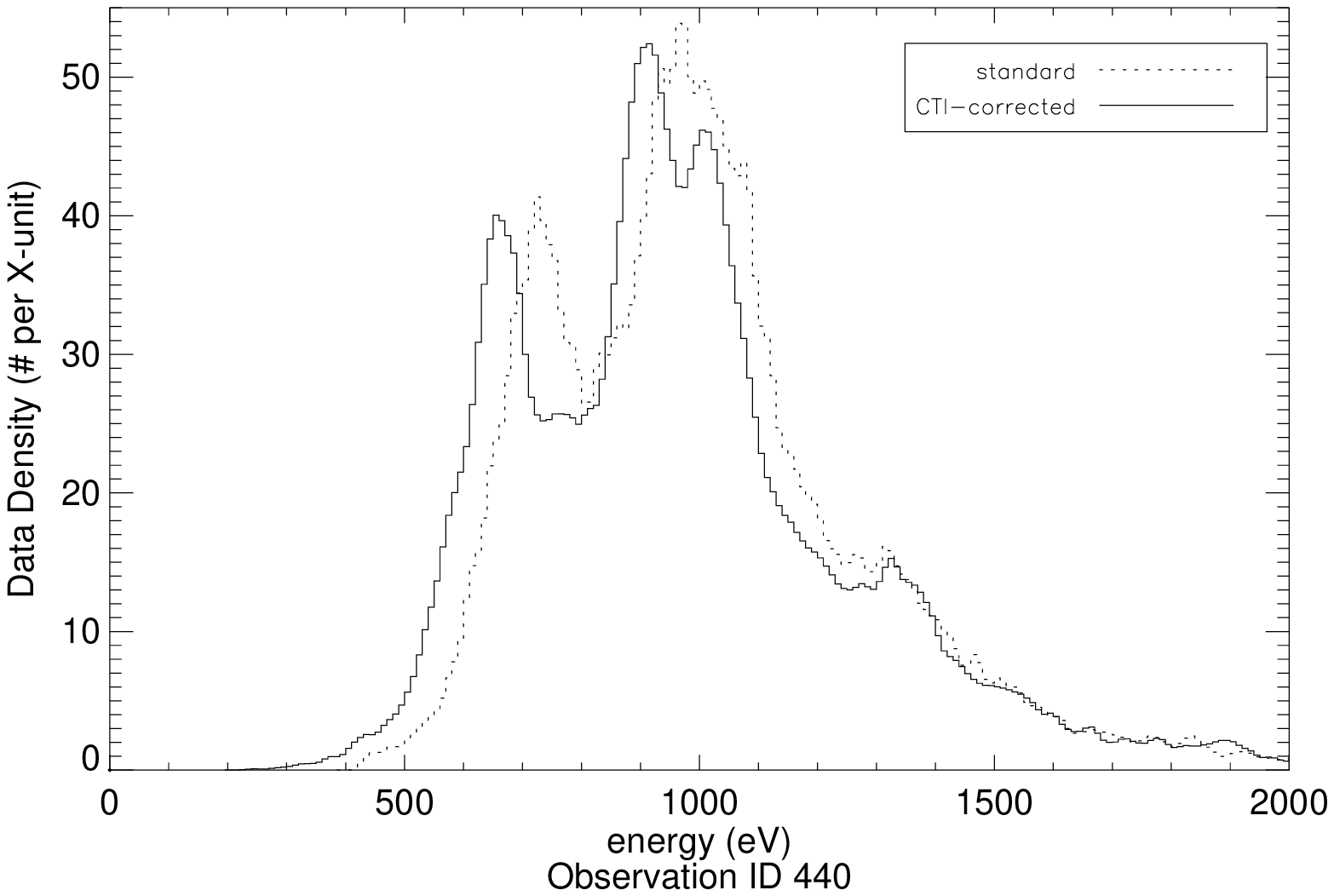,width=3.25in }
	\epsfig{file=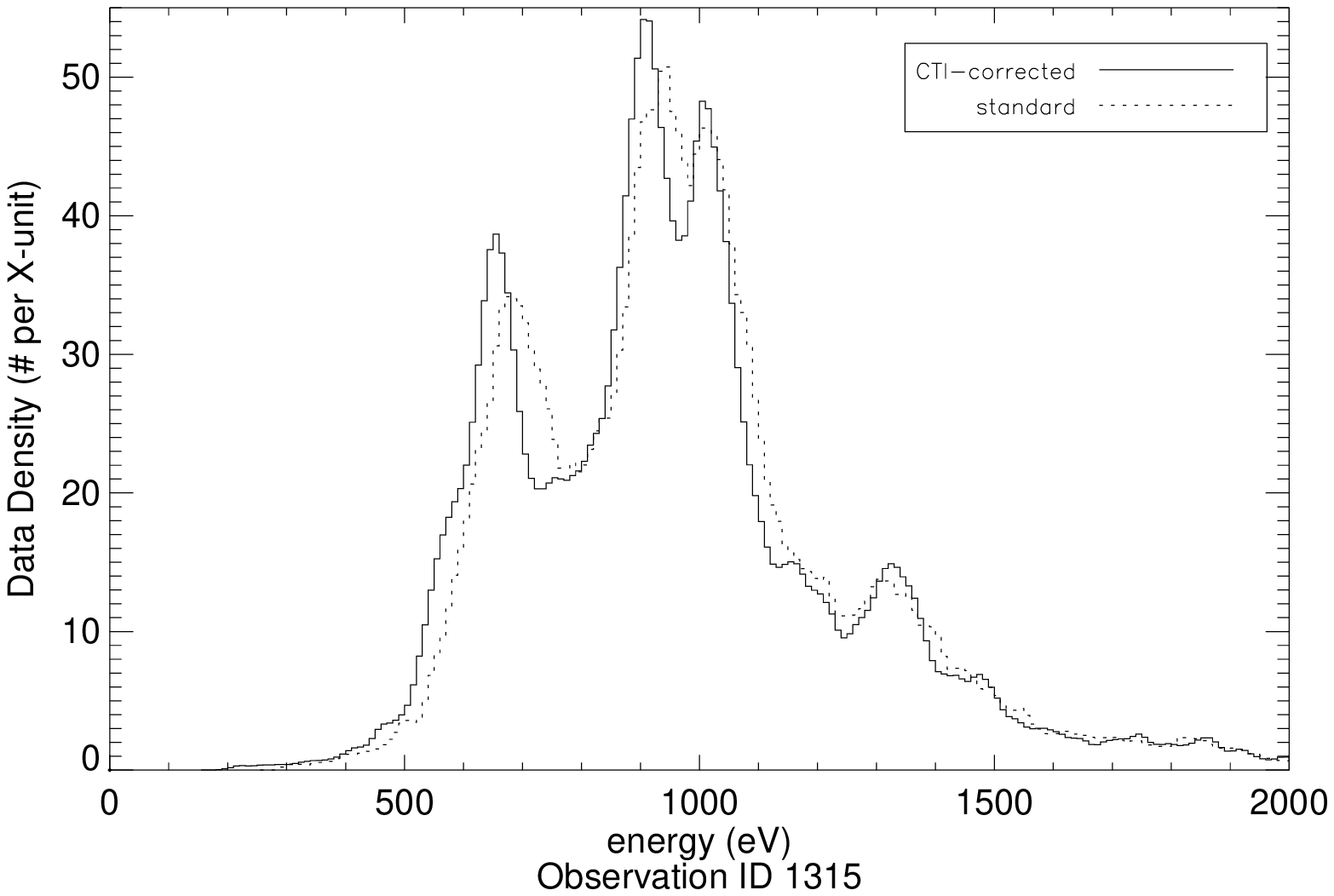,width=3.25in } }}
\centerline{\epsfig{file=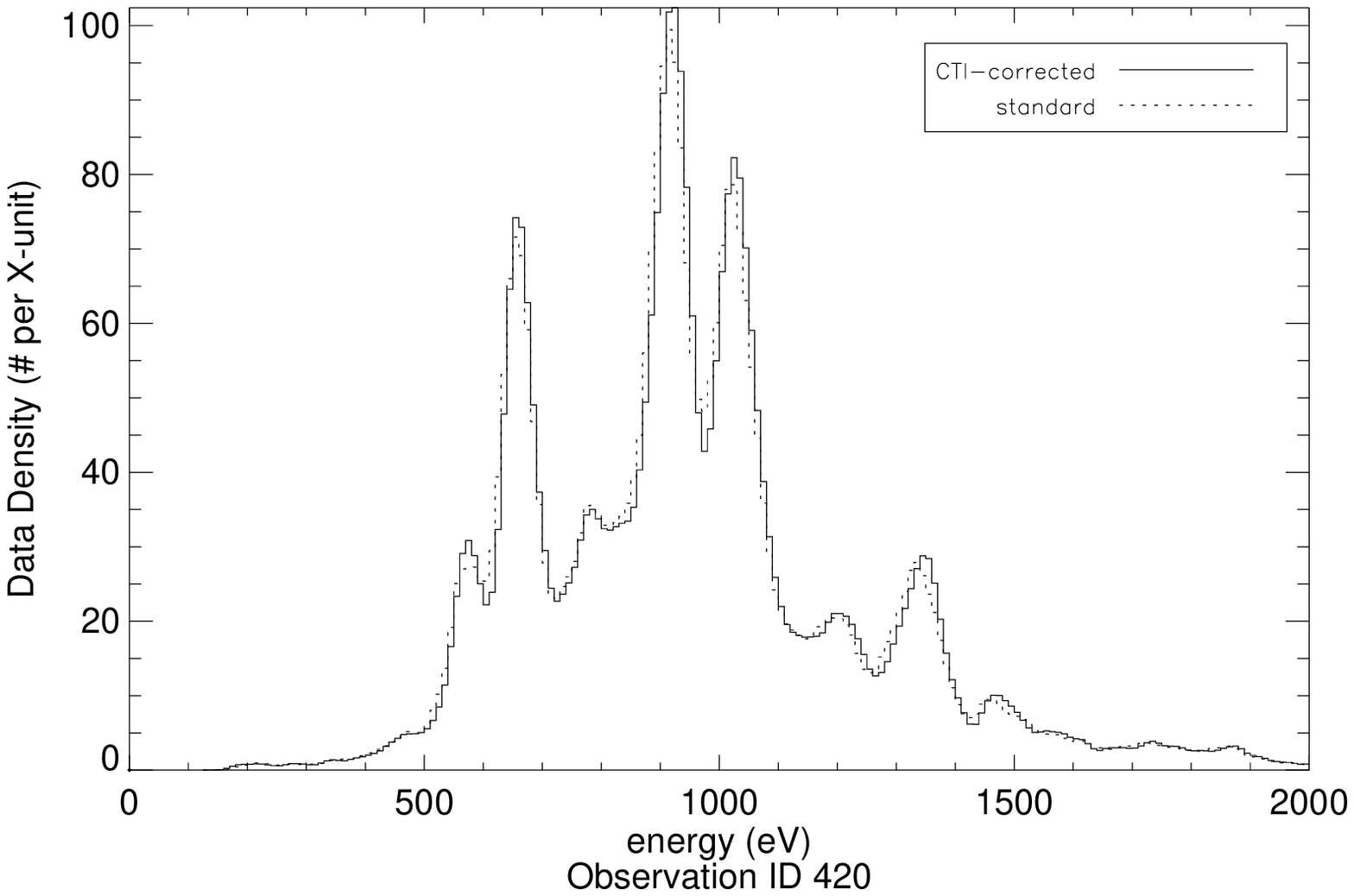,width=3.25in } } 
\caption{\protect \small E0102 spectra on I3 at $-$120C, comparing
standard processing with CTI-corrected data at several off-axis
positions.  Observation 440:  0.5 arcmin off-axis; 1315:  4 arcmin
off-axis; 420:  7 arcmin off-axis.  }

%all g02346 filtered, binsize 10 eV, gaussian smoothing of width 1 bin
 
\normalsize
\label{fig:I3-e0102}
\end{figure}

Since E0102 has been observed at several off-axis positions on I3, we
can compare its standard and CTI-corrected spectra as a function of the
position-dependent spectral resolution.  Figure~\ref{fig:I3-e0102}
shows these spectra, with the spectral resolution increasing with
off-axis angle (the I-array aimpoint is at the top of the I3 chip,
where CTI-induced spectral resolution degradation is at its worst).
All spectra used ASCA g02346 grade filtering, a binsize of 10~eV, and
were smoothed with a Gaussian of width 1 bin. 

%-------------------------------------------------------------------------  
%-------------------------------------------------------------------------

Near the aimpoint, the CTI corrector changes the low-energy gain
substantially and improves the spectral resolution.  At 4 arcmin
off-axis (about half-way up the I3 chip) the spectral lines are
noticeably narrower.  The CTI corrector applies a more modest gain
change and still improves the spectral resolution compared to standard
processing.  At 7 arcmin off-axis (only 100 rows from the bottom of the
chip, where CTI is minimal), standard and CTI-corrected spectra are
very similar in gain and spectral resolution.  The line energies
between the three observations are comparable in CTI-corrected data.

\clearpage

%==========================================================================
\section{Summary}
%==========================================================================

We have developed a phenomenological Monte Carlo model of CTI in both
front- and back-illuminated ACIS CCDs and used it to devise a CTI
corrector for ACIS observations of celestial X-ray sources.  The PSU
CTI corrector has been tuned for the ACIS CCDs most often used for
imaging:  I0--I3, S2, and S3.  This tuning uses a composite dataset of
observations of the ACIS External Calibration Source.  Both major focal
plane temperatures ($-$110C, September 1999 -- 29 January 2000, and $-$120C,
29 January 2000 -- present) are calibrated.

The CCD simulator and CTI corrector IDL source code are provided to
the community\footnote{http://www.astro.psu.edu/users/townsley/cti/}.
Also available are the parameter files that instantiate the simulator
and corrector tuning for each device, the modified QEU files, and
a suite of response matrices for use with CTI-corrected data.

The corrector works on a Level 1 event list from ACIS observations
using the full-frame readout mode.  The $3 \times 3$ or $5 \times 5$
pixel event islands must be included in order for the corrector to
modify the event grade.  The user is then required to perform the other
filtering steps (grade filtering, applying Good Time Intervals, etc.)
to recover the CTI-corrected Level 2 event list.

Refinements of the model since the original report of this work (in
early 2000) \cite{townsley00} have led to several improvements in CTI-corrected
data:
\begin{itemize}
\item Modeling the charge loss per pixel transfer as three lines improves
the fidelity of corrected event amplitudes at high energies without compromising the low-energy results.  
\item Removing column-to-column gain variations (to the degree possible with
this model) improves the energy resolution on both FI and BI CCDs.
\item Including charge trailing and shielding within a single event 
makes it possible to compensate for CTI-induced grade morphing
at all event locations.
\item Using a more extensive set of calibration observations of the soft
SNR E0102-72.3, we have improved the model tuning in the spectral range
0.5--1.5~keV.
\end{itemize}

Since this corrector does not account for precursor events, we cannot
totally eliminate the row-dependent energy resolution of the FI
devices.  Thus position-dependent response matrices are still necessary
for accurate FI spectral fitting.  By regularizing the gain and grade
distributions across these CCDs, though, we have reduced the number of
matrices needed for typical analysis tasks to fewer than 10.  For BI
devices, the improvement is even better; since S3 does not exhibit
position-dependent energy resolution, a single S3 response matrix is
usually adequate for CTI-corrected data.  This is a great
simplification for users, especially those working on extended targets
such as supernova remnants or clusters of galaxies.

Lowering the ACIS focal plane temperature to
$-$120C achieved dramatic reduction in the effects of CTI on FI devices.
This reduced the FI spectral resolution degradation and its
row-dependence, moderated the gain changes across the device, and
reduced the charge trailing, resulting in much less high-energy quantum
efficiency loss at the top of FI devices.  Since a large fraction of
ACIS data were obtained at $-$120C, we have concentrated on those data in
this paper.  It is important to remember, however, that the early days
of the mission included observations of some very exciting targets and
those data are now publicly available.  So for the sake of archival
research, we have invested substantial effort in tuning the CTI
corrector at $-$110C.

The CTI corrector has been used for a variety of {\em Chandra} investigations,
including:
\begin{itemize}
\item Resolving the X-ray background emission in the {\em Chandra} Deep Field North,
{\em e.g.} \cite{hornschemeier00}, \cite{hornschemeier01}, \cite{brandt01}
\item Star formation regions, {\em e.g.} \cite{tsuboi01}, \cite{garmire00}
\item Supernova remnants, {\em e.g.} \cite{burrows00}
\item The starburst galaxy M82 \cite{griffiths00}
\item A search for low-luminosity active galactic nuclei \cite{ho01}.
\end{itemize}
All parameter files, source code, and instructions comprising the PSU
CTI corrector and the PSU CCD simulator are available on the Web\footnote{http://www.astro.psu.edu/users/townsley/cti/}.
We encourage the use of these tools and welcome feedback. 

Other software mitigation techniques for CTI are under development at
the {\em Chandra} X-ray Center and at MIT, with support by {\em
Chandra} Project Science at Marshall Space Flight Center and the rest
of the {\em Chandra} team.  The MIT/ACIS group also is working to
devise hardware mitigation techniques that may further reduce the
effects of the radiation damage.  Prigozhin {\em et al.}
\cite{prigozhin00a}, \cite{prigozhin00b} are developing a more physical
model of the charge traps that is likely to replace this
phenomenological model.  We are confident that these efforts will
eventually lead to a more robust CTI correction scheme for ACIS data.

%==========================================================================
\ack
Financial support for this effort was provided by NASA contract
NAS8-38252 to G.~P.~G., the ACIS Principal Investigator.  This work
made use of the NASA Astrophysics Data System.  We appreciate the time
and gracious response of the anonymous referee.

The External Calibration Source data were provided by Allyn Tennant ($-$110C)
and the CXC ($-$120C).  We are very grateful for their efforts to organize
and compile this large body of information.

The extensive data analysis necessary to tune the CTI corrector and
simulator for each CCD and temperature would not have been practical
without the help of many members of the PSU/ACIS team.  We thank Xinyu
Dai, Sarah Gallagher, Konstantin Getman, Varsha Gupta, Ann
Hornschemeier, Bulent Kiziltan, Karen Lewis, Sangwook Park, Divas
Sanwal, Michael Sipior, and Yohko Tsuboi for their hard labor and
George Chartas and Yoshitomo Maeda for their ideas.

We thank our MIT/ACIS colleagues, especially Catherine Grant, Mark
Bautz, and Gregory Prigozhin, for many helpful discussions and
suggestions regarding CCD modeling and for their lucid explanations of
the device physics of radiation damage and charge traps.  We thank the
CXC for providing a forum for discussion of CTI and other ACIS
calibration issues and we especially thank Paul Plucinsky for his help
with the analysis of E0102-72.3 data, for testing our data products,
and for reminding us, now many years ago, that CTI was a force to
reckon with.
%==========================================================================

\bibliography{xraysim}
\bibliographystyle{elsart-num}

\end{document}